\documentclass[pra,aps,citeautoscript,twocolumn]{revtex4-2}

\usepackage{physics}
\usepackage{amsfonts}
\usepackage{subcaption}
\usepackage{tikz}
\usetikzlibrary{calc,positioning}
\usepackage{quantikz}
\usepackage{amsmath}
\usepackage{amssymb}
\usepackage{bm}
\usepackage{hyperref}

\usepackage[version=3]{mhchem}
\usepackage[left=1.5cm, right=1.5cm, top=1.785cm, bottom=2.0cm]{geometry}
\usepackage{mathptmx}
\usepackage{graphicx}
\usepackage[format=plain,justification=justified,singlelinecheck=false,font={stretch=1.125,small,sf},labelfont=bf,labelsep=space]{caption}
\usepackage{float}
\usepackage[english]{babel}

\usepackage[left=1.5cm, right=1.5cm, top=1.785cm, bottom=2.0cm]{geometry}

\usepackage{array}
\usepackage[T1]{fontenc}
\usepackage[utf8]{inputenc} 
\usepackage[english]{babel}
\usepackage{graphicx}
\definecolor{cream}{RGB}{222,217,201}

\begin{document}
\title{Unified adiabatic and diabatic excited-state description \emph{via} the ensemble-variational quantum eigensolver}

\author{Christophe Soule$^a$}
\author{Bruno Senjean$^a$}
\author{Benjamin Lasorne$^{a}$}
\email{benjamin.lasorne@umontpellier.fr}
\affiliation{
$^a$
ICGM, Univ Montpellier, CNRS, ENSCM, Montpellier, France
}

\begin{abstract}
Within the present noisy intermediate-scale quantum-computing era, hybrid quantum-classical-processor algorithms have emerged as promising avenues for tackling electronic-structure eigenproblems.
Among them, the so-called ensemble-variational quantum eigensolver has been designed to treat ground and excited states on an equal footing and proven effective in capturing features such as conical intersections and avoided crossings between two electronic states, as we recently demonstrated for formaldimine. 
We also showed on that example how the underlying ensemble-variational principle was prone to provide a quasi-diabatic representation “for free”. 
To date, this method has been limited to computing only two eigenstates of a Hamiltonian.
The aim of the present paper is to show how and under what conditions this can be generalized to models that involve three coupled electronic states or more. 
Our approach relies on designing a parameterized basis transformation that can directly be implemented on a quantum computer for further post-treatment. 
This nontrivial step is accompanied by the development of quantum circuits specifically adapted to the several states of interest. 
An algebraic optimization strategy for the parameters of the basis transformation is formulated to obtain the target eigenstates as well as the optimally diabatic states under various objective flavors of the ensemble-variational principle. Our approach was tested for addressing the first three coupled electronic states of the H$_4^+$ molecular ion as a proof of principle, with three electrons in four spatial orbitals, along various geometries.
\end{abstract}

\maketitle

\section{Introduction}

Non-radiative photochemical processes occur in both natural systems, such as animal vision~\cite{johnson_local_2015} and the photostability of nucleic acids following UV absorption~\cite{improta_quantum_2016} as well as in artificial systems, for instance in the case of photoswitches~\cite{berkovic_spiropyrans_2000}. During these ultrafast processes, typically occurring on the timescale of tens to hundreds of femtoseconds, non-radiative transitions can arise in systems exhibiting strong nonadiabatic couplings (NACs) between several adiabatic electronic states. NACs describe the coupling between the electronic and nuclear motion~\cite{domcke_conical_2004}. When these coupling become large, the Born--Oppenheimer approximation (BOA) is no longer valid, and we require an approach that goes beyond it~\cite{domcke_conical_2004, baer_beyond_2006}.

The study of molecular dynamics provides valuable insight into these processes and helps achieve a deeper understanding of their underlying mechanisms. In principle, a faithful description of their dynamics requires a full quantum treatment of both electronic and nuclear degrees of freedom, as electron–nuclear dynamics can evolve on comparable energy and timescales in ultrafast regimes~\cite{gatti_applications_2017}. A fully quantum framework defines the field of `molecular quantum dynamics', which is based on the vibronic time-dependent Schrödinger equation (TDSE). The numerical solution of this equation has led to the development of various quantum dynamical methods, including wavepacket propagation techniques such as the `multiconfiguration time-dependent Hartree' (MCTDH) method~\cite{meyer_multi-configurational_1990, beck_multiconfiguration_2000, worth_quantics_2020}. In parallel, and with the aim of capturing the essential features of the dynamics of these processes at reduced computational cost, mixed quantum–classical approaches have been developed. Among these approaches, Tully’s 'trajectory surface hopping' (TSH)~\cite{tully_molecular_1990} is one of the most widely used and extensively studied methods~\cite{subotnik_understanding_2016, barbatti_newton-x_2022}. It is based on an ensemble of classical nuclear trajectories evolving on potential energy surfaces (PESs), with stochastic transitions between electronic states, and a quantum treatment of the electronic degrees of freedom.

Each of these families of approaches is connected in practice to the electronic-structure problem, requiring the evaluation of electronic properties as functions of the nuclear geometry. The solution of the electronic-structure problem provides the eigenstates of the electronic Hamiltonian at each nuclear geometry, thereby defining the adiabatic representation. However, when two adiabatic electronic states approach degeneracy at a given nuclear geometry, as occurs near conical intersections, which play a central role in non-radiative processes, the NACs can become very large or formally singular. This behavior has consequences for the formulation and implementation of nonadiabatic dynamics methods, depending on the nature of the underlying approach.

TSH, which adopts a pointwise description of nuclear motion is most often implemented in the adiabatic representation~\cite{tully_mixed_1998, mai_nonadiabatic_2018}, where nuclear dynamics is driven by the gradients of the PES, while NACs between electronic states govern population transfer. In this framework, these couplings enter as transition amplitudes along individual trajectories. In practice, regions where they become singular are rarely encountered along most trajectories and therefore do not pose a significant problem for the overall dynamics. Nevertheless, various methodological developments have been proposed to improve the robustness of surface hopping in such situations~\cite{crespo-otero_recent_2018}. In contrast, wavepacket-based methods are fully quantum and inherently non-local with respect to the nuclear coordinates. They therefore involve the evaluation of multidimensional Hamiltonian matrix elements over nuclear wavefunctions defined in the full nuclear coordinate space.  As a result, operators that are singular or rapidly varying with respect to nuclear coordinates may lead to significant numerical difficulties in the evaluation of local derivatives and multidimensional integrals. This provides a key rationale for the fact that the adiabatic representation is often not the most convenient choice in these approaches, and has therefore led to a preference to use so-called (quasi)diabatic representations.

In principle, a diabatic representation corresponds to an electronic basis in which the NACs vanish. A possible way to construct such a representation consists in applying a geometry-dependent unitary transformation of the electronic basis, commonly referred to as the adiabatic-to-diabatic transformation. However, such a transformation is not unique and, in general, cannot be defined globally over the full nuclear configuration space~\cite{baer_beyond_2006}. This has led to the development of approximate schemes~\cite{richings_practical_2015,vie21:084302, zhang_diabatization_2021,shu_diabatic_2022}, commonly referred to as quasi-diabatic representations, in which the NACs are minimized rather than made to vanish.

Recently, some of the present authors have shown that
a quantum algorithm, the so-called `state-average orbital-optimized variational quantum eigensolver' (SA-OO-VQE)~\cite{yalouz_state-averaged_2021,yalouz_analytical_2022,beseda2024state}, had a natural propensity to achieve
the adiabatic-to-diabatic transformation ``for free''~\cite{illesova_transformation-free_2025}.
Excited-state quantum algorithms
have seen growing interest since the last few years, with developments
based on variational deflation approaches~\cite{higgott2019variational,jones2019variational,jouzdani2019method,ibe2020calculating,chan2021molecular,quiroga2025quantum},
quantum subspace methods~\cite{motta2024subspace,ruiz2022accessing,cortes2022quantum,de2026new,gentinetta2026quantum,tkachenko2024quantum,berthusen2024multi,motta2020determining,ollitrault2020quantum,kim2023two,asthana2023quantum,bhatia2026quantum,mcclean2017hybrid,colless2018computation,castellanos2024quantum,motta2023quantum,akande2025variational,klymko2022real}, and others~\cite{wang2023electronic,wang2024quantum,cadi2024folded,endo2020calculation,yoshikura2023quantum,nykanen2024toward,tilly2020computation,xie2022orthogonal,klymko2022real,shen2023real,parrish2019quantumfilter,cohn2021quantum,stroschein2025ground}.
Extensions to the calculations of NACs and analytical gradients have also been derived~\cite{yalouz_analytical_2022,tamiya2021calculating,omiya2022analytical,obrien2022efficient}, thus enabling the implementation of
hybrid quantum-classical (in the sense of quantum and classical computers) nonadiabatic molecular dynamics~\cite{koridon2024hybrid,hirai2022non,gil2025sharc,gandon2024nonadiabatic,li2026efficient,sangiogo2025nonadiabatic,bultrini2023mixed}.
However,
they all focus on the preparation of
adiabatic states, in contrast to Ref.~\citenum{illesova_transformation-free_2025},
where almost optimal quasi-diabatic states were
obtained for the two-state formaldimine molecule.
In this work, we improve the robustness of our
algorithm to obtain optimal quasi-diabatic states, by combining our rotational-procedure to obtain adiabatic states in Ref.~\citenum{yalouz_analytical_2022} with the minimization of the diabatic-descriptor derived in Ref.~\citenum{illesova_transformation-free_2025}.
A three-state case of the H$_4^+$ molecule is studied within the `full configuration interaction' (FCI) picture in a minimal basis.
As no active space is considered, orbital optimization is not needed and we only
focus on the state-average VQE method, also called ensemble-VQE.
The paper is organised as follows.
First, we provide the concepts and methods in Sec.~\ref{section:methods},
where we describe the multi-state electronic-structure problem (Sec.~\ref{subsec:objectives}) and
the ensemble-VQE approach (Sec.~\ref{subsec:evqe}).
The derivations to prepare the adiabatic states and the optimal quasi-diabatic ones are
explained in Secs.~\ref{subsec:eigenstates}
and \ref{subsec:opt_diab}, respectively.
Computational details are given in Sec.~\ref{sec:comp_det}, where
the model system is described together with the details of the implementation.
Finally, results are shown
in Sec.~\ref{sec:results}, with successful preparation of the adiabatic and optimal quasi-diabatic states in Secs.~\ref{subsec:eigenstates_results} and \ref{subsec:diabatic_results}, respectively.
We conclude in Sec.~\ref{sec:conclu}.

\section{Concepts and Methods}\label{section:methods}

\subsection{General objective: the multi-state electronic-structure problem}\label{subsec:objectives}

For any set of nuclear coordinates, denoted by \(\bm{R}\), the electronic Hamiltonian, \(\hat{H}^{\mathrm{el}}[\bm{R}]\), is simply defined upon removing the nuclear kinetic-energy operator from the first-quantized molecular Hamiltonian.
It can be viewed as depending locally on \(\bm{R}\), which thus plays the role of an external parameter.
In the context of this work, the internuclear Coulomb repulsion, $h_\mathrm{nuc}(\bm{R})$, is included within \(\hat{H}^{\mathrm{el}}[\bm{R}]\).
For any value of \(\bm{R}\), and provided a one-body basis of spatial orbitals, $\lbrace \bm{r} \mapsto \varphi_p(\bm{r} ; \bm{R})\rbrace $, where $\bm{r}$ denote the electron coordinates, this Hamiltonian, within the nonrelativistic, spin-free, second-quantization framework, is expressed as~\cite{helgaker_molecular_2000}
\begin{equation}
\begin{aligned}
\label{eq:hamiltonian_second_quantization}
\hat{H}^{\mathrm{el}}[\bm{R}] = &
\sum_{pq} h_{pq}(\bm{R}) \, \hat{E}_{pq}
 + \frac{1}{2} \sum_{pqrs} g_{pqrs}(\bm{R}) \, \hat{e}_{pqrs} 
+ h_\mathrm{nuc}(\bm{R}) \, \hat{I} \quad,
\end{aligned}
\end{equation}
where \(h_{pq}(\bm{R})\) are the one-electron integrals, including the kinetic energy of the electrons and their Coulomb attraction to the nuclei, and \( g_{pqrs}(\bm{R}) \) are the two-electron integrals, describing the Coulomb repulsion between electrons.
The one- and two-body spin-free excitation operators are defined as \(\hat{E}_{pq} = \sum_\sigma \hat{a}^\dagger_{p\sigma} \hat{a}_{q\sigma}\) and \(\hat{e}_{pqrs} = \sum_{\sigma,\tau} \hat{a}^\dagger_{p\sigma} \hat{a}^\dagger_{r\tau} \hat{a}_{s\tau} \hat{a}_{q\sigma}\), where \(\hat{a}^\dagger_{p\sigma}\) (\(\hat{a}_{p\sigma}\)) denotes the creation (annihilation) operator of an electron with spin \(\sigma\) in spatial orbital \(p\). \(\hat{I}\) corresponds to the electronic identity operator.

Within this framework, solving the electronic-structure problem consists in finding the eigensolutions, \(\lbrace  |\Psi_n ; \bm{R}\rangle \rbrace _{n \geq 0}\) and \(\lbrace  E_n(\bm{R}) \rbrace _{n \geq 0}\), to the electronic time-independent Schrödinger equation (TISE) at any \(\bm{R}\),
\begin{equation}
\label{eq:electronic_schrodinger}
\hat{H}^{\mathrm{el}}[\bm{R}]\,|\Psi_n ; \bm{R}\rangle
= E_n(\bm{R})\, |\Psi_n ; \bm{R}\rangle \quad.
\end{equation}

The so-called adiabatic electronic eigenstates, \(|\Psi_n ; \bm{R}\rangle\), thus depend parametrically on the nuclear coordinates \(\bm{R}\) and their corresponding eigenvalues, \(E_n(\bm{R})\), ordered according to \(E_n(\bm{R}) \leq E_{n+1}(\bm{R})\), define continuous functions of \(\bm{R}\). 
Namely, the functions \(\bm{R} \mapsto E_n(\bm{R})\) are to be identified to the adiabatic PESs for the motion of the nuclei within the Born--Oppenheimer picture.

Let us recall here some subtlety that is not as innocent as it may seem when translating formal theory into practical algorithms.
The second-quantized Hamiltonian given in Eq.~\eqref{eq:hamiltonian_second_quantization} is a faithful representation of the first-quantized one if and only if the one-body orbital basis is complete (say, within the `complete basis set' -- CBS -- limit) and if and only if we restrict ourselves to the corresponding $N_\mathrm{el}$-electron Hilbert sector within the total second-quantized fermionic Fock space of spin-space antisymmetric many-body wavefunctions (note that the value of $N_\mathrm{el}$ is preserved by the Hamiltonian, which provides a block-structured representation).

Since, in practice, we can only work with a finite basis of \(N_\mathrm{orb}\) spatial orbitals,
\( \lbrace \varphi_p(\bm{r} ; \bm{R})\rbrace _{p=1}^{N_\mathrm{orb}}\), this implies that, for a given value of \(N_\mathrm{orb}\), giving rise to \(N = 2 N_\mathrm{orb}\) spin-orbitals (using a spin-restricted picture), and a given value of \(N_\mathrm{el}\), we can form up to $N_{\mathrm{det}} = \binom{N}{N_{\mathrm{el}}}$ Slater determinants within the $2^N$-dimensional fermionic Fock space (where $\sum_{N_{\mathrm{el}}=0}^N \binom{N}{N_{\mathrm{el}}} = 2^N$).
Upon projection onto the finite span of the spin-restricted Slater determinants corresponding to $N = 2 N_\mathrm{orb}$ and $N_\mathrm{el}$, both first- and second-quantized approaches share the same approximate eigensolutions within the FCI approach.
They will be considered in what follows as ``variationally exact'' within the $N_\mathrm{det}$-dimensional FCI Hilbert space.

\subsection{Method: the ensemble-variational quantum eigensolver}
\label{subsec:evqe}

Current quantum computers belong to the `noisy intermediate-scale quantum' (NISQ) era~\cite{preskill_quantum_2018}, which is characterized by their susceptibility to various sources of noise that can affect computational accuracy, especially as quantum circuit depth increases. As a result, fully quantum algorithms such as `quantum phase estimation' (QPE)~\cite{aspuru-guzik_simulated_2005}, which are in principle capable of solving the electronic-structure problem, remain limited in practice due to the large circuit depths that they require. In this context, hybrid quantum-classical variational algorithms, such as the 'variational quantum eigensolver' (VQE)~\cite{peruzzo_variational_2014}, have emerged as particularly attractive approaches, as they enable accurate results while requiring shallower quantum circuits~\cite{kandala_hardware-efficient_2017}.
Such a hybrid approach leverages the relative strengths of both quantum and classical computing: the quantum device avoids the classical limitation of storing an exponentially-large parameterized wavefunction, allowing the exploration of a broader Hilbert space (the $2^N$-dimensional fermionic Fock space being spanned by $N$ qubits only), while the parameter optimization is efficiently handled classically. However, the parameter landscape in VQE often suffers from local minima and barren plateaus, which may sometimes undermine any expected potential quantum advantage~\cite{bittel2021training,larocca2025barren,cerezo2025does}.

VQE was originally developed to compute the ground state of molecular systems. The ensemble-VQE method extends VQE~\cite{nakanishi_subspace-search_2019} within a state-averaged framework, allowing the simultaneous calculation of multiple electronic states by treating them on an equal footing. The present work focuses on generalizing the approach to the particular three-state case. 
For simplicity, the explicit dependence on \(\bm{R}\) is omitted from now on, as the method is applied separately for each molecular geometry.

\paragraph{Workflow of ensemble-VQE.}
Starting from a numerical one-body basis of real `atomic orbitals' (AOs) \(\lbrace \chi_\mu(\bm{r})\rbrace _{\mu=1}^{N_\mathrm{orb}}\), 
a set of `molecular orbitals' (MOs) \( \lbrace \varphi_p(\bm{r})\rbrace _{p=1}^{N_\mathrm{orb}}\) is constructed as `linear combinations of atomic orbitals' (LCAO). 
Within an FCI approach (our focus herein), the LCAO coefficients are typically obtained from an initial Hartree--Fock calculation, and the resulting canonical MOs are then used to represent the electronic Hamiltonian \(\hat{H}^\mathrm{el}\) in second quantization (see Eq.~\eqref{eq:hamiltonian_second_quantization}). 
Yet, any unitary change of MO basis is perfectly acceptable as well, since the FCI eigensolutions are gauge-invariant (we shall come back to this aspect later on).
The algorithm can be divided into the following steps:

{\it 1. Defining the qubit-based representations of the electronic Hamiltonian and of the  Slater determinants.}
To implement the electronic Hamiltonian \(\hat{H}^\mathrm{el}\) within a hybrid quantum-classical approach, the Jordan--Wigner transformation is applied to map the second-quantized  Hamiltonian onto the qubit space, where it is usually written as a `linear combination of unitaries' (LCU); namely, Pauli strings~\cite{jordan1928p}.
In this, the $N_\mathrm{det}$ Slater determinants with $N_\mathrm{el}$ electrons in $N$ spin-orbitals are one-to-one mapped to ordered $N$-lists made of $N - N_\mathrm{el}$ zeros (unoccupied spin-orbitals) and $N_\mathrm{el}$ ones (occupied orbitals), referred to as bitstrings or 'occupation number vectors' (ONVs) and physically realized as a subset of the span of the $2^N$ Fock states of an $N$-qubit register: $\ket{11100\cdots0}$, $\ket{11010\cdots0}$, $\ket{10110\cdots0}$, etc. 

{\it 2. Initialization.}
The second step of the method consists in preparing three orthonormal initial states on the quantum computer, denoted \( \ket{\Phi^0_{\mathrm {A}}} \), \( \ket{\Phi^0_{ \mathrm {B}}} \), and \( \ket{\Phi^0_{ \mathrm {C}}}\) (the model, or ``guess'').
Each of them correspond to a single Slater determinant (encoded as a basic ONV).
This can be generalized to `configuration state functions' (CSFs) of given total spin $S$.
The idea behind is to select chemically intuitive electronic configurations (according to the \emph{Aufbau} principle), that are expected to have dominant contributions in the first three targeted eigenstates of  \(\hat{H}^\mathrm{el}\), \(\ket{\Psi_0}\), \(\ket{\Psi_1}\), and \(\ket{\Psi_2}\). Thus, the number of different quantum circuits corresponds to the number of states of the ensemble, here three.

{\it 3. State preparation.}
A parameterized unitary operator, \(\hat{U}(\bm{t})\), referred to as the `ansatz' is then applied to each of these initial states, leading to multideterminantal states, \(\ket{\Phi_{\mathrm {A}}(\bm{t})} = \hat{U}(\bm{t}) \ket{\Phi^0_{\mathrm {A}}} \), \(\ket{\Phi_{\mathrm {B}}(\bm{t})} = \hat{U}(\bm{t})\ket{\Phi^0_{\mathrm {B}}} \), and \(\ket{\Phi_{\mathrm {C}}(\bm{t})} = \hat{U}(\bm{t})  \ket{\Phi^0_{\mathrm {C}}} \), that depend on the ensemble-variational parameters \(\bm{t}\). 
Here, we used as an ansatz the Trotterized -- also called disentangled -- `generalized unitary coupled-cluster single double' (GUCCSD)~\cite{lee_generalized_2019}.
It must be understood that $\hat{U}(\bm{t})$ should allow us, in principle, to navigate through the full span of the numerically available FCI Hilbert space (assuming a perfectly-expressible ansatz).
In addition, the fact that it is defined from the onset as a unitary operator ensures that the resulting states, \(\ket{\Phi_{\mathrm {A}}(\bm{t})}\), \(\ket{\Phi_{\mathrm {B}}(\bm{t})}\), and \(\ket{\Phi_{\mathrm {C}}(\bm{t})}\), remain orthonormal at all stages of the optimization process (\emph{i.e.}, for any value of $\bm{t}$).

{\it 4. Quantum measurements.}
To estimate the expectation values of the electronic Hamiltonian \( \hat{H}^{\mathrm{ el}}\) with respect to the prepared states, denoted \( H_{\mathrm {AA}}(\bm{t}) = \bra{\Phi_{\mathrm A}(\bm{t})} \hat{H}^{\mathrm{ el}}\ket{\Phi_{\mathrm A}(\bm{t})} \),
\(H_{\mathrm {BB}}(\bm{t}) = \bra{\Phi_{\mathrm {B}}(\bm{t})} \hat{H}^{\mathrm{ el}}\ket{\Phi_{\mathrm {B}}(\bm{t})}\), and \( H_{\mathrm {CC}}(\bm{t}) = \bra{\Phi_{\mathrm {C}}(\bm{t})} \hat{H}^{\mathrm{ el}}\ket{\Phi_{\mathrm {C}}(\bm{t})}\), respectively,
statistical sampling is performed through a series of shots on the quantum computer (or deterministically using a classical state-vector emulator, as used in the present work).
    
{\it 5. Classical optimization.}
The previous steps are intended to have been performed on a quantum computer for a given set of parameters \(\bm{t}\) provided by a classical computer.
Herein, the parameters have been initialised to zero. Then, in order to determine the subspace of minimal ensemble-energy, we invoke the ensemble extension of the Rayleigh--Ritz variational principle~\cite{gross_rayleigh-ritz_1988,A_K_Theophilou_1979},
\begin{equation}
    \label{eq:ensemble_energy_vp}
     E^{\mathrm{ens}}(\bm{t}) = H_{\mathrm {AA}}(\bm{t})+ H_{\mathrm {BB}}(\bm{t}) + H_{\mathrm {CC}}(\bm{t}) \geq E_0+  E_1 +  E_2 \quad,
\end{equation}
such that
\begin{equation}
    \label{eq:ensemble_energy_min}
     E^{\mathrm{ens}}(\bm{t}^{*})  = \min_{\bm{t}} \lbrace E^{\mathrm{ens}}(\bm{t})  \rbrace\nonumber \quad.
\end{equation}
In this, \(E_0\), \(E_1\), and \(E_2\) are the, supposedly-exact, lowest three eigenvalues of the Hamiltonian \(\hat{H}^{\mathrm {el}}\).
Within the present context (FCI; see discussion above), we can assume that they are numerically accessible in principle, which implies that the equality can be reached: $E^{\mathrm{ens}}(\bm{t}^{*}) = E_0 + E_1 + E_2$, and it is our actual objective at stationarity.
In more general situations, the exact solution only serves as a lower bound for the optimization process driven by the variational principle.
In any case, the minimization of the functional \(E^{\mathrm{ens}}(\bm{t})\) is performed using a classical optimization algorithm. 
To perform each iteration, the quantum algorithm evaluates the function \(E^{\mathrm{ens}}(\bm{t})\) for new sets of parameters that are classically updated. 
These parameters are then used to prepare the corresponding parameterized quantum circuits, and the process is repeated until ensemble-VQE convergence is reached at \(\bm{t}^*\).

\paragraph{A note on classical \emph{vs.} quantum computing.}
It must be understood that, in the above setting, the ensemble-energy minimization procedure essentially aims at mimicking the result of a $(3 \times 3)$-block-diagonalization of the many-body Hamiltonian $(N_\mathrm{det} \times N_\mathrm{det})$-matrix.
Indeed, the minimum of the objective function, $E^{\mathrm{ens}}(\bm{t}^{*})$, is simply the subtrace of the target block-matrix, and is gauge-invariant through any arbitrary unitary/orthogonal transformation within the target subspace.
Hence, this produces, from a model subspace of states, \(\lbrace \ket{\Phi^0_I}\rbrace _{I \in \lbrace \mathrm {A},\mathrm {B},\mathrm {C}\rbrace }\), an optimal set of states, \(\lbrace \ket{\Phi_I(\bm{t}^*)}\rbrace _{I \in \lbrace \mathrm {A},\mathrm {B},\mathrm {C}\rbrace }\), that span the eigensubspace of minimal ensemble-energy, denoted \(\mathcal{V}_{*}^{\mathrm{ens}}\), without being, \textit{a priori}, detailed eigenstates of \(\hat{H}^{\mathrm {el}}\).
They are only so up to some additional unitary change-of-basis $(3 \times 3)$-matrix (post- or retro-variational objectives; see Sec.~\ref{subsubsec:transformation}).

The states \(\lbrace \ket{\Phi_I(\bm{t})}\rbrace _{I \in \lbrace \mathrm {A},\mathrm {B},\mathrm {C}\rbrace }\) that are prepared on the quantum computer by the unitary operator \(\hat{U}(\bm{t})\) from the ``guess'' are to be viewed as implicit CI expansions over the numerical basis of Slater determinants, \(\lbrace \ket{\Phi^0_I}\rbrace _{I=1}^{N_\mathrm{det}}\), 
\begin{eqnarray}
\label{eq:u_phi0}
    \ket{\Phi_{I}(\bm{t})} &=& \hat{U}(\bm{t}) \ket{\Phi^0_I} \equiv \sum_{J=1}^{N_\mathrm{det}} O_{JI}(\bm{t}) \ket{\Phi^0_J} \quad, \nonumber\\
    O_{JI}(\bm{t}) &=& \braket{\Phi^0_J}{\Phi_{I}(\bm{t})} = \bra{\Phi^0_J}\hat{U}(\bm{t})\ket{\Phi^0_I} \quad.
\end{eqnarray}
Let us stress out that a classical-computing perspective would require the explicit evaluation of the elements of the CI-coefficient (overlap) $\bm{O}(\bm{t})$-matrix (in practice, here, the $(N_\mathrm{det} \times 3)$-block of its first three columns), while a quantum-computing approach circumvents the storage of them into memory and rather relies on implementing the direct action of the $\hat{U}(\bm{t})$-operator onto the model states (which is represented by sequences of parameterized elementary unitary gates acting on qubit registers in the quantum computer).
This probably is the most critical difference between both computational strategies.

\paragraph{Taking the spin in to account.}
Since we are not considering spin-orbit coupling herein, the eigenstates of \(\hat{H}^{\mathrm {el}}\) are also eigenstates of \(\hat{S}^2\) and \(\hat{S}_z\). 
A single Slater determinant is always an eigenstate of \(\hat{S}_z\), but it is not generally an eigenstate of \(\hat{S}^2\). 
Therefore, an arbitrary linear combination of Slater determinants is not necessarily an eigenstate of \(\hat{S}^2\) either. 
Even when we are only considering ensemble-variationally-optimal ``pre-eigenstates'', \(\lbrace \ket{\Phi_I(\bm{t}^*)}\rbrace _{I \in \lbrace \mathrm {A},\mathrm {B},\mathrm {C}\rbrace }\), we want them to be pure-spin states so that further mixing them post-variationally within the target subspace will eventually produce pure-spin eigenstates too.

Now, although a spin-adapted version of GUCCSD can conserve \(\langle\hat{S}^2\rangle \) in theory, practical implementations using Trotterization may introduce some spin contamination~\cite{tsuchimochi_spin-projection_2020}. Therefore, a penalty term defined as
\begin{equation}
\label{eq:spin_penalty}
     \lambda \sum_{I \in \lbrace \mathrm {A},\mathrm {B},\mathrm {C}\rbrace }  \left| \langle \Phi_I(\bm{t}) | \hat{S}^2 | \Phi_I(\bm{t}) \rangle - S_I(S_{I}+1) \right| \quad,
\end{equation}
where \(\lambda\) is some real positive number and  \(S_I \) is the targeted value of the spin quantum number for the state \( I \),
can be added to the objective function \(E^{\mathrm{ens}}(\bm{t})\) to promote the generation of eigenstates of the \(\hat{S}^2\) operator, as shown in the constrained VQE algorithm~\cite{ryabinkin_constrained_2019}. In our numerical experiments, setting \(\lambda = 1\) generally leads to a converged solution without the need for extensive parameter tuning. In some cases, fine-tuning \(\lambda\) may be necessary to reach convergence. As an alternative, one can minimize the objective while enforcing a constraint directly using a Lagrangian approach~\cite{Nocedal2006Numerical}, where Lagrange multipliers are used to incorporate the constraint explicitly. The `sequential least squares quadratic programming' (SLSQP) algorithm~\cite{kraft1988slsqp}, implemented in SciPy~\cite{2020SciPy-NMeth}, provides a practical way to handle constrained minimization, typically without requiring explicit penalty parameter calibration. Although SLSQP can handle an equality constraint, we observed that enforcing it numerically can reduce the accuracy of the convergence of \(E^{\mathrm{ens}}(\bm{t})\) toward the minimal ensemble-energy \(E^{\mathrm{ens}}(\bm{t}^{*})\).
Upon enforcing the constraint as an inequality instead, 
we obtained better convergence properties. We therefore define the constraint as
\begin{equation}
\label{eq:spin_constraint}
     \sum_{I \in \lbrace \mathrm {A},\mathrm {B},\mathrm {C}\rbrace }  \left| \langle \Phi_I(\bm{t}) | \hat{S}^2 | \Phi_I(\bm{t}) \rangle - S_I(S_{I}+1) \right|  \leq \epsilon_S \quad,
\end{equation}
where \(\epsilon_S\) is a small positive real number specifying the tolerance on the constraint.

\subsection{Solving for eigenstates}\label{subsec:eigenstates}

From now on, we shall use the breve symbol, $\breve{\cdots}$, over a matrix to signal that it is a $(3 \times 3)$-block within a ``full'' $(N_\mathrm{det} \times N_\mathrm{det})$-matrix (\emph{i.e.}, a restriction of it into the model subspace, into the target subspace, or the representation of a transformation between both).

The target-space representation of \(\hat{H}^{\mathrm {el}}\) is a real-symmetric matrix \(\bm{\breve{H}}(\bm{t}^*) \in \mathbb{S}_3(\mathbb{R})\) 
(the set of  real-symmetric \((3 \times 3)\)-matrices),
since the MO basis \(\bm{\varphi}\) can be assumed to be real under normal circumstances. 
Its elements  \( \lbrace  H_{IJ}( \bm{t}^*) \rbrace _{I,J \in \lbrace \mathrm {A},\mathrm {B},\mathrm {C}\rbrace } \) are defined as
\begin{equation}
\label{eq:matrice_projection_hamiltonian}
    H_{IJ}(\bm{t}^*) = \langle \Phi_I(\bm{t}^*) | \hat{H}^{\mathrm {el}} | \Phi_J(\bm{t}^*) \rangle \quad.
\end{equation}

Because the ensemble-variationally-optimal states \(\lbrace \ket{\Phi_I(\bm{t}^*)}\rbrace _{I \in \lbrace \mathrm {A},\mathrm {B},\mathrm {C}\rbrace }\) are not, \textit{a priori}, eigenstates of \(\hat{H}^{\mathrm {el}} \), 
the off-diagonal elements of \(\bm{\breve{H}}(\bm{t}^*)\) can be nonzero. Determining the target eigenstates of \(\hat{H}^{\mathrm {el}} \) then amounts to the final diagonalization of $\bm{\breve{H}}(\bm{t}^*)$. 

At this stage, two different approaches can be distinguished. The first consists in evaluating each matrix element on the quantum computer, followed by a classical diagonalization to obtain the eigenvalues and eigenstates of the system, as done in most quantum subspace methods~\cite{motta2024subspace}. 
The second approach, adopted in the present work and following the one developed in Ref.~\citenum{yalouz_analytical_2022}, consists in obtaining the eigenstates directly on the quantum computer. This strategy opens the way to various post-processing critical tasks, such as gradient evaluations, the computation of NACs, or the measurement of other physical observables.

\subsubsection{Diagonalizing \emph{via} an orthogonal similarity transformation: toward adiabatic eigensolutions}
\label{subsubsec:transformation}

For any matrix \( \bm{\breve{H}} \in \mathbb{S}_3(\mathbb{R})\), there exists a nonunique orthogonal matrix \(\bm{\breve{Q}} \in \mathrm{O}(3)\) (the group of orthogonal \((3 \times 3)\)-matrices)~\cite{golub_matrix_2013} such that 
\begin{equation}
    \bm{\breve{H}} \mapsto \bm{\breve{H}}^{\prime} = \bm{\breve{Q}}^\top  \bm{\breve{H}}  \bm{\breve{Q}} \quad,
\end{equation}
where $\bm{\breve{H}}^{\prime}$ is real-diagonal.
Such an orthogonal similarity transformation map applied to the variationally-optimized matrix 
$\bm{\breve{H}}(\bm{t}^*)$ leads to a diagonal matrix $\bm{\breve{H}}^{\prime}(\bm{t}^*)$ the elements of which, $\lbrace  H^{\prime}_{IJ}( \bm{t}^*) \rbrace _{I,J \in \lbrace \mathrm {A},\mathrm {B},\mathrm {C}\rbrace }$, are defined as
\begin{equation}
    H^{\prime}_{IJ}( \bm{t}^*) =  \langle \Phi^{\prime}_I(\bm{t}^*) | \hat{H}^{\mathrm {el}} | \Phi^{\prime}_J(\bm{t}^*) \rangle \quad,
\end{equation}
where the post-transformed eigenbasis within the target subspace, $\lbrace \ket{ \Phi^{\prime}_{I}(\bm{t}^*)}\rbrace _{I \in \lbrace \mathrm {A},\mathrm {B},\mathrm {C}\rbrace }$, satisfies
\begin{eqnarray}
   \ket{ \Phi^{\prime}_I(\bm{t}^*)} &=& \sum_{J \in \lbrace \mathrm {A},\mathrm {B},\mathrm {C}\rbrace } Q_{JI}(\bm{t}^*) \ket{\Phi_J(\bm{t}^*)} \quad, \nonumber \\ Q_{JI}(\bm{t}^*) &=& \braket{\Phi_J(\bm{t}^*)}{\Phi_I^{\prime}(\bm{t}^*)} \quad.
\end{eqnarray}
Now, from Eq.~(\ref{eq:u_phi0}), we also have
\begin{eqnarray}
\label{eq:change_of_basis_with_rotation}
   \ket{ \Phi^{\prime}_I(\bm{t}^*)} &=& \sum_{J \in \lbrace \mathrm {A},\mathrm {B},\mathrm {C}\rbrace } Q_{JI}(\bm{t}^*) \hat{U}(\bm{t}^*) \ket{\Phi^0_J} \nonumber \\
   &=& \hat{U}(\bm{t}^*) \sum_{J \in \lbrace \mathrm {A},\mathrm {B},\mathrm {C}\rbrace } Q_{JI}(\bm{t}^*) \ket{\Phi^0_J} \quad,
\end{eqnarray}
where we used the linearity of $\hat{U}(\bm{t}^*)$ to move it left, back in front of the sum.
This brings an incentive for introducing a pre-transformed basis of the model subspace, \(\lbrace \ket{ \Phi^{0\prime}_{I}(\bm{t}^*)}\rbrace _{I \in \lbrace \mathrm {A},\mathrm {B},\mathrm {C}\rbrace }\), such that
\begin{equation}
    \ket{\Phi^{0\prime}_I(\bm{t}^*)} =  \sum_{J \in \lbrace \mathrm {A},\mathrm {B},\mathrm {C}\rbrace } Q_{JI}(\bm{t}^*) \ket{\Phi^0_J}, \quad Q_{JI}(\bm{t}^*) = \bra{\Phi^0_J}\ket{\Phi^{0\prime}_I(\bm{t}^*)}  ,
\end{equation}
which will be important later on when implementing this operation on quantum circuits.
It is a basis of the model subspace indeed, but its very definition depends implicitly on the pre-knowledge of $\bm{t}^*$, \emph{i.e.}, on the retroactive ensemble-variational optimization of the target subspace.

It may seem at first sight that, when moving $\hat{U}(\bm{t}^*)$ to the left, we have illegally swapped the order of two noncommuting operations.
This is not the case. 
Let us re-express the above in matrix form, within the spirit of classical-computing CI.
For consistency, we must introduce the following orthogonally-augmented $(N_\mathrm{det} \times N_\mathrm{det})$-matrix,
\begin{equation}
\bm{Q} = 
\begin{pmatrix}
\bm{\breve{Q}} & \bm0 \\
\bm0 & \bm1
\end{pmatrix} \quad.
\end{equation}
The classical CI approach assumes
\begin{equation}
   \ket{\Phi^{\prime}_I(\bm{t}^*)}
   = \sum_{J=1}^{N_\mathrm{det}}
   [\bm{C}(\bm{t}^*)]_{JI}
   \ket{\Phi^0_J}
   \quad,
\end{equation}
where $\bm{C}(\bm{t}^*)$ is the orthogonal CI-coefficient matrix of the eigenstates with respect to the numerical basis of Slater determinants (or CSFs if relevant).
In practice, here, we are only interested in the $(N_\mathrm{det} \times 3)$-block of the first three columns of $\bm{C}(\bm{t}^*)$.
The ``post-variational'' approach (determine $\bm{t}^*$ variationally such that applying $\hat{U}(\bm{t}^*)$ to the model will block-diagonalize, and post-transform the result to diagonalize) corresponds to
\begin{equation}
   \bm{C}(\bm{t}^*) = \bm{O}(\bm{t}^*) \bm{Q}(\bm{t}^*)
   \quad.
\end{equation}
Alternatively, the ``retro-variational'' approach (determine $\bm{t}^*$ variationally as above and keep it; then, pre-transform the model such that applying $\hat{U}(\bm{t}^*)$ to the result will diagonalize) corresponds to
\begin{equation}
   \bm{C}(\bm{t}^*) = \bm{Q}(\bm{t}^*) \bm{O}^{\prime}(\bm{t}^*)
   \quad.
\end{equation}
Both approaches are formally equivalent, simply because $\bm{O}^{\prime}(\bm{t}^*) = \bm{Q}^\top(\bm{t}^*) \bm{O}(\bm{t}^*) \bm{Q}(\bm{t}^*)$.
While the former seems more natural, the latter is required in practice if one wants to keep the eigenstates within the quantum circuit.

From the above, we can obtain pointwise eigensolutions at any $\bm{R}$: $\lbrace \ket{\Phi^{\prime}_I(\bm{t}^*)}\rbrace _{I \in \lbrace \mathrm {A},\mathrm {B},\mathrm {C}\rbrace }$ and $\lbrace  H^{\prime}_{IJ}( \bm{t}^*) \rbrace _{I,J \in \lbrace \mathrm {A},\mathrm {B},\mathrm {C}\rbrace }$ where $H^{\prime}_{I \neq J}( \bm{t}^*) = 0 $.
The adiabatic identification and connection, $\lbrace \ket{\Psi_n}\rbrace _{n \in \lbrace 0,1,2\rbrace } \longleftrightarrow \lbrace \ket{\Phi^{\prime}_I (\bm{t}^*)}\rbrace _{I \in \lbrace \mathrm {A},\mathrm {B},\mathrm {C}\rbrace }$ and $\lbrace E_n\rbrace _{n \in \lbrace 0,1,2\rbrace } \longleftrightarrow \lbrace  H^{\prime}_{II}( \bm{t}^*) \rbrace _{I \in \lbrace \mathrm {A},\mathrm {B},\mathrm {C}\rbrace }$, hence $\lbrace \mathrm {A},\mathrm {B},\mathrm {C}\rbrace \longleftrightarrow\lbrace 0,1,2\rbrace $, has to be finalized globally according to some conventional increasing-energy order at any value of $\bm{R}$: $E_0 \leq E_1 \leq E_2$.

In addition, we have assumed so far that \(\bm{\breve{Q}} \in \mathrm{O}(3)\).
Yet, by replacing \(\bm{\breve{Q}}\) with \(-\bm{\breve{Q}}\), one can impose to always choose for convenience \(\bm{\breve{Q}} \in \mathrm{SO}(3)\), the special-orthogonal group of rotation \((3\times3)\)-matrices (with determinant \(+1\)). This choice is not restrictive for the purpose of diagonalization and may appear as sensible in order to help maintaining consistency across different molecular geometries \(\bm{R}\) as regards the potential introduction of arbitrary sign changes in the states (\textit{i.e.}, \(\pm 1\) real-phase state factors).
Note that we are not concerned by the concepts nor the implications of the Berry/Longuet--Higgins geometrical/topological phases in the present work~\cite{baer_beyond_2006}.

\subsubsection{Parameterization of the similarity transformation}
\label{subsubsec:implementation_similarity_transformation}

In order to implement the retro-variational approach, it is necessary to express the transformation in Eq.~\eqref{eq:change_of_basis_with_rotation}, associated with a rotation matrix \( \bm{\breve{Q}} \), in a parameterized form that can be translated into quantum gates.  The mathematical procedure to achieve this, inspired by Ref.~\citenum{lasorne_use_2014}, consists in expressing the rotation matrix \( \bm{\breve{Q}} \) as a product of three rotation matrices, each parameterized by an angle. 

Rodrigues' formula provides an explicit expression for the rotation matrix \(\mathcal{R}_{\bm{n}}(\alpha)\), representing a rotation by an angle \(\alpha\) around a unit vector \(\bm{n} \in \mathbb{R}^3\) (see Ref.~\citenum{bernstein_matrix_2009}),
\begin{equation}
\mathcal{R}_{\bm{n}}(\alpha) = \mathbb{I}_3 + \sin(\alpha) \bm{K}_{\bm{n}} + (1 - \cos(\alpha)) \bm{K}_{\bm{n}}^2 \quad,
\end{equation}
where \( \bm{K}_{\bm{n}} \) denotes the  $(3 \times 3)$-matrix representation of the linear map \(\hat{\bm{k}}_{\bm{n}} : \bm{v} \mapsto \bm{n} \times \bm{v} \), which is defined by the cross product of the unit vector \( \bm{n}\) with a vector  \(\bm{v} \in \mathbb{R}^3 \).

One can show that any rotation matrix \(\bm{\breve{Q}} \in \mathrm{SO}(3)\) can then be expressed as a product of three elementary rotation matrices \(\mathcal{R}_{\bm{n}_i}\), with mutually orthogonal axes satisfying \(\bm{n}_1 \perp \bm{n}_2\) and \(\bm{n}_2 \perp \bm{n}_3\) (see Ref.~\citenum{piovan_coordinate-free_2012}). Thus, combining this result with the real decomposition for real-symmetric matrices leads to the following corollary: for three unit vectors \( \bm{n}_1, \, \bm{n}_2, \, \bm{n}_3 \in \mathbb{R}^3 \)  such that \( \bm{n}_1 \perp \bm{n}_2 \) and \( \bm{n}_2 \perp \bm{n}_3 \), there exist three angles \( (\theta^{*}, \phi^{*}, \psi^{*}) \in [-\pi, \pi[^3 \) that diagonalize the symmetric matrix  \( \bm{\breve{H}}(\bm{t}^*) \) \emph{via} the change-of-basis matrix defined as
\begin{equation}
\label{eq:paramterized_rotation_matrix}
    \mathcal{R}_{\bm{n}_1, \bm{n}_2, \bm{n}_3}(\theta^{*}, \phi^{*}, \psi^{*}) = \mathcal{R}_{\bm{n}_1}(\theta^{*})\, \mathcal{R}_{\bm{n}_2}(\phi^{*})\, \mathcal{R}_{\bm{n}_3}(\psi^{*}) \quad.
\end{equation}

In the present study, the \( \bm{x}, \bm{z}, \bm{y} \)  axes were respectively used to define the elementary rotation matrices \( \mathcal{R}_{\bm{n_1}}, \mathcal{R}_{\bm{n_2}}, \mathcal{R}_{\bm{n_3}} \).

We now consider the following parameterized similarity transformation,
\begin{equation}
\begin{aligned}
 \mathbb{S}_3(\mathbb{R}) \times \left[-\pi;\pi\right[^{3}  &\to \mathbb{S}_3(\mathbb{R}) \\
(\bm{\breve{H}} , \theta, \phi, \psi) &\mapsto \bm{\breve{H}}^{\prime}(\theta, \phi, \psi) \quad,
\end{aligned}
\end{equation}
where
\begin{equation}
\begin{aligned}
\label{eq:similarity_transformation}
\bm{\breve{H}}^{\prime}(\theta, \phi, \psi) =  \mathcal{R}_{\bm{x} , \bm{z}, \bm{y}} \left(\theta, \phi, \psi \right) ^\top \bm{\breve{H}}  \,  \mathcal{R}_{\bm{x} , \bm{z} , \bm{y}}\left(\theta, \phi, \psi \right) \quad.
\end{aligned}
\end{equation}

Applying the change of basis associated with this transformation to \( \bm{\breve{H}}(\bm{t}^*) \) yields the target states \(\ket{\Phi^{\prime}_I(\bm{t}^*, \theta, \phi, \psi)}_{I \in \lbrace \mathrm {A},\mathrm {B},\mathrm {C}\rbrace }\) expressed as
 \begin{equation}
 \label{eq:state_change_of_basis_rotation}
\begin{aligned}
\ket{\Phi^{\prime}_I(\bm{t}^*, \theta, \phi, \psi)}
&= \hat{U}(\bm{t}^*) \ket{\Phi^{0\prime}_I(\theta, \phi, \psi)} 
\end{aligned}
\end{equation}   
where
\begin{widetext}
\begin{subequations} \label{eq:paramterized_change_of_basis_rotation}
\begin{align}
& \ket{\Phi^{0\prime}_{\mathrm {A}}(\theta, \phi, \psi)} = 
\cos \phi \cos \psi \ket{\Phi^0_{\mathrm {A}}} 
 + (\sin \phi \cos \psi \cos \theta + \sin \psi \sin \theta) \ket{\Phi^0_{\mathrm {B}}} 
 + (\sin \phi \sin \theta \cos \psi - \sin \psi \cos \theta) \ket{\Phi^0_{\mathrm {C}}} \quad, \\
& \ket{\Phi^{0\prime}_{\mathrm {B}}( \theta, \phi, \psi)} =
- \sin \phi \ket{\Phi^0_{\mathrm {A}}} 
+ \cos \phi \cos \theta \ket{\Phi^0_{\mathrm {B}}} 
 + \sin \theta \cos \phi \ket{\Phi^0_{\mathrm {C}}} \quad, \\
& \ket{\Phi^{0\prime}_{\mathrm {C}}(\theta, \phi, \psi)} =
\sin \psi \cos \phi \ket{\Phi^0_{\mathrm {A}}} 
 + (\sin \phi \sin \psi \cos \theta - \sin \theta \cos \psi) \ket{\Phi^0_{\mathrm {B}}} 
 + (\sin \phi \sin \psi \sin \theta + \cos \psi \cos \theta) \ket{\Phi^0_{\mathrm {C}}} \quad.
\end{align}
\end{subequations}
\end{widetext}
For the sake of notational simplicity, we did not indicate the implicit dependence of the pre-transformed model states \(\ket{\Phi^{0\prime}_I(\theta, \phi, \psi)}_{I \in \lbrace \mathrm {A},\mathrm {B},\mathrm {C}\rbrace }\) on $\bm{t}^*$ (which actually occurs \emph{via} the implicit dependence of the values of the three rotation angles on the pre-knowledge of $\bm{t}^*$).
As already stated, the quantum circuits able to generate these linear combinations are relatively easy to design only if the original model states are trivial in terms of bitstrings (\emph{i.e.}, Slater determinants or CSFs), as shown for instance in Ref.~\citenum{yalouz_analytical_2022}
and further discussed in Sec.~\ref{sec:comp_det}
for the three-state case presented in the present study.

\subsubsection{Various flavors of the eigenproblem objective}
\label{subsubsec:solving_for_eigenstates}

Obtaining the eigenstates relies on determining the angles \((\theta, \phi, \psi)\) for which the matrix \(\bm{\breve{H}}^{\prime}(\bm{t}^*,\theta, \phi, \psi)\), denoted \(\bm{\breve{H}}^{\prime}(\theta, \phi, \psi)\) hereafter for simplicity, is diagonal. From this perspective, it might appear natural, \textit{a priori}, to minimize a function constructed from the off-diagonal elements of \(\bm{\breve{H}}^{\prime}(\theta, \phi, \psi)\). However, this approach would entail designing specific circuits for each of these terms~\cite{nakanishi_subspace-search_2019}, which renders this approach more costly. Thus, the approach adopted here, inspired by that proposed for a two-state system~\cite{yalouz_analytical_2022}, relies on constructing an objective function that depends solely on the diagonal elements of the matrix.

\paragraph{Approach based on the conservation of the Frobenius norm.}
The Frobenius norm of a matrix is invariant under similarity transformations.
It follows that
\begin{equation}
   \sum_{\substack{I, J \in \lbrace \mathrm {A},\mathrm {B},\mathrm {C}\rbrace  \\ I \ne J}}H^{\prime}_{IJ}(\theta, \phi, \psi )^{2}
    =   \|\bm{\breve{H}}(\bm{t}^*)\|^{2}_{\mathrm{F}}  - \sum_{I \in \lbrace \mathrm {A},\mathrm {B},\mathrm {C}\rbrace } H^{\prime}_{II}( \theta, \phi, \psi)^{2} .
\end{equation}
Thus, minimizing the above left-hand side is equivalent to minimizing the following function,
\begin{equation}
   f^{\mathrm{F}}(\theta, \phi, \psi) = - \sum_{I \in \lbrace \mathrm {A},\mathrm {B},\mathrm {C}\rbrace } H^{\prime}_{II}( \theta, \phi, \psi)^{2} \quad.
   \label{objective_func_f_f}
\end{equation}
Given the existence of a triplet \((\theta^{*}, \phi^{*}, \psi^{*})\) that diagonalizes the matrix, the function \(f^{\mathrm{F}}(\theta, \phi, \psi)\) possesses a global minimum. 
However, it depends on \((\theta, \phi, \psi)\) through sine and cosine functions, which are neither convex nor concave over their entire domains. 
As a result, \(f^{\mathrm{F}}(\theta, \phi, \psi)\) is, \textit{a priori}, not convex, implying that the search for a global minimum could be complicated by the presence of local minima.

\paragraph{Approach based on the relationship between partial derivatives and off-diagonal terms.}
Since the goal is to identify the extrema of an objective function, it can be advantageous to construct this function such that its partial derivatives with respect to each variable are, up to a multiplicative factor, proportional to the off-diagonal elements of the matrix \(\bm{\breve{H}}^{\prime}(\theta, \phi, \psi)\). Under these conditions, the vanishing of the partial derivatives is equivalent to the vanishing of these off-diagonal elements,
thus leading to a two-step strategy.
In the first step, two rotations are applied to eliminate two of the three off-diagonal terms.
The two resulting relations correspond to
\begin{subequations}
\begin{align}
& \frac{\partial  H^{\prime}_{\mathrm{BB}}( \theta, \phi, 0)}{\partial \theta}= 2 \cos(\phi) H^{\prime}_{\mathrm{BC}}( \theta, \phi,0) \quad, \\
& \frac{\partial H^{\prime}_{\mathrm{BB}}(\theta, \phi, 0)  }{\partial \phi} = -2 H^{\prime}_{\mathrm{AB}}(\theta, \phi, 0) \quad.
\end{align}
\end{subequations}
It consists in seeking an extremum in
\((\theta, \phi)\) of the function \(H^{\prime}_{\mathrm{BB}}(\theta, \phi, 0)\) (or its negative) in order to cancel the off-diagonal terms. It can be noted that the second condition may be satisfied for \(\phi \equiv  \frac{\pi}{2} \pmod{\pi}\) without the term \( H^{\prime}_{\mathrm{BC}}(\theta, \phi, 0)\) being, \emph{a priori}, zero, which requires a separate treatment for this particular case. Denoting by \((\theta^*, \phi^*)\) the pair that ensures the cancellation of the partial derivatives, one obtains at the end of this step a matrix \( \bm{\breve{H}}^{\prime}(\theta^{*}, \phi^{*}, 0) \) such that
\begin{equation}
   H^{\prime}_{\mathrm{AB}}(\theta^{*}, \phi^{*}, 0) = H^{\prime}_{\mathrm{BC}}(\theta^{*}, \phi^{*}, 0) = 0 \quad.
\end{equation}

In a second step, the third rotation is performed to eliminate the remaining term while preserving the vanishing of the first two.
The application of the third rotation \(\mathcal{R}_{\bm{y}}(\psi)\) preserves the cancellation of the two off-diagonal elements \(H^{\prime}_{\mathrm{AB}}   \) and \( H^{\prime}_{\mathrm{BC}} \),
\begin{equation}
    H^{\prime}_{\mathrm{AB}}(\theta^{*}, \phi^{*}, \psi) = H^{\prime}_{\mathrm{BC}}(\theta^{*}, \phi^{*}, \psi) =0  \quad.
\end{equation}
Furthermore, it yields a third relation
\begin{equation}
\frac{\partial H^{\prime}_{\mathrm{CC}}(\theta^*, \phi^*, \psi)}{\partial \psi} = 2H^{\prime}_{\mathrm{AC}}(\theta^*, \phi^*, \psi) \quad.
\end{equation}
The second step therefore consists in seeking an extremum with respect to 
\(\psi\) of the function \(H^{\prime}_{\mathrm{CC}}(\theta^*, \phi^*, \psi)\) (or its negative) in order to cancel the off-diagonal term \(H^{\prime}_{\mathrm{AC}}(\theta^*, \phi^*, \psi)\).
Relying on these analytical conditions for the cancellation of the partial derivatives, this approach does not appear to be affected by numerical issues associated with the presence of local minima. Nevertheless, it is less straightforward to generalize than the approach based on norm conservation.

\paragraph{Approach based on the ensemble-variational principle}
Since a similarity transformation preserves the trace of a matrix, the states 
\(\lbrace \ket{ \Phi^{\prime}_I(\bm{t}^*, \theta, \phi, \psi)}\rbrace _{I \in \lbrace \mathrm {A},\mathrm {B},\mathrm {C}\rbrace }\) remain within the target subspace associated with the minimal ensemble-energy for any choice of the triplet \((\theta, \phi, \psi)\). Thus,  the ensemble extension of the Rayleigh--Ritz variational principle~\cite{gross_rayleigh-ritz_1988} with different weights can be applied into this subspace to obtain the triplet \((\theta^{*}, \phi^{*}, \psi^{*})\) that diagonalizes the matrix. Following this idea, we can minimize the objective function \( f^{\mathrm{w}}(\theta, \phi, \psi)\) defined as
\begin{equation}
   f^{\mathrm{w}}(\theta, \phi, \psi) =  \sum_{I \in \lbrace \mathrm {A},\mathrm {B},\mathrm {C}\rbrace } w_I H^{\prime}_{II}( \theta, \phi, \psi) \quad,
    \label{objective_func_f_w}
\end{equation}
with \(w_\mathrm{A}, w_\mathrm{B}, w_\mathrm{C} \) three positive real numbers ordered such that: \( w_\mathrm{A} > w_\mathrm{B}  > w_\mathrm{C}   > 0 \).
As with the norm-conservation approach, the objective function does not appear to be convex \textit{a priori}. Nevertheless, the objective function \(f^{\mathrm{w}}(\theta, \phi, \psi)\) is likely to be numerically more stable (a perturbative argument based on small variations is provided in Appendix~\ref{si_subsubsec:various_flavors_eigenproblem_objective}).
This approach should be as straightforward to generalize as the norm-conservation approach but choosing adequate values of the weights remains critical from a practical and numerical perspective (see Refs.~\citenum{hong2023refining,ding2024ground,rajamani2025equi}).

\subsection{Achieving a quasi-diabatic representation}
\label{subsec:opt_diab}

Following the framework developed in Ref.~\citenum{illesova_transformation-free_2025}, we explore the capability of the ensemble-VQE method to produce a quasi-diabatic representation, here for the case of three states.
As for the eigenstates, the objective is to obtain the target diabatic states on the quantum computer.
In the case of two states, our previous investigations have shown that the method naturally yields quasi-optimal diabatic states through the block-diagonalization procedure.
Herein, we investigate this feature for three states and, as suggested in Ref.~\citenum{illesova_transformation-free_2025}, consider strategies to enforce diabatic optimality when necessary.
One of our approaches is based on the same change-of-basis machinery as developed in the previous section for solving the eigenstates.

\paragraph{Diabatic molecular orbitals and Slater determinants.}
The model states, \(\lbrace \ket{\Phi^0_I}\rbrace _{I \in \lbrace \mathrm {A},\mathrm {B},\mathrm {C}\rbrace }\), which are Slater determinants (or CSFs if relevant), can be made ``as diabatic as possible'' through a suitable least-changing choice of the underlying MO basis with respect to $\bm{R}$. 
This construction relies on the concept of `diabatic orbitals', as first defined by 
Werner \emph{et al.}~\cite{werner_adiabatic_1988, simah_photodissociation_1999} and further described in Ref.~\citenum{illesova_transformation-free_2025}. 
In this, a reference geometry \(\bm{R}^{0}\) is chosen, together with its canonical Hartree--Fock MOs \( \lbrace \varphi_p(\bm{r} ; \bm{R}^{0})\rbrace _{p=1}^{N_\mathrm{orb}}\). 
At any other given geometry \(\bm{R}\), the corresponding canonical Hartree--Fock MOs \( \lbrace \varphi_p(\bm{r} ; \bm{R})\rbrace _{p=1}^{N_\mathrm{orb}}\) are transformed using an orthogonal Procrustes procedure (based on similar grounds as the symmetric Löwdin orthogonalization), 
yielding the so-called `diabatic' MO basis \( \lbrace \xi_p(\bm{r} ; \bm{R}, \bm{R}^{0})\rbrace _{p=1}^{N_\mathrm{orb}}\)
that maximizes its orbital-resolved overlaps with \( \lbrace \varphi_p(\bm{r} ; \bm{R}^{0})\rbrace _{p=1}^{N_\mathrm{orb}}\), and which thus depends parametrically on the current geometry \(\bm{R}\) but also on the reference geometry \(\bm{R}^{0}\).

\paragraph{Least-transformation diabaticity criterion.}
If we were solving the corresponding block-diagonalization problem with a classical-computing approach, our aim would be to determine the first three columns of the many-body overlap $(N_\mathrm{det} \times N_\mathrm{det})$-matrix 
\(\bm{O}(\bm{t}^*)\), which is orthogonal.
The model-to-target $(3 \times 3)$-submatrix has already been denoted \(\bm{\breve{O}}(\bm{t}^*)\).
Of course, it has no reason to be orthogonal but it will be expected (see below) to be as close as possible to the \((3 \times 3)\)-identity-matrix if one wants to transfer as much diabatic character as possible from the model subspace to the target subspace.
We shall further define the remaining \(((N_\mathrm{det} - 3) \times 3)\)-submatrix as \(\bm{X}(\bm{t}^*)\).

Let us specifically denote \(\bm{\breve{O}}_\star := \bm{\breve{O}}(\bm{t}_\star)\) and 
\(\bm{X}_\star := \bm{X}(\bm{t}_\star)\) as the overlap submatrices corresponding to the ``optimal diabatization'' of the states belonging to the target subspace of minimal ensemble-energy, \(\mathcal{V}_{*}^{\mathrm{ens}}\), such that the corresponding parameter \(\bm{t}_\star\) fulfills
\begin{equation}
\bm{t}_{\star} = \arg\min_{\bm{t} \in \mathcal{T}^{\mathrm{ens}}} 
\Bigl\lVert \bm{\breve{O}}(\bm{t}) - \bm{1} \Bigr\rVert_\mathrm{F}    \quad,
\end{equation}
where
\begin{equation}
\mathcal{T}^{\mathrm{ens}} =
\Bigl\lbrace  \bm{t} \;\Big|\; \lbrace  \ket{\Phi_I(\bm{t})} \rbrace _{I \in \lbrace \mathrm {A},\mathrm {B},\mathrm {C}\rbrace } \in \mathcal{V}_*^{\mathrm{ens}} \Bigr\rbrace     \quad.
\end{equation}
The manifold $\mathcal{T}^{\mathrm{ens}}$ is \emph{a priori} not made of a single value of $\bm{t}^*$ (reflecting the aforementioned gauge freedom of the block-diagonalization), and $\bm{t}_{\star}$ is some particular value of $\bm{t}^*$, which has to be determined through some additional constraint.
Since the overlap matrix \(\bm{O}(\bm{t})\) is orthogonal for any $\bm{t}$, it follows that
\begin{equation}
\bm{\breve{O}}^\top(\bm{t}) \bm{\breve{O}}(\bm{t}) + \bm{X}^\top(\bm{t}) \bm{X}(\bm{t}) = \bm{1} \quad.    
\end{equation}
Hence, requiring \(\bm{\breve{O}}_\star\) to be close to the identity matrix automatically implies that \(\bm{X}_\star\) be close to the zero matrix.
Following Ref.~\citenum{illesova_transformation-free_2025}, the ``optimal diabatization'' can thus be viewed as requiring the following condition to be satisfied as closely as possible for any geometry \(\bm{R}\),
\begin{equation}
\begin{pmatrix}
\bm{\breve{O}}_\star \\
\bm{X}_\star
\end{pmatrix}
\approx
\begin{pmatrix}
\bm1 \\
\bm0
\end{pmatrix} \quad.
\end{equation}
The optimal diabatic states thus simply correspond to the least-transformed orthonormal projections of the model states \(\lbrace \ket{\Phi^0_I}\rbrace _{I \in \lbrace \mathrm {A},\mathrm {B},\mathrm {C}\rbrace }\) onto the target subspace \(\mathcal{V}_{*}^{\mathrm{ens}}\)\cite{pacher1988approximately, cederbaum1989block, werner_adiabatic_1988, simah_photodissociation_1999, illesova_transformation-free_2025}.

\paragraph{Descriptors of diabaticity.}
Our objective is to obtain a particular value of $\bm{t}^*$ such that \(\bm{\breve{O}}(\bm{t}^*)\) is as close as possible to $\bm1$. 
In order to quantify the deviation from this condition, two descriptors were introduced in Ref.~\citenum{illesova_transformation-free_2025}.
Their definitions are recalled below.
They are both based on the `singular value decomposition' (SVD) of \(\bm{\breve{O}}(\bm{t}^*)\), which reads
\begin{equation}
\label{eq:overlap-svd}
\bm{\breve{O}}(\bm{t}^*) = \bm{\breve{U}}(\bm{t}^*) \bm{\breve{\Sigma}}(\bm{t}^*)\bm{\breve{W}}^\top(\bm{t}^*) \quad,
\end{equation}
where \(\bm{\breve{U}}(\bm{t}^*)\) and  \(\bm{\breve{W}}(\bm{t}^*)\) are two orthogonal matrices, and \(\bm{\breve{\Sigma}}(\bm{t}^*) = \text{diag}(\sigma_1, \sigma_2 , \sigma_3)\)
with \( \sigma_1 \geq \sigma_2 \geq \sigma_3 \geq 0 \) (real-diagonal and positive semi-definite).
Hence, we have the following reverse polar decomposition,
\begin{equation}
\bm{\breve{O}}(\bm{t}^*)= \bm{\breve{O}}_\star(\bm{t}^*)\bm{\breve{B}}(\bm{t}^*) \quad,
\end{equation}
where the respective real-symmetric and orthogonal matrix components are
\begin{subequations}
\begin{align}
&\bm{\breve{O}}_\star(\bm{t}^*) = \bm{\breve{U}}(\bm{t}^*) \bm{\breve{\Sigma}}(\bm{t}^*) \bm{\breve{U}}^\top(\bm{t}^*) \quad, \\
&\bm{\breve{B}}(\bm{t}^*) = \bm{\breve{U}}(\bm{t}^*) \bm{\breve{W}}^\top(\bm{t}^*) \quad.
\end{align}
\end{subequations}

Our first descriptor, \(d\), quantifies the maximal diabaticity that can be achieved in the target subspace from a given model susbspace.
It is defined as~\cite{illesova_transformation-free_2025}
\begin{equation}
d(\bm{t}^*) = \norm{\bm{\breve{O}}_\star(\bm{t}^*)-\bm1}_\mathrm{F} = \norm{\bm{\breve{\Sigma}}(\bm{t}^*) -\bm1}_\mathrm{F}  \quad.
\end{equation}
Its deviation from zero (expected to be finite but small enough) indicates the extent to which the model and target necessarily differ when considering finite-dimensional subpaces.
It indirectly reflects the existence of the so-called `irremovable CSF contributions to the NACs' (due to the moderate but unavoidable variations with $\bm{R}$ of the MO-LCAO-coefficients and AO-overlaps involved in the MOs that underlie the Slater determinants). 
Per construction, its amplitude is globally minimized (and varies as smoothly as possible with $\bm{R}$) when one uses a diabatic MO basis, \( \lbrace \xi_p(\bm{r} ; \bm{R}, \bm{R}^{0})\rbrace _{p=1}^{N_\mathrm{orb}}\).
Its average value over $\bm{R}$ is essentially incompressible but depends slightly on the choice of the reference geometry \( \bm{R}^0 \).
In addition, let us note that any brutal increase of $d$ along continuous variations of $\bm{R}$ just indicates that the model subspace has become inadequate as the best guess of the target subspace.

Our second descriptor, \(r\), quantifies the residual deviation of the actual block-diagonalization from diabatic optimality, once the target subspace of minimal ensemble-energy has been determined.
It is defined as~\cite{illesova_transformation-free_2025}
\begin{equation}
r(\bm{t}^*) = \norm{\bm{\breve{B}}(\bm{t}^*)-\bm1}_\mathrm{F} = \norm{\bm{\breve{U}}(\bm{t}^*) - \bm{\breve{W}}(\bm{t}^*)}_\mathrm{F}   \quad.
\end{equation}

In particular, this descriptor tells us how close \(\bm{\breve{O}}(\bm{t}^*)\) is to the  real-symmetric matrix $\bm{\breve{O}}_\star(\bm{t}^*)$ and thus to being optimal as regards diabaticity.
It is an indirect measure of the CI-contributions to the NACs and shares with them the property of being fully removable within the target subspace, as will be illustrated below.

In the case of two states for the formaldimine molecule, our previous results have shown that the block-diagonalization procedure could yield quasi-optimal diabaticity~\cite{illesova_transformation-free_2025}. A possible explanation is that the ensemble-VQE algorithm behaves ``as lazily as possible'' when optimizing the variational parameters \(\bm{t}\), particularly when a gradient-based optimizer is employed. 
The idea is that the optimizer follows the steepest-descent path and thus tends to follow the ``shortest path'' in parameter space while converging to the target subspace. 
In particular, this reasoning appears to explain the small deviations from optimality in the weights of the initial states \( \ket{\Phi^{0}_I}\) within the corresponding optimized states \(\ket{\Phi_I(\bm{t}^*)}\). 
Nevertheless, it seems that this does not guarantee that the deviation arising from the mixing of the other initial states \(\ket{ \Phi^{0}_{J\neq I}}\) into \(\ket{\Phi_I(\bm{t}^*)}\) remains small, especially as the number of states and the size of the basis grow. Therefore, it seems useful, if not compulsory, to consider more systematic strategies to achieve optimal diabaticity in general situations.

\paragraph{Achieving optimal diabaticity.}

A way to achieve optimal diabaticity, \textit{i.e.}, to turn the value of the removable diabatic descriptor $r(\bm{t}^*)$ to zero, is to incorporate it directly as a penalty term in the objective function. 
However, unlike in the case of spin, our numerical experiments indicate that a direct choice of \(\lambda = 1\) does not yield satisfactory results, especially for the convergence of the ensemble-energy.
Reducing the value of \(\lambda \) can improve the results, but calibrating it to achieve a converged solution appears more challenging. Instead, one can follow the same strategy as in the spin case and minimize the objective function while enforcing the constraints directly using a Lagrangian approach~\cite{Nocedal2006Numerical}. Following the same rationale as for the spin constraint, this condition is enforced as an inequality defined as
\begin{equation}
\label{eq:r_constraint}
     r(\bm{t}^*)  \leq \epsilon_r \quad,
\end{equation}
where \(\epsilon_r\) is a small positive real number specifying the tolerance on the constraint.
In any case, this direct approach worked here because our ansatz was expressible enough, and $\bm{t}_\star$ could be obtained as some specific value of $\bm{t}^*$ within the $\mathcal{T}^{\mathrm{ens}}$-manifold that optimally achieves block-diagonalization so as to produce $\mathcal{V}_*^{\mathrm{ens}}$.

Alternatively, this can be reformulated as an orthogonal Procrustes problem (see also Ref.~\citenum{richings2020a} for some recent and similar approach to diabatization),
\begin{equation}
\label{eq:r_procrustes}
r_{\star} = \min_{ \bm{\breve{Q}}(\bm{t}^*) \in \mathrm{O}(3)}
\norm{\bm{\breve{U}}(\bm{t}^*) -\bm{\breve{Q}}(\bm{t}^*) \, \bm{\breve{W}}(\bm{t}^*) }_{\mathrm{F}} \quad .
\end{equation}
The optimal solution corresponds to $r_{\star} = 0$ and is obtained from $\bm{\breve{O}}(\bm{t}^*)$ and its SVD (Eq.~\eqref{eq:overlap-svd}) as
\begin{align}
\bm{\breve{Q}}_\star(\bm{t}^*) =   \bm{\breve{U}}(\bm{t}^*) \bm{\breve{W}}^\top(\bm{t}^*) = \bm{\breve{B}}(\bm{t}^*) \quad.
\label{eq:q_star}
\end{align}
Hence, upon following the same post- or retro-variational strategy as for the determination of the eigenstates but with a different objective, we can determine the optimal diabatic states from the pre-optimal target ones (those that have been pre-determined by some ensemble-variationally-optimal value of $\bm{t}^*$) upon using an additional \((\theta, \phi, \psi)\)-parameterized rotation (Eq.~\eqref{eq:paramterized_rotation_matrix}).
This is the strategy that we used herein.
Note that we also implemented the alternative use of a rotoreflection for situations where  \(\det\left(\bm{Q}_{\star}(\bm{t}^*) \right) = -1\) may be preferred, but this has never been encountered in the present work.
While we believe that there are good reasons for that, further investigating the generality of this property is left for future work.

In summary, we can separate the different steps of our objective, \textit{i.e.}, the preliminary ensemble-variational determination of the subspace of minimal ensemble-energy (given by some $\bm{t}^*$), followed by the explicit determination of the rotation \((\theta_\star, \phi_\star, \psi_\star)\)-angles that are required for subsequent optimal diabatization.
At each iteration step for the current values of $(\theta, \phi, \psi)$ and for some pre-obtained value of $\bm{t}^*$, one measures the actual overlap matrix $\bm{\breve{O}}(\bm{t}^*, \theta, \phi, \psi)$ on the quantum computer.
Its SVD is then performed on the classical computer so as to get the value of the objective $\norm{\bm{\breve{U}}(\bm{t}^*, \theta, \phi, \psi) - \bm{\breve{W}}(\bm{t}^*, \theta, \phi, \psi)}_\mathrm{F}$ to be minimized to zero with respect to variations of $(\theta, \phi, \psi)$ until convergence to $(\theta_\star, \phi_\star, \psi_\star)$.
The advantage of this approach is that it somewhat expands the parametrized-circuit expressibility beyond that of the original ansatz $\hat{U}(\bm{t})$.
The drawback of this approach compared to the direct one is that we have to effectively implement a change of basis, which may become challenging as the number of states increases.

\section{Computational Details}
\label{sec:comp_det}

\paragraph{Model system and molecular-symmetry considerations.}
The molecular ion H$_4^+$ was experimentally observed for the first time in 1984, as a dissociation product of H$_5^+$ in a collision cell during mass spectrometry analyses~\cite{kirchner_first_1984}. 
A subsequent theoretical study identified ten stable geometrical configurations for it~\cite{jiang_geometrical_1998}.
The tetrahedral \ce{T_d} configuration of H$_4^+$
-- for which Cartesian coordinates are given in Table~\ref{tab:coordinate} --
although not corresponding to a stable geometry according to the Jahn--Teller theorem, provides a relevant starting point for the present study.
This system will serve as a toy-model in what follows.

\begin{table}[h!]
\begin{tabular}{cccc}
Atom & \(x\) & \(y\) & \(z\)\\
H   &               0.000000000000  &  0.000000000000  &  1.142278716718   \\
H   &               0.807713026596  &  0.000000000000  &  0.000000000000\\
H   &               -0.403856513298  &  0.699500000000  &  0.000000000000\\
H   &              -0.403856513298 &   -0.699500000000  &  0.000000000000\\
\end{tabular}
\caption{Cartesian coordinates (in \AA) of the H$_4^+$ molecular ion in the optimized \ce{T_d} geometry obtained at the FCI/STO-3G level of theory.}
\label{tab:coordinate}
\end{table}

This high-symmetry geometrical configuration features one \ce{a_1} orbital and a set of three degenerate \ce{t_2} orbitals, resulting in a threefold degeneracy of the lowest \ce{^2T_2} electronic states, which correspond to three spin doublets (\(S=1/2\)).
A gradual symmetry descent can be performed, passing through the symmetries \ce{C_{3v}}, \ce{C_s}, and \ce{C_1}. 
We examined many of such multidimensional descriptions and did not experience any particular difficulty as regards our various procedures (see the Appendices~\ref{si_subsubsec:distortion_cs} and \ref{si_subsubsec:distortion_c3v}).
Below, we specifically focus on results obtained for \ce{C_1} geometries in order to address the most general situation (no symmetry).

\begin{figure}
\begin{subfigure}{0.7\columnwidth}
\centering
    \includegraphics[width=\columnwidth]{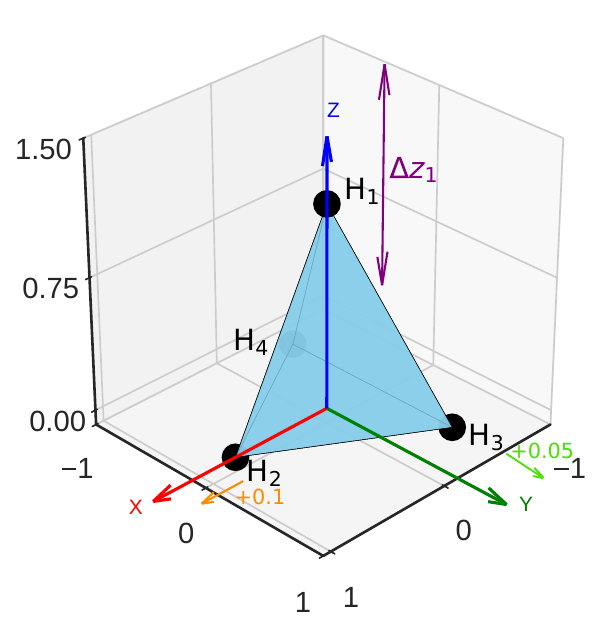}
    \caption{}
   \label{fig:H4_geometry}
  \end{subfigure}
  \hfill
  \begin{subfigure}{\columnwidth}
  \centering
    \includegraphics[width=\columnwidth]{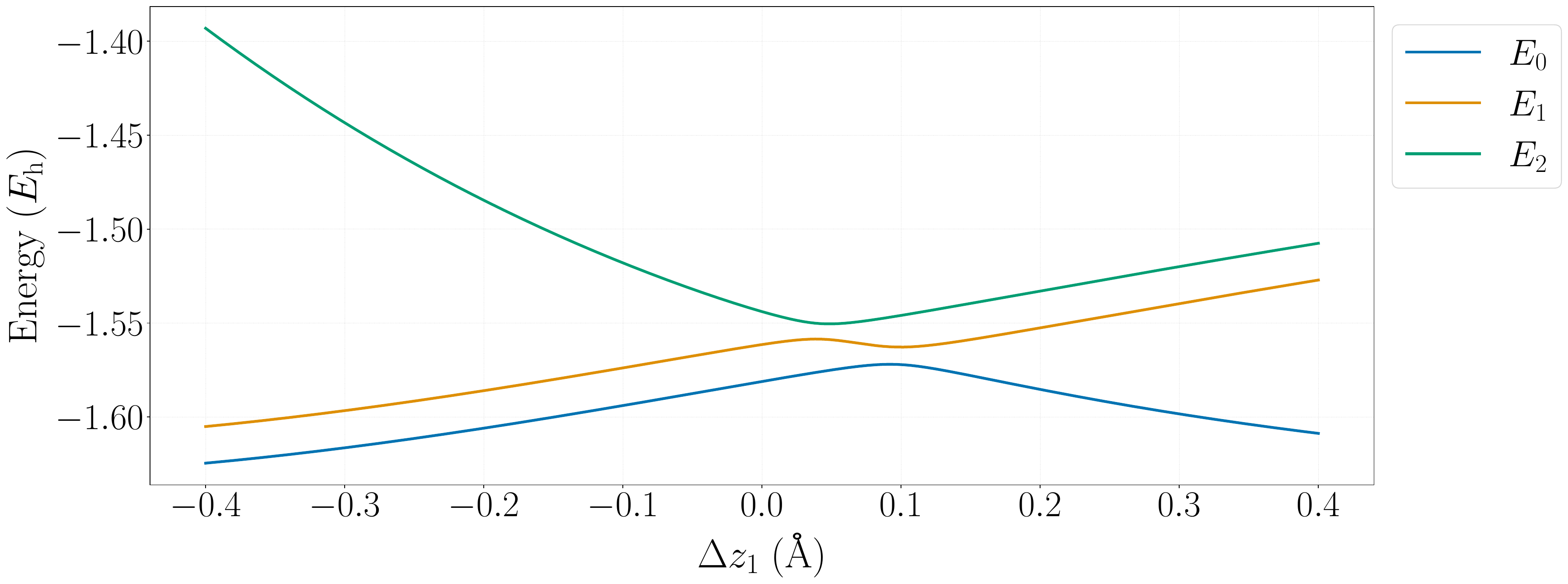}
        \caption{}
  \label{fig:FCI_PES}
  \end{subfigure}
  \caption{(a) Distortion from the tetrahedral \(\ce{T_d}\)  geometry of H$_4^+$ to \(\ce{C_1}\) geometries.
  (b) Corresponding  adiabatic potential energies of the first three eigenstates, \(  E_0 ,E_1, E_2 \), along $\Delta z_1$ (from $\Delta x_2 = 0.1~\textrm{\AA}$ and $\Delta y_3 = 0.05~\textrm{\AA}$) obtained from a 'restricted open-shell Hartree--Fock' (ROHF) FCI/STO-3G calculation with Psi4~\cite{turney_psi4_2012}.}
\end{figure}

The H$_4^+$ molecular ion was initially placed in a tetrahedral configuration.
The origin of the Cartesian coordinate system was chosen at the center of the circumscribed circle of one of the faces lying in the \(xy\)-plane. Three nuclei (\ce{H2}, \ce{H3}, \ce{H4}) were positioned in this plane, while nucleus \ce{H1} was placed on the \(z\)-axis at the vertex opposite this face, as shown in Fig.~\ref{fig:H4_geometry}. 
``Variationally exact'' adiabatic potential-energy curves along a \(\ce{C_1}\)-coordinate are shown in Fig.~\ref{fig:FCI_PES}.

The Cartesian coordinates of nucleus \(i\) in the optimized \ce{T_d} geometry (see Table~\ref{tab:coordinate}) are denoted \(x^{\ce{T_d}}_i, y^{\ce{T_d}}_i, z^{\ce{T_d}}_i \). Let \(x_i, y_i, z_i \) be the Cartesian coordinates of nucleus \(i\) in the geometry under consideration. The corresponding Cartesian displacements of nucleus \(i\) with respect to the optimized \ce{T_d} geometry are defined as
\begin{equation}
\Delta x_i = x_i - x_i^{\ce{T_d}} \quad, \qquad
\Delta y_i = y_i - y_i^{\ce{T_d}} \quad, \qquad
\Delta z_i = z_i - z_i^{\ce{T_d}}\quad. 
\end{equation}

\paragraph{Ensemble-VQE considerations.}
The Trotterized GUCCSD ansatz was employed with two repetitions because a standard single repetition did not appear to provide satisfactory
convergence of the ensemble-energy.
Heuristically speaking, the twofold Trotterized repetition seemed to be required here in order to obtain a fully-expressible ansatz.
This critical aspect will be further analyzed in more detail in future publications.
In any case, the ansatz was parameterized so as to preserve the total $z$-spin component ($M_S = 1/2)$, corresponding to two \(\alpha\)-spin electrons and one \(\beta\)-spin electron. Its parameters \(\boldsymbol{t}\) were initially set to zero and subsequently optimized by minimizing the function
defined in Eq.~\eqref{eq:ensemble_energy_vp}. The optimization was carried out using the aforementioned SLSQP algorithm~\cite{kraft1988slsqp} as implemented in the SciPy library~\cite{2020SciPy-NMeth}, with a maximum of 500 iterations and a 
convergence criterion of \(10^{-10}\). 
The tolerances \(\epsilon_S\) and \(\epsilon_r\) for the constraints given in Ineq.~\eqref{eq:spin_constraint} and Ineq.~\eqref{eq:r_constraint} were both set to \(10^{-8}\).
For the eigenstate determination 
and diabatic optimization, the angles \((\theta, \phi, \psi)\) were initially set to zero and then optimized upon minimizing the relevant objective function for the chosen approach.
Such optimizations were performed under the same iteration limit, with a 
convergence threshold of \(10^{-10}\).
Reference eigenstates \(\lbrace \ket{\Psi_n}\rbrace _{n \in \lbrace 0,1,2\rbrace }\), together with their eigenenergies, were obtained \emph{via} classical FCI calculations using Psi4.
The ensemble-VQE calculations were made using the open-source code available in Refs.~\citenum{beseda2024state,SAOOVQE}.

\paragraph{Implementation of the change of basis under a similarity transformation}

The implementation of the change of basis described in Eq.~\eqref{eq:paramterized_change_of_basis_rotation} on a quantum computer is illustrated in this section. Here, we adopt a convention for the ONV representation in which all \(\alpha\)-spin orbitals are ordered before \(\beta\)-spin orbitals.
Our reference states correspond respectively to
\(\ket{\Phi^{0}_{\mathrm {A}}} = \ket{11001000}\), \(\ket{\Phi^{0}_{\mathrm {B}}} = \ket{10101000}\), and \(\ket{\Phi^{0}_{\mathrm {C}}} = \ket{10011000}\).
Starting from an eight-qubit system initialized in the state \(\ket{10001000}\), a series of quantum gates is successively applied to obtain the parameterized states
\( \lbrace  \ket{\Phi^{0\prime}_I(\theta, \phi, \psi)} \rbrace _{I \in \lbrace A, B, C \rbrace  } \) such as described in Eq.~\eqref{eq:paramterized_change_of_basis_rotation}. In the following circuits, only the three qubits \(q_2, \, q_3, \, q_4\), on which the quantum gates act, are shown. 

\begin{figure*}
\centering
\resizebox{0.7\textwidth}{!}{
\begin{quantikz}[row sep=0.1cm, column sep=0.5cm, line width=0.1mm]
\lstick{  \(q_2 : \ket{0}\)}       & \qw               &  \gate{X} & \gate{X}  &  \ctrl{2}         & \gate{X}   & \qw      & \qw       & \qw               & \qw       & \qw     &  \qw    \\
\lstick{  \(q_3 : \ket{0}\)}       & \gate{R_{\bm{y}}(2\psi)} & \ctrl{-1} & \qw       & \qw               & \qw        & \ctrl{1} & \gate{X}  & \gate{R_{\bm{y}}(2\theta)} & \ctrl{1}  & \qw     &  \qw \\
\lstick{  \(q_4 : \ket{0}\)}       & \qw               & \qw      & \qw       & \gate{R_{\bm{y}}(2\phi)} &  \ctrl{-2} & \gate{X} & \ctrl{-1} & \ctrl{-1}         & \gate{X}  & \gate{Z} & \qw     \\
\\
\\
\\
\\
\\
\\
\\
\lstick{  \(q_2 : \ket{0}\)}       & \gate{R_{\bm{y}}(2\phi)} & \ctrl{1} & \qw       & \qw                  & \qw         & \gate{Z}       & \qw    & \qw & \qw & \qw       & \qw\\
\lstick{ \(q_3 : \ket{0}\)}       & \qw               & \gate{X} & \gate{X}  & \ctrl{1}             & \gate{X}    & \qw            & \qw    & \qw & \qw & \qw       & \qw \\
\lstick{ \(q_4 : \ket{0}\)}     & \qw               & \qw      & \qw       & \gate{R_{\bm{y}}(2\theta)}  & \ctrl{-1}   & \qw            & \qw    & \qw & \qw & \qw       & \qw \\
\\
\\
\\
\\
\\
\\
\\
\lstick{  \(q_2 : \ket{0}\)}      & \gate{R_{\bm{y}}(2\psi)} & \ctrl{2} & \qw       & \ctrl{1}          & \gate{X}   & \qw       & \qw                 & \qw       & \qw      & \qw       & \qw   \\
\lstick{  \(q_3 : \ket{0}\)}      & \qw               & \qw      & \qw       & \gate{R_{\bm{y}}(2\phi)} & \ctrl{-1}  & \gate{X}  & \ctrl{1}            & \gate{X}  & \qw   & \qw       & \qw   \\
\lstick{  \(q_4 : \ket{0}\)}      & \qw               & \gate{X} & \gate{X}  & \qw               & \qw        & \ctrl{-1} & \gate{R_{\bm{y}}(2\theta)} & \ctrl{-1}  & \qw  & \qw       & \qw     
\end{quantikz}
}
\caption{Quantum circuits for preparing the state \(\ket{\Phi^{0\prime}_{\mathrm {A}}(\theta, \phi, \psi)}\), \(\ket{\Phi^{0\prime}_{\mathrm {B}}(\theta, \phi, \psi)}\), and \(\ket{\Phi^{0\prime}_{\mathrm {C}}(\theta, \phi, \psi)}\) from the initial state \(\ket{10001000}\). The circuits are shown from top to bottom in the order A, B, C.}
\label{fig:circuit_phi_prime}
\end{figure*}

In the context of achieving optimal diabaticity, if \(\det\left(\bm{Q}_{\star}(\bm{t}^*) \right) = -1\), the quantum circuit for preparing the state \(\ket{\Phi^{0\prime}_{\mathrm {C}}(\theta, \phi, \psi)}\) from \(\ket{10001000}\) is the same as that shown in Fig.~\ref{fig:circuit_phi_prime}, with an additional \(Z\) gate applied to qubit \(q_1\).
In the quantum circuits shown in Fig.~\ref{fig:circuit_phi_prime}, quantum gates are applied from left to right, following the usual convention. The Pauli-\(X\), Pauli-\(Z\), and rotation \( R_{\bm{y}}(\theta) \) quantum gates used in these circuits follow standard definitions (see Ref.~\citenum{nielsen_quantum_2012}). In the case of a controlled gate, the filled circle (\( \bullet \)) denotes the control qubit, while the connected gate acts on the target qubit, applied conditionally on the control qubit being in the state $\ket{1}$. Details regarding the sequence of quantum operations implemented in the circuits are provided in the Appendix~\ref{app:circuit}.

\section{Results and Discussion}
\label{sec:results}

\subsection{Solving for eigenstates}
\label{subsec:eigenstates_results}

\begin{figure}

\begin{subfigure}{\columnwidth}
\centering
    \includegraphics[width=\columnwidth]{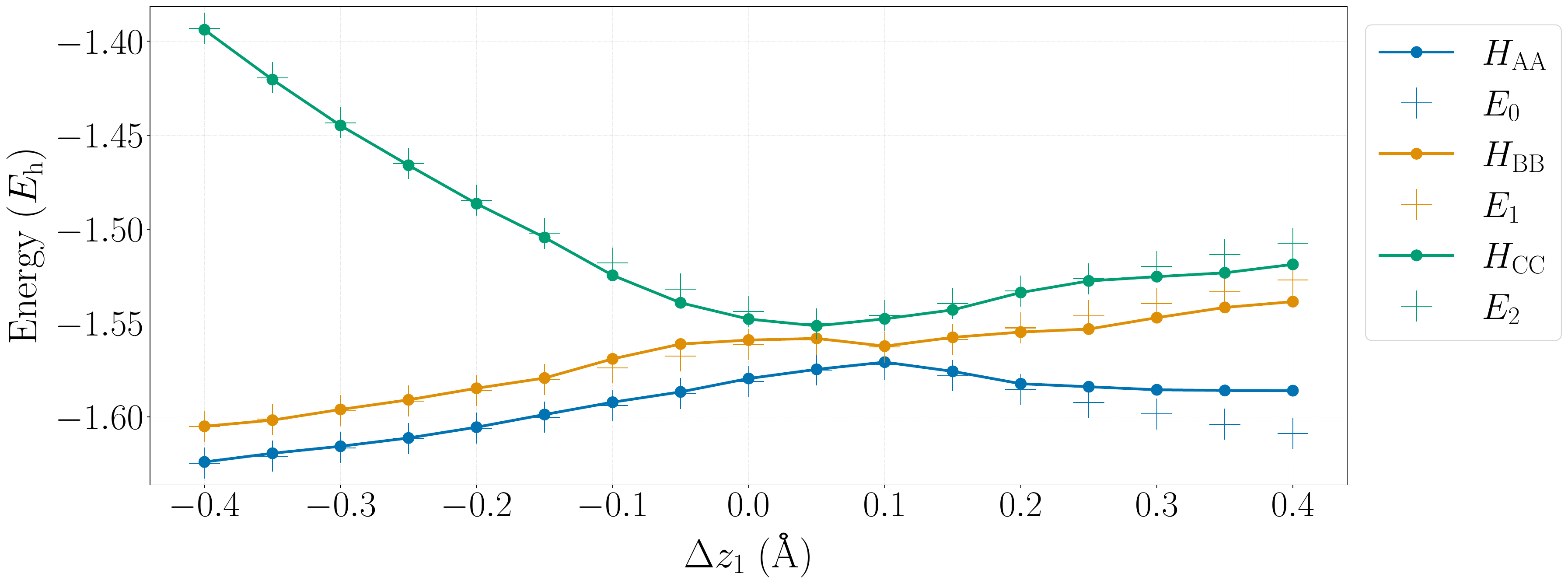}
    
   \caption{After determining the subspace of minimal ensemble-energy}
    \label{fig:solving_for_eigenstates_block_diagonalization}
  \end{subfigure}
  \hfill
  \begin{subfigure}{\columnwidth}
  \centering
   \includegraphics[width=\columnwidth]{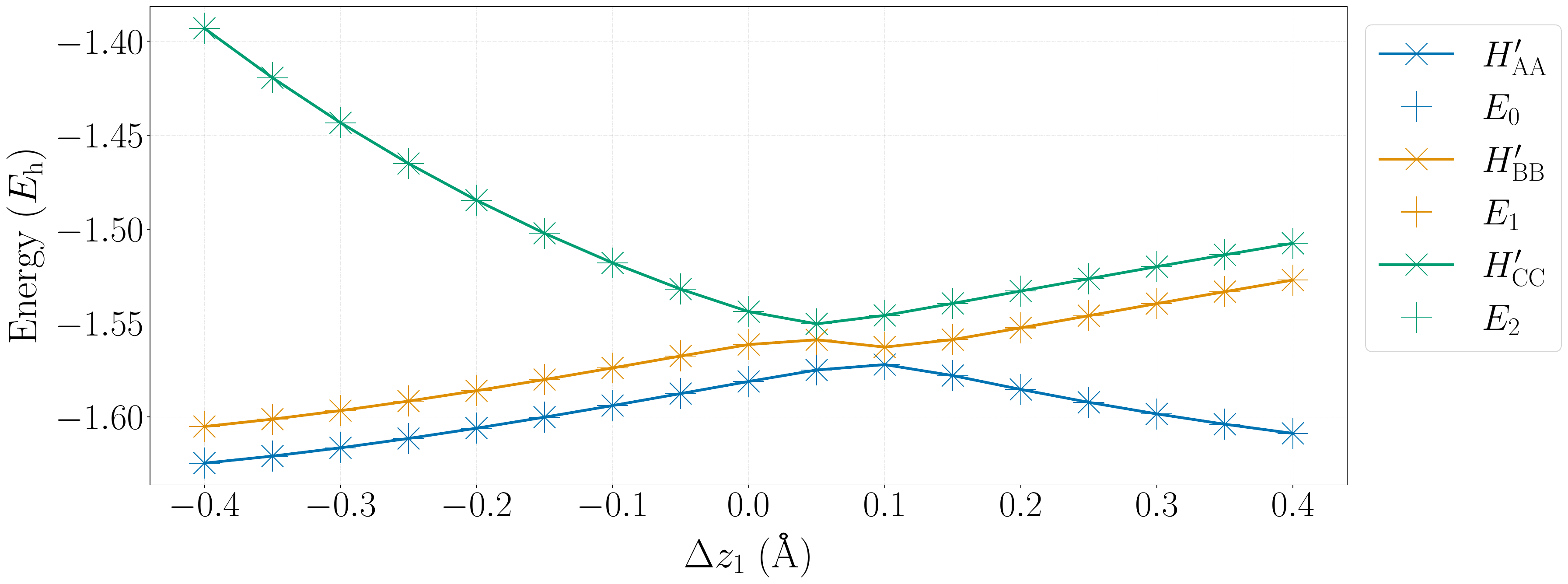}
    
    \caption{After eigenstate resolution}
     \label{fig:solving_for_eigenstates_diagonalization}
  \end{subfigure}
  \caption{ Energies of the ensemble-VQE states: (a) \( H_{\mathrm{AA}}, \, H_{\mathrm{BB}}, \, H_{\mathrm{CC}} \) after determining the subspace of minimal ensemble-energy; (b) \( H^{\prime}_{\mathrm{AA}}, \, H^{\prime}_{\mathrm{BB}}, \, H^{\prime}_{\mathrm{CC}} \) after eigenstate resolution (see Fig.~\ref{fig:circuit_phi_prime} and Eq.~(\ref{eq:similarity_transformation})). FCI energies of the first three eigenstates: \( E_0 , \,E_1, \, E_2 \). Canonical MOs are used.}
\end{figure}

Using the ensemble-VQE approach, the diagonal ensemble-variational energies, \( H_{\mathrm{AA}}, \, H_{\mathrm{BB}}, \, H_{\mathrm{CC}} \),
are shown in Fig.~\ref{fig:solving_for_eigenstates_block_diagonalization}.
They differ slightly from those of the first three eigenstates, \( E_0 , \,E_1, \, E_2 \).
The retro-variational ones, \( H^{\prime}_{\mathrm{AA}}, \, H^{\prime}_{\mathrm{BB}}, \, H^{\prime}_{\mathrm{CC}} \), fully match the eigenenergies as expected.
Interestingly enough, none of them have been post-reordered by hand, which is consistent with having used canonical MOs here for defining the model states according to the \emph{Aufbau} principle; however, this behavior cannot always be guaranteed and may depend on context.
For illustration purposes, a successful optimization deliberately started from a ``wrong guess'' (doubly excited configurations) is shown in the Appendix~\ref{app:initial_guess}.
The procedure is robust enough (and the ansatz expressible enough) that it did not end up at a local minimum.
Indeed, it yields correct values of the targeted eigenenergies at convergence, but now sorted erratically along consecutive molecular geometries.

\subsection{Achieving a quasi-diabatic representation}
\label{subsec:diabatic_results}

\subsubsection{After determining the subspace of minimal ensemble-energy}

\begin{figure}
\begin{subfigure}{\columnwidth}
\centering
   \includegraphics[width=\columnwidth]{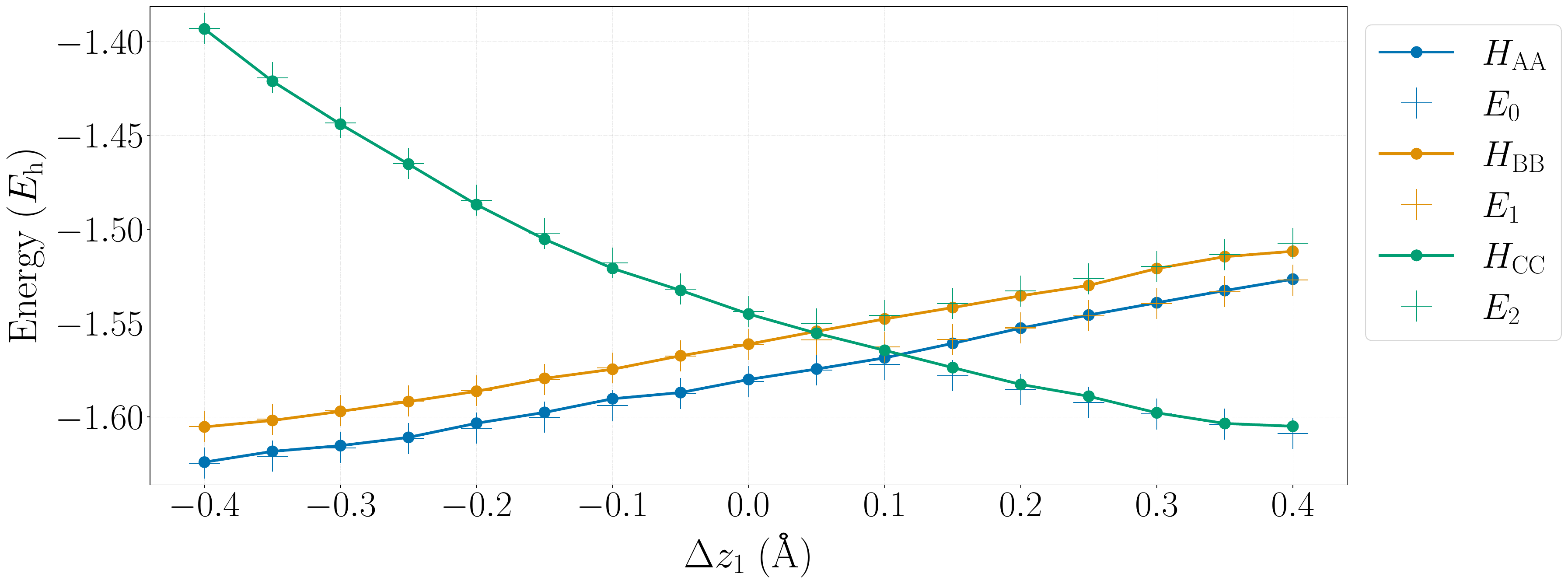}
    \caption{}
    \label{fig:diabatization_block_diagonalization_minus_zero_one}
  \end{subfigure}
  \hfill
\begin{subfigure}{\columnwidth}
\centering
   \includegraphics[width=\columnwidth]{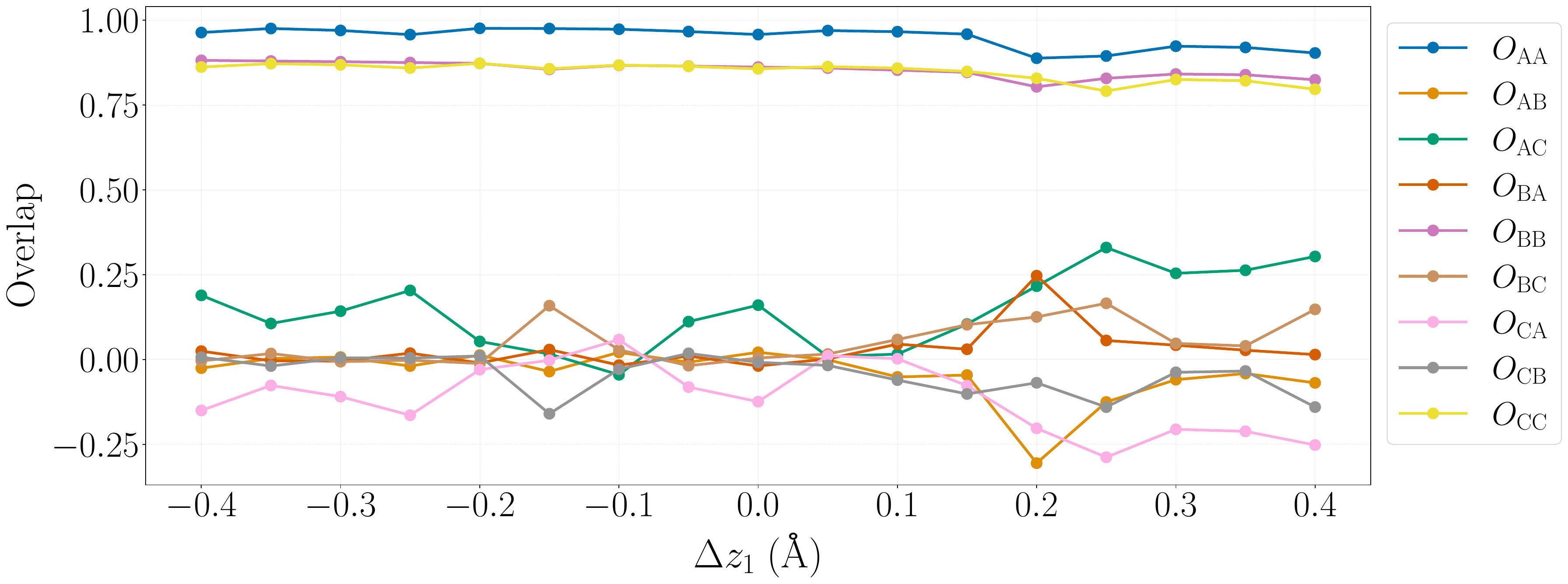}
    \caption{}
     \label{fig:o_matrix_minus_zero_one}
  \end{subfigure}
  \caption{(a) Energies of the ensemble-VQE states, \( H_{\mathrm{AA}}, \, H_{\mathrm{BB}}, \, H_{\mathrm{CC}} \), and FCI energies of the first three eigenstates, \( E_0 , \,E_1, \, E_2 \). (b) CI-coefficient (overlap) submatrix elements of the ensemble-VQE states $\lbrace \ket{\Phi_I({\bm{t}^*})}\rbrace_{I\in\lbrace \text{A},\text{B},\text{C}\rbrace}$. Diabatic orbitals are used.}
  \label{fig:diabatization_block_diagonalization}
\end{figure}

In what follows, we chose a $\ce{C_s}$ reference geometry, $\bm{R}^0$, for defining the diabatic orbitals~\cite{illesova_transformation-free_2025}, such that $\Delta x_2^0 = 0.1~\textrm{\AA}$ and $\Delta z_1^0 = -0.1~\textrm{\AA}$.
This small distortion from \ce{T_d} allows us to clearly discriminate the three singly-occupied orbitals (\ce{t_2}-degenerate in \ce{T_d}), hence to fix the choice of the three singly-excited model states that will specify the nature of the targeted diabatic representation. 
Our results are shown in Fig.~\ref{fig:diabatization_block_diagonalization_minus_zero_one} after the determination of the subspace of minimal ensemble-energy.
Although the energies of the states seem to follow a diabatic behavior, in contrast to Fig.~\ref{fig:solving_for_eigenstates_block_diagonalization} where they do not cross each others, they are not really smooth functions of the deformation coordinate.
Not surprisingly, this irregular behavior
is also observed for the elements of the overlap matrix along the deformation coordinate, as shown in 
Fig.~\ref{fig:o_matrix_minus_zero_one}.
Yet, the relatively large values of the diagonal terms indicate that the optimization procedure leads, as expected, to target states that remain relatively close to the initial model states (the ``guess''). 

\begin{figure}
    \begin{subfigure}{\columnwidth}
\centering
   \includegraphics[width=\columnwidth]{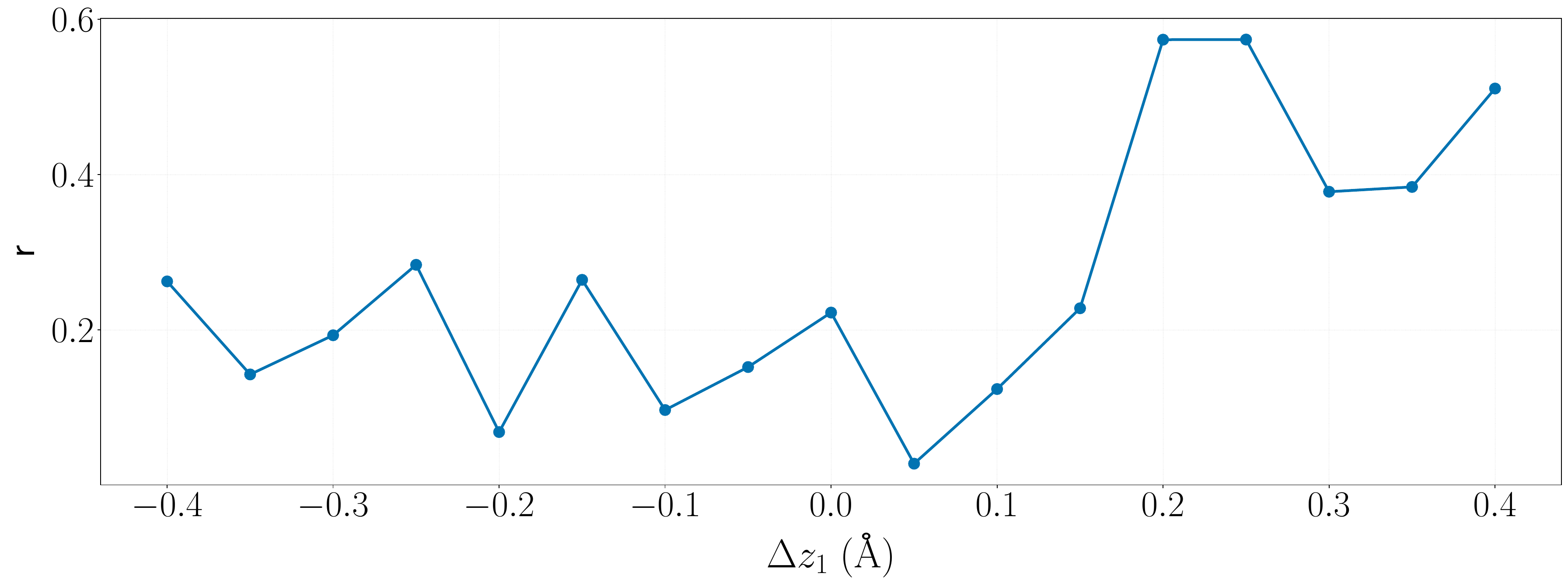}
    \end{subfigure}
    \hfill
    \begin{subfigure}{\columnwidth}
\centering
   \includegraphics[width=\columnwidth]{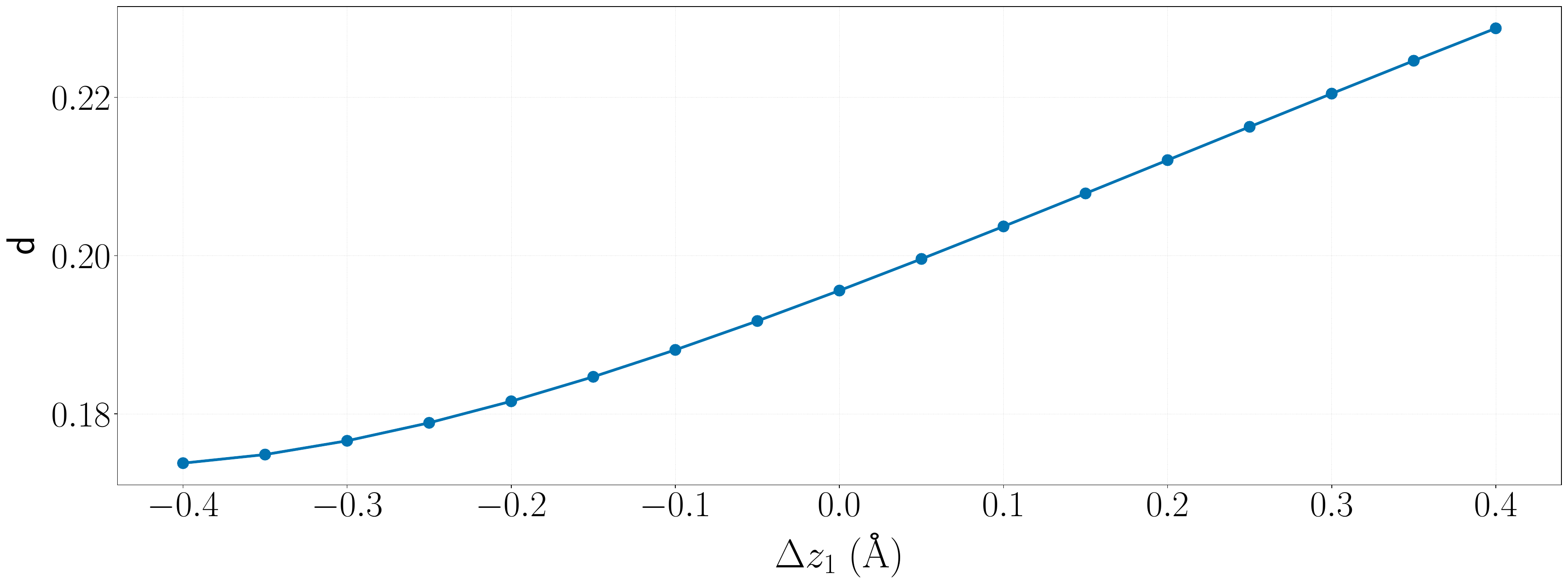}
  \end{subfigure}
  \caption{Diabaticity descriptors $r$ and $d$.}
    \label{fig:descriptors}
\end{figure}

In both cases, the overlap matrices are not symmetric, and for certain pairs of off-diagonal elements, such as \(O_{\mathrm{AC}}\) and \(O_{\mathrm{CA}}\), the values appear to be approximately opposite at some points along the deformation coordinate. The overall structure appears to correlate with the behavior of the \(r\) descriptor
in Fig.~\ref{fig:descriptors}.
For instance, Fig.~\ref{fig:o_matrix_minus_zero_one} reveals that two out of the three pairs of off-diagonal elements remain close to zero when \(\Delta z_1 \leq 0.15~\textrm{\AA}\), which appears to be consistent with the relatively small values of the \(r\) descriptor in this region.
Regarding the descriptors in Fig.~\ref{fig:descriptors}, it can be observed that the descriptor \(r\) is not smooth at all,
in contrast to the two-state case of the formaldimine molecule studied in Ref.~\citenum{illesova_transformation-free_2025}.
Further numerical tests indicate that a similar irregular behavior of the descriptor \(r\) is always observed, irrespective of the choice made for the reference geometry \(\bm{R}^{0}\) used to construct the diabatic MO basis (see the Appendix~\ref{app:impact_geometry}).
This may reflect some sensitivity of the results to the optimization procedure used in the determination of the subspace of minimal ensemble-energy, potentially due to the relatively large number of parameters in the ansatz. 
Regarding the descriptor \(d\), it is smooth
on the entire range of $\Delta z_1$.
This may be due, on the one hand, to the fact that the diabatic orbitals were constructed to vary smoothly with respect to \(\bm{R}\)-variations. On the other hand, \(d\) is invariant under a change of basis within the subspace of minimal ensemble-energy, and is therefore not sensitive to the way the states converge to this subspace.
Additional numerical experiments indicate that the value of \(d\) depends slightly on small variations in the reference geometry \(\bm{R}^{0}\) (see the Appendix~\ref{app:impact_geometry}).

Overall, while the ensemble-VQE in Ref.~\citenum{illesova_transformation-free_2025}
led to a quasi-diabatic representation ``for free'', a more robust implementation was required for the three-state case studied herein. 
In the following, we achieve optimal diabaticity following the strategy described in Sec.~\ref{subsec:opt_diab}, {\textit{i.e.}}, upon using the parameterized-circuit rotations aimed at minimizing the descriptor \(r\).

\subsubsection{After descriptor \(r\) minimization}

\begin{figure}

\begin{subfigure}{\columnwidth}
\centering
    \includegraphics[width=\columnwidth]{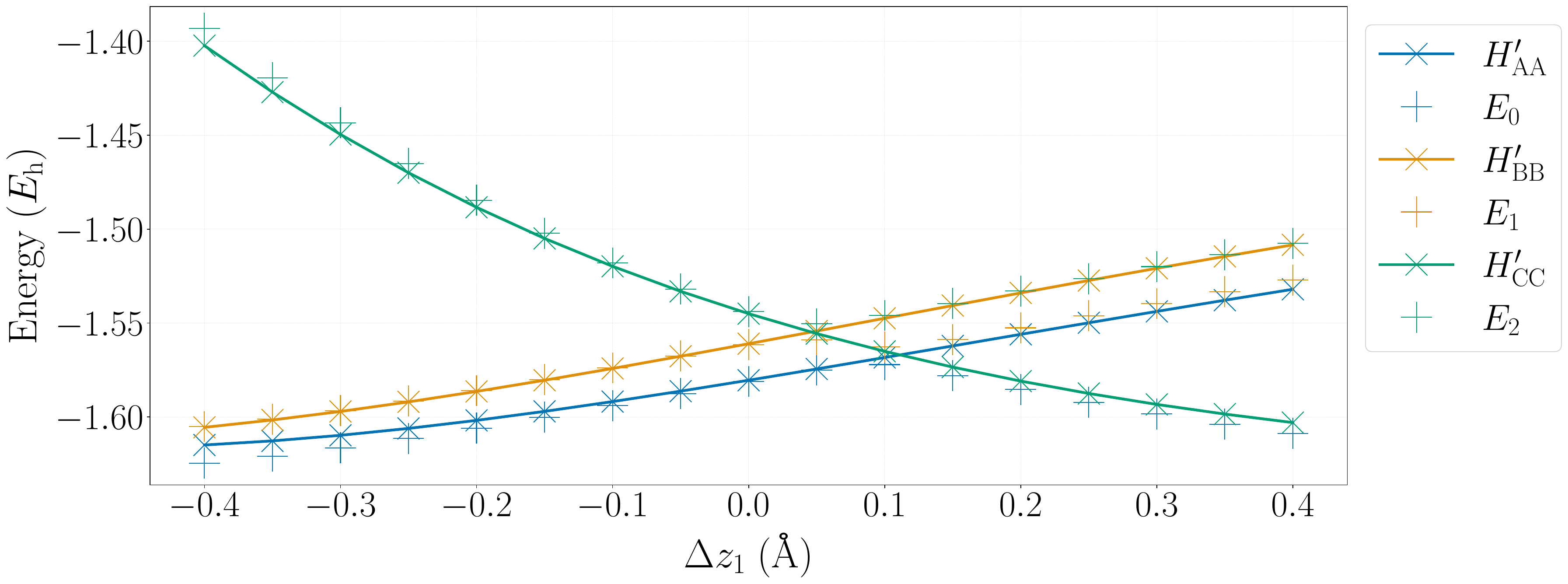}
   \caption{}
   \label{fig:energies_optimally_diabatic}
  \end{subfigure}
  \hfill
  \begin{subfigure}{\columnwidth}
  \centering
   \includegraphics[width=\columnwidth]{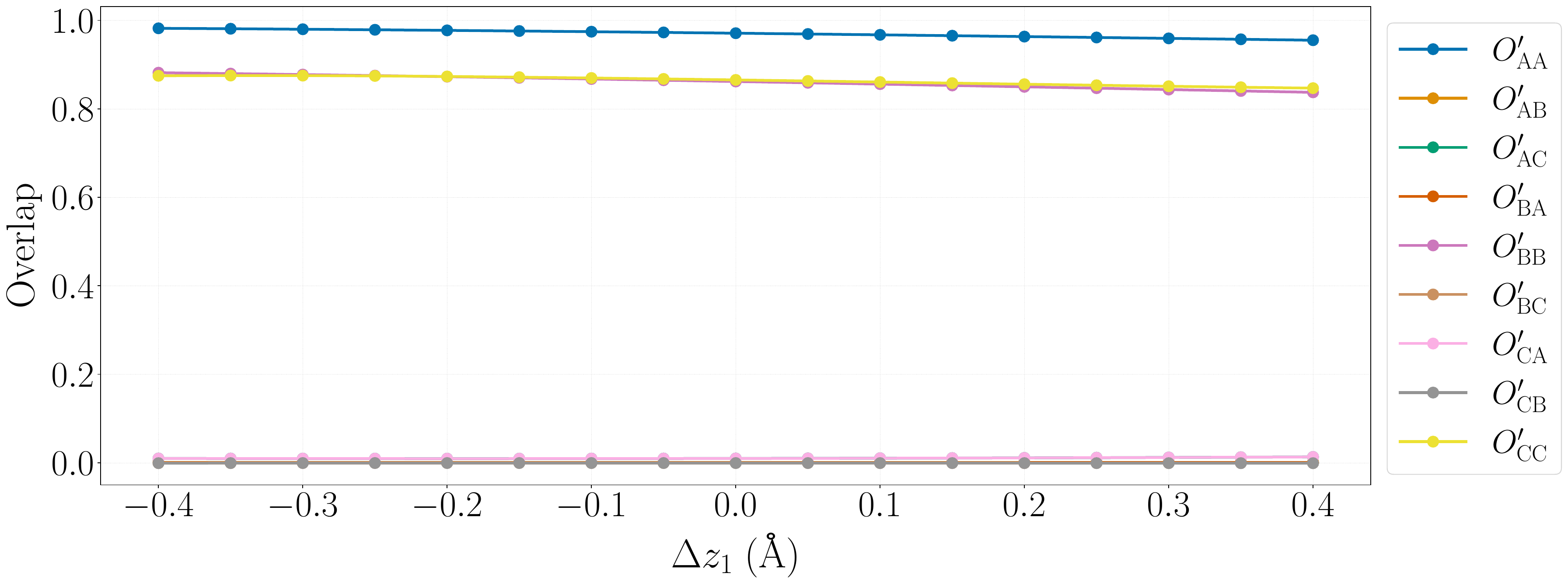}
    \caption{}
  \label{fig:o_star_matrix_minus_zero_one}
  \end{subfigure}
  \caption{(a) Energies of the ensemble-VQE states, \( H^{\prime}_{\mathrm{AA}}, \, H^{\prime}_{\mathrm{BB}}, \, H^{\prime}_{\mathrm{CC}} \), and FCI energies of the first three eigenstates, \( E_0 , \,E_1, \, E_2 \). (b) CI-coefficient (overlap) submatrix elements of the ensemble-VQE states $\lbrace \ket{\Phi'_I({\bm{t}^*,\theta_\star, \phi_\star, \psi_\star)})}\rbrace_{I\in\lbrace \text{A},\text{B},\text{C}\rbrace}$. Diabatic orbitals are used.}
  \label{fig:diabatization_r_minimisation}
\end{figure}

From Fig.~\ref{fig:energies_optimally_diabatic}, it is clear that the diabatic energies obtained after the minimization of the descriptor \(r\) by a change of basis
(as discussed at the end of Sec.~\ref{subsec:opt_diab}) are smooth over the entire geometry profile, as expected.
The corresponding values of \(r\) are close to zero (it remains below \(2 \times 10^{-8}\) for all values of \(\Delta z_1\)), suggesting that the solution is numerically close to the optimal diabatic limit.
Indistinguishable results are obtained by minimizing $r(\bm{t})$ without a change of basis,
{\it i.e.}, when it is incorporated into the objective function via a Lagrangian formulation, as also explained in Sec.~\ref{subsec:opt_diab}.
From Fig.~\ref{fig:o_star_matrix_minus_zero_one}, it can be observed that the overlap matrix is effectively symmetric, with elements varying only slightly along the deformation coordinate. 
It can be noted that the diagonal elements of the overlap matrix obtained after the determination of the subspace of minimal ensemble-energy in Fig.~\ref{fig:o_matrix_minus_zero_one} were relatively close to their optimal values, which seems to support the notion that the optimizer followed a relatively direct path toward the subspace of minimal ensemble-energy, without fully exploring the variational space. 
Consistent with the slight dependence of the descriptor \(d\) on small variations of the reference geometry \(\bm{R}^{0}\) used to construct the diabatic MO basis, the optimized diabatic states also vary slightly with the choice of \(\bm{R}^{0}\) (see the Appendix~\ref{app:impact_geometry}).

\begin{figure}
 \begin{subfigure}{\columnwidth}
\centering
    \includegraphics[width=\columnwidth]{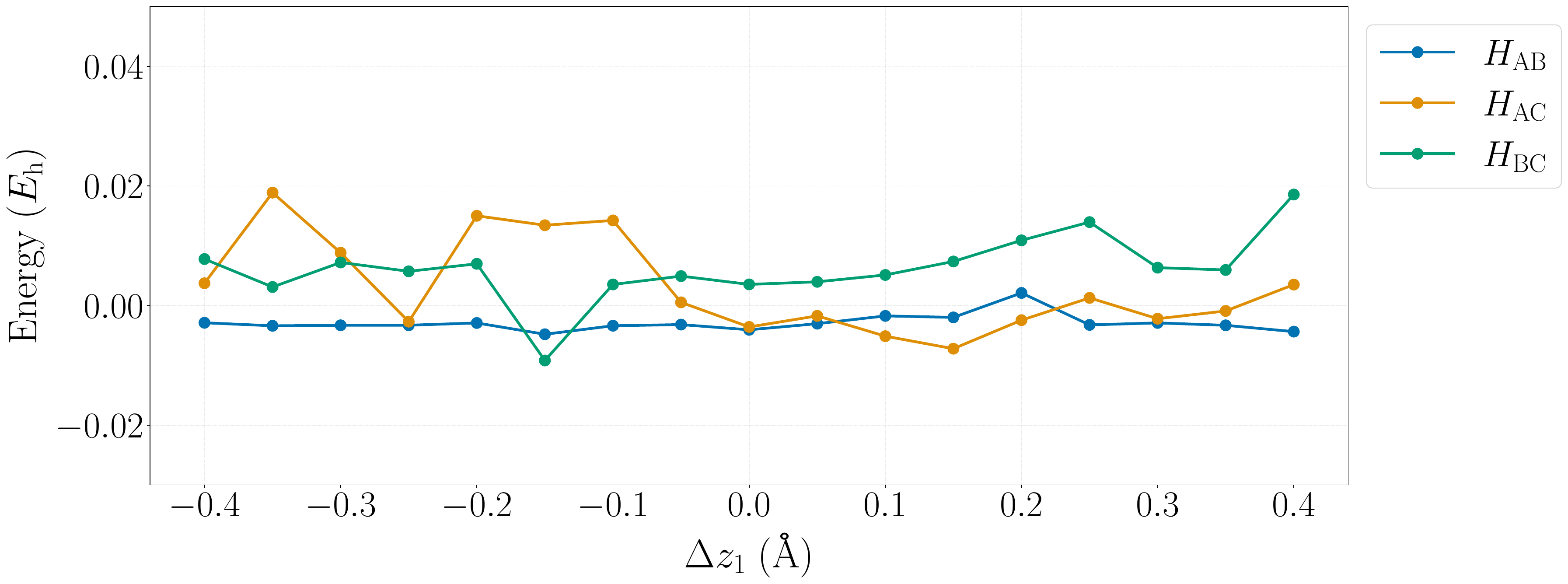}
    \caption{}
  \end{subfigure}
  \hfill
\begin{subfigure}{\columnwidth}
\centering
    \includegraphics[width=\columnwidth]{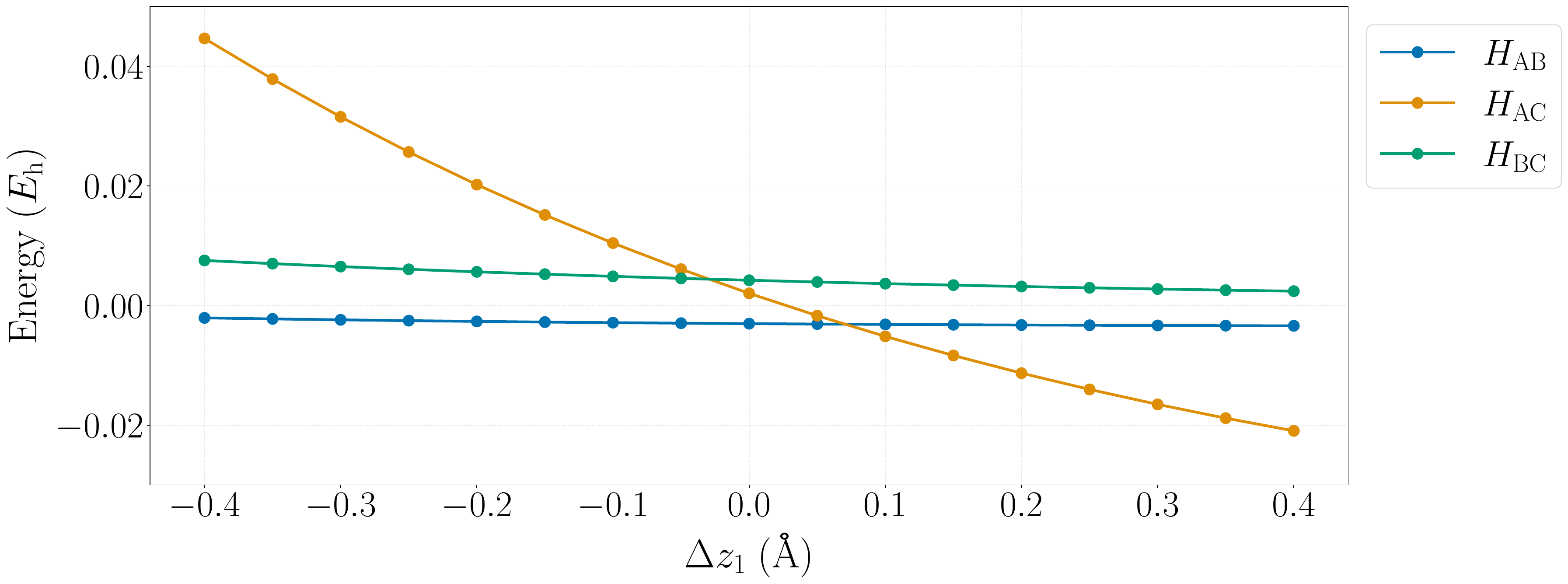}
    \caption{}
  \end{subfigure}
   \caption{(a) Off-diagonal Hamiltonian matrix elements in the basis of the ensemble-VQE states \(\lbrace \ket{\Phi_I(\boldsymbol{t}^*)} \rbrace _{I \in \lbrace \mathrm {A},\mathrm {B},\mathrm {C}\rbrace }\) (see Figs.~\ref{fig:diabatization_block_diagonalization_minus_zero_one} and~\ref{fig:o_matrix_minus_zero_one}). (b) Off-diagonal Hamiltonian matrix elements in the basis of the ensemble-VQE states \( \lbrace \ket{\Phi^{\prime}_I(\boldsymbol{t}^*, \theta_\star, \phi_\star, \psi_\star)} \rbrace _{I \in \lbrace \mathrm {A},\mathrm {B},\mathrm {C}\rbrace }  \) (see Figs.~\ref{fig:energies_optimally_diabatic} and~\ref{fig:o_star_matrix_minus_zero_one}). Diabatic orbitals are used.}
 \label{fig:off_diagonal_hamiltonian_matrix_c1_optimally_diabatic}
\end{figure}

Finally,
as can be seen in Fig.~\ref{fig:off_diagonal_hamiltonian_matrix_c1_optimally_diabatic},
the off-diagonal Hamiltonian matrix elements 
before the change-of-basis
are not smooth in contrast to Ref.~\citenum{illesova_transformation-free_2025}, despite using diabatic orbitals, suggesting again that the optimal diabaticity is more difficult to achieve and requires additional care.
After the change of basis, the off-diagonal Hamiltonian matrix elements become smooth and are still nonzero, which is consistent with the \ce{C_1} point group where no symmetry consideration enforces them to vanish.
Note that we also checked that
two out of the three off-diagonal Hamiltonian matrix elements become zero when considering the \ce{C_s} point group, and
the three off-diagonal Hamiltonian matrix elements are zero at \(\Delta z_1 = 0.0\) when considering the \ce{T_d} point group, due to symmetry considerations (see the Appendices~\ref{si_subsubsec:distortion_cs} and \ref{si_subsubsec:distortion_c3v}).

\section{Conclusions and Outlooks}
\label{sec:conclu}

In photochemistry,
the democratic description of quasi-degenerate states -- such as the states
involved in conical intersections or avoided crossings --
is crucial and can be achieved
by using the equi-weighted ensemble-VQE algorithm~\cite{yalouz_state-averaged_2021}.
However, the latter only allows for a block-diagonalization of the Hamiltonian, and
preparing the eigenstates
requires an extra quantum circuit to implement the rotation between the many-body basis states, dependent on the initial model states.
This additional step was designed and
successfully implemented for
two states in our previous work~\cite{yalouz_analytical_2022},
when the model states are the Hartree--Fock one and a spin-singlet singly-excited CSF.
In the present study,
we applied the algorithm to a three-state case
(the H$_4^+$ molecular ion)
where the model states are spin-doublet Slater determinants differing by a single electron excitation.
Together with the design of specific quantum circuits to perform rotations among these states,
we also derived several generally applicable objective functions,
thus extending the approach to any number of states.
In addition, being inspired by the recent finding
that ensemble-VQE has the propensity to
prepare quasi-diabatic states~\cite{illesova_transformation-free_2025},
we developed
a rigorous strategy to
prepare optimal quasi-diabatic states
with no additional cost compared to the adiabatic ones, while still avoiding any challenging adiabatic-to-diabatic transformation based on NAC integration.
In conclusion, the extensions
provided in the present work pave the way toward
excited-state quantum dynamics, 
such as ($i$) mixed quantum-classical trajectory-based methods (known as `nonadiabatic molecular dynamics') using knowledge of the analytical gradients and NACs within the adiabatic basis,
or ($ii$) nuclear wavepacket methods, beyond the BOA,
for which diabatic states are to be preferred in order to build a practical Hamiltonian matrix representation.

\section*{Author Contributions}

The three co-authors contributed equally to this work.

\section*{Acknowledgments}

This work benefited through the QARES project from State support managed by the ANR under the France 2030 program (ANR-23-PETQ-0006) and from the European Union through the QLASS project (EU Horizon91
Europe grant agreement 101135876).

\appendix

\section{Various flavors of the eigenproblem objective}
\label{si_subsubsec:various_flavors_eigenproblem_objective}
To analyze the stability of the functions \(f^{\mathrm{F}}\) defined Eq.~\ref{objective_func_f_f} and \(f^{\mathrm{w}}\) defined Eq.~\ref{objective_func_f_w}, we consider small variations \(\lbrace \delta H'_{II}\rbrace _{I \in \lbrace A,B,C\rbrace }\) in the diagonal components \(\lbrace  H'_{II}\rbrace _{I \in \lbrace A,B,C\rbrace }\), resulting from perturbations of the rotation angles \((\theta,\phi,\psi)\). We define \(\delta f = f(H + \delta H) - f(H)\), leading to
\begin{subequations}
\begin{align}
& \delta f^{\mathrm{F}} = - \sum_{I \in \lbrace A,B,C\rbrace } \left( 2 H'_{II} \, \delta H'_{II} + (\delta H'_{II})^2 \right) \quad, \\
& \delta f^{\mathrm{w}} = \sum_{I \in \lbrace A,B,C\rbrace } w_I \, \delta H'_{II} \quad.
\end{align}
\end{subequations}

While \(f^{\mathrm{w}}\) varies directly with perturbations in the energies, \(f^{\mathrm{F}}\) varies proportionally to the energies themselves. This analysis can be related to classical sensitivity analysis, which aims at estimating the variation of an objective function induced by small perturbations of the underlying parameters~\cite{laporte_numerical_2003}.
A more detailed analysis of such aspects will be provided in future work. 

\section{Molecular-orbital representation}

\begin{figure}[H]
  \centering
  \begin{subfigure}[b]{0.17\columnwidth}
    \includegraphics[width=\columnwidth]{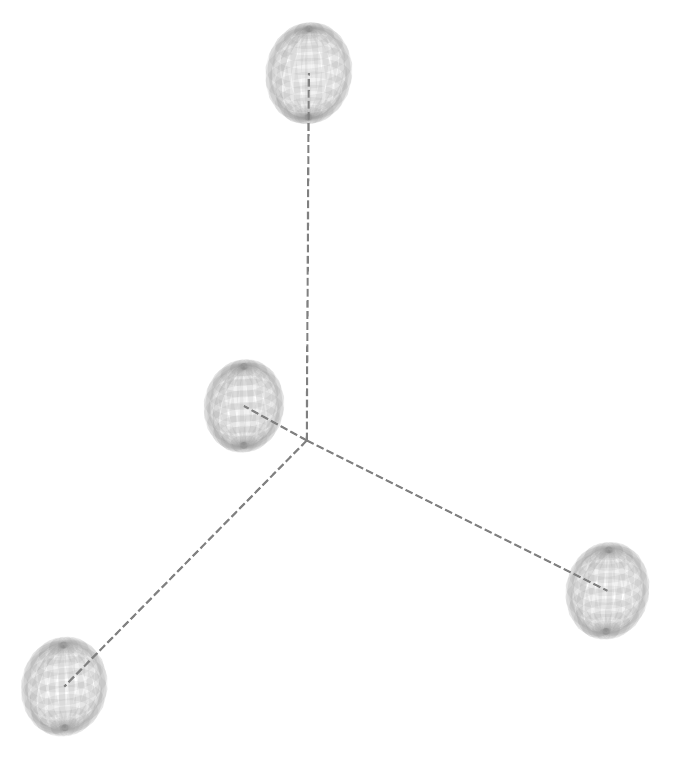}
    \caption{\ce{a_1} \\orbital}
  \end{subfigure}
  \hfill
  \begin{subfigure}[b]{0.17\columnwidth}
    \includegraphics[width=\columnwidth]{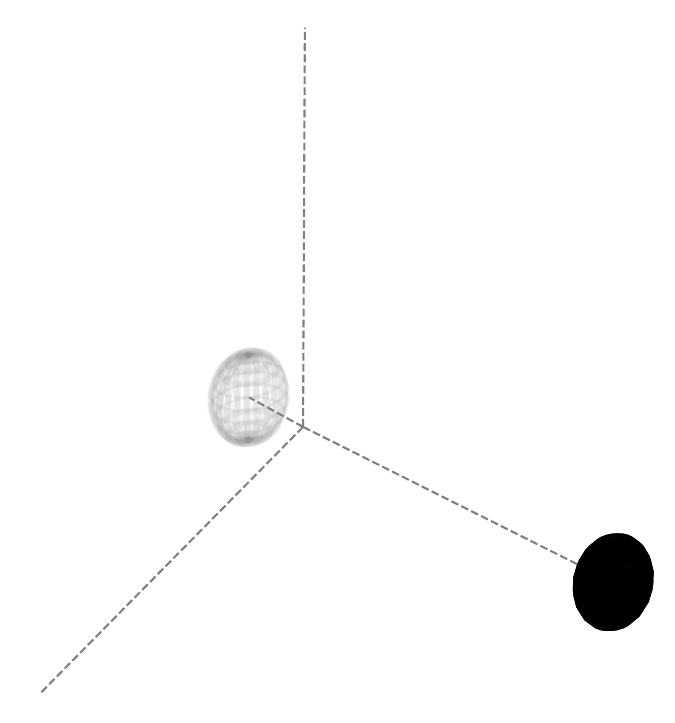}
    \caption{\ce{t_2}  \\orbital}
  \end{subfigure}
  \hfill
  \begin{subfigure}[b]{0.17\columnwidth}
    \includegraphics[width=\columnwidth]{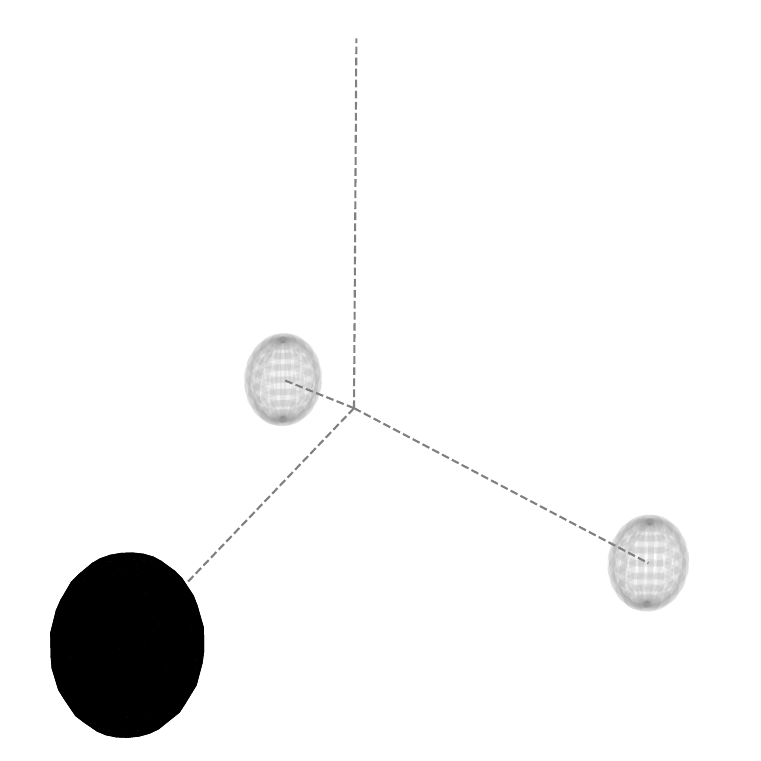}
    \caption{\ce{t_2} \\orbital}
  \end{subfigure}
  \hfill
  \begin{subfigure}[b]{0.17\columnwidth}
    \includegraphics[width=\columnwidth]{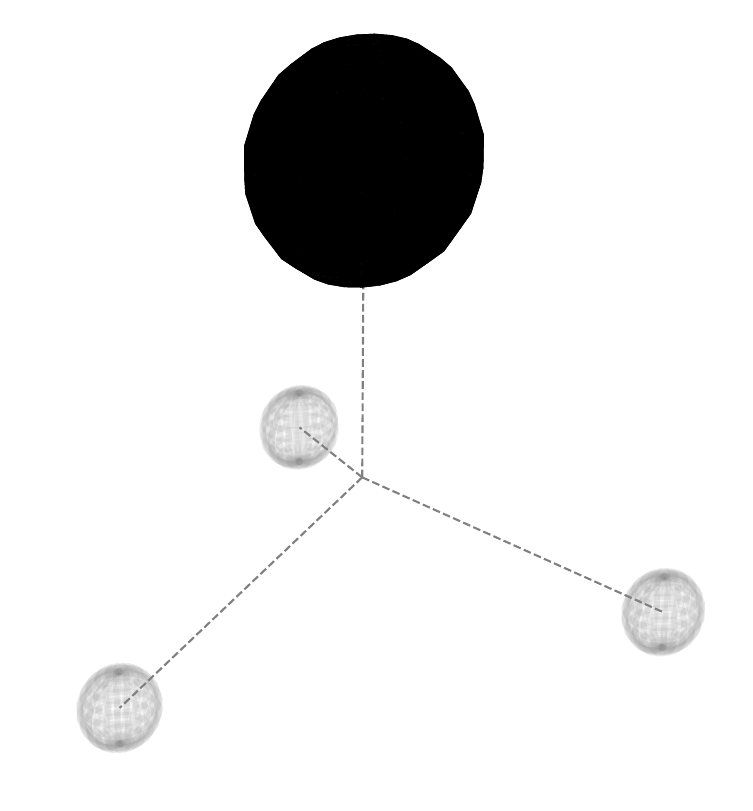}
    \caption{\ce{t_2} \\orbital}
  \end{subfigure}
  \caption{Molecular-orbital representation of H$_4^+$ in the tetrahedral geometry}
\end{figure}

\section{Implementation of the change of basis under a similarity transformation}
\label{app:circuit}

In the following, we describe the sequence of quantum operations implemented in the circuits, shown in Table~\ref{fig:circuit_phi_prime}, for preparing the states \(\ket{\Phi^{0\prime}_{\mathrm {A}}(\theta, \phi, \psi)}\), \(\ket{\Phi^{0\prime}_{\mathrm {B}}(\theta, \phi, \psi)}\), and \(\ket{\Phi^{0\prime}_{\mathrm {C}}(\theta, \phi, \psi)}\) of the $a_1^2t_2^1$-open-shell H$_4^+$ system from the initial state \(\ket{10001000}\) corresponding to various ways of adding one extra $\alpha$-electron (into the $t_2$-manifold) to the $a_1^2$-closed-shell H$_4^{2+}$ system.
The latter is first prepared by applying \(X\) gates to qubits 1 and 5.
Here, \(X_{(i)}\), \(Z_{(i)}\), \(R_{\bm{y}}(\theta)_{(i)}\) denote \(X\), \(Z\), \(R_{\bm{y}}(\theta)\) quantum gates with target qubit \(i\). Moreover, \(CX_{(i,j)}\) and \(CR_{\bm{y}}(\theta)_{(i,j)}\) denote controlled-\(X\) and controlled-\(R_{\bm{y}}(\theta)\) quantum gates, respectively, with control qubit \(i\) and target qubit \(j\).

\subsection{Quantum circuit for preparing the state \(\ket{\Phi^{0\prime}_{\mathrm {A}}(\theta, \phi, \psi)}\)}

\subsubsection*{\(R_{\mathbf{y}}(2 \psi)_{(3)}\)}
\begin{equation}
    \cos (\psi) \ket{10001000}  + \sin (\psi) \ket{10101000}  
\end{equation}

\subsubsection*{\( CX_{(3,2)}\)}
\begin{equation}
    \cos (\psi) \ket{10001000}  + \sin (\psi) \ket{11101000}  
\end{equation}

\subsubsection*{\(X_{(2)}\)}
\begin{equation}
    \cos (\psi) \ket{11001000}  + \sin (\psi) \ket{10101000}   
\end{equation}

\subsubsection*{\(CR_{\mathbf{y}}(2 \phi)_{(2,4)}\)}
\begin{eqnarray}
  && \cos(\phi)  \cos (\psi) \ket{11001000}  + \sin(\phi) \cos (\psi) \ket{11011000} \nonumber \\
    &&+ \sin (\psi) \ket{10101000}  
\end{eqnarray}

\subsubsection*{\(CX_{(4,2)}\)}
\begin{eqnarray}
  &&   \cos(\phi)  \cos (\psi) \ket{11001000}  + \sin(\phi) \cos (\psi) \ket{10011000}  \nonumber \\
    && + \sin (\psi) \ket{10101000}  
\end{eqnarray}

\subsubsection*{\(CX_{(3,4)}\)}
\begin{eqnarray}
   &&   \cos(\phi)  \cos (\psi) \ket{11001000}  + \sin(\phi) \cos (\psi) \ket{10011000} \nonumber \\
   && + \sin (\psi) \ket{10111000}    
\end{eqnarray}

\subsubsection*{\(CX_{(4,3)}\)}
\begin{eqnarray}
    && \cos(\phi)  \cos (\psi) \ket{11001000}  + \sin(\phi) \cos (\psi) \ket{10111000} \nonumber \\
    && + \sin (\psi) \ket{10011000} 
\end{eqnarray}

\subsubsection*{\(CR_{\mathbf{y}}(2 \theta)_{(4,3)}\)}
\begin{equation}
\begin{aligned}
& \cos(\phi)  \cos (\psi) \ket{11001000}  - \sin(\theta) \sin(\phi) \cos (\psi) \ket{10011000}\\
& +  \cos(\theta) \sin(\phi) \cos (\psi) \ket{10111000} 
 + \cos(\theta) \sin (\psi) \ket{10011000}   \\
 & + \sin(\theta) \sin (\psi) \ket{10111000}    
\end{aligned}
\end{equation}

\subsubsection*{\(CX_{(3,4)}\)}
\begin{equation}
\begin{aligned}
& \cos(\phi)  \cos (\psi) \ket{11001000}  - \sin(\theta) \sin(\phi) \cos (\psi) \ket{10011000}  \\
&+  \cos(\theta) \sin(\phi) \cos (\psi) \ket{10101000} + \cos(\theta) \sin (\psi) \ket{10011000}  \\
&  + \sin(\theta) \sin (\psi) \ket{10101000}    
\end{aligned}
\end{equation}

\subsubsection*{\(Z_{(4)}\)}
\begin{equation}
\begin{aligned}
& \cos(\phi)  \cos (\psi) \ket{11001000} \\
&+ \left( \sin(\theta) \sin(\phi) \cos (\psi) -  \cos(\theta) \sin (\psi)\right)  \ket{10011000}   \\
 & + \left(  \cos(\theta) \sin(\phi) \cos (\psi)    + \sin(\theta) \sin (\psi) \right) \ket{10101000}    
\end{aligned}
\end{equation}

\subsection{Quantum circuit for preparing the state \(\ket{\Phi^{0\prime}_{\mathrm {B}}(\theta, \phi, \psi)}\)}

\subsubsection*{\(R_{\mathbf{y}}(2 \phi)_{(2)}\)}
\begin{equation}
    \cos (\phi) \ket{10001000}  + \sin (\phi) \ket{11001000}  
\end{equation}

\subsubsection*{\(CX_{(2,3)}\)}
\begin{equation}
   \cos (\phi) \ket{10001000}  + \sin (\phi) \ket{11101000}  
\end{equation}

\subsubsection*{\(X_{(3)}\)}
\begin{equation}
   \cos (\phi) \ket{10101000}  + \sin (\phi) \ket{11001000}  
\end{equation}

\subsubsection*{\(R_{\mathbf{y}}(2 \theta)_{(3,4)}\)}
\begin{eqnarray}
 &&  \cos (\theta) \cos (\phi) \ket{10101000} +  \sin (\theta) \cos (\phi) \ket{10111000} \nonumber \\
   && + \sin (\phi) \ket{11001000}  
\end{eqnarray}

\subsubsection*{\(CX_{(4,3)}\)}
\begin{eqnarray}
  && \cos (\theta) \cos (\phi) \ket{10101000} +  \sin (\theta) \cos (\phi) \ket{10011000} \nonumber \\
   && + \sin (\phi) \ket{11001000}  
\end{eqnarray}

\subsubsection*{\(Z_{(2)}\)}
\begin{eqnarray}
   &&\cos (\theta) \cos (\phi) \ket{10101000} +  \sin (\theta) \cos (\phi) \ket{10011000}\nonumber \\
   &&  - \sin (\phi) \ket{11001000}  
\end{eqnarray}

\subsection{Quantum circuit for preparing the state \(\ket{\Phi^{0\prime}_{\mathrm {C}}(\theta, \phi, \psi)}\)}

\subsubsection*{\(R_{\mathbf{y}}(2 \psi)_{(2)}\)}
\begin{equation}
    \cos (\psi) \ket{10001000}  + \sin (\psi) \ket{11001000}  
\end{equation}

\subsubsection*{\( CX_{(2,4)}\)}
\begin{equation}
 \cos (\psi) \ket{10001000}  + \sin (\psi) \ket{11011000}  
\end{equation}

\subsubsection*{\( X_{(4)}\)}
\begin{equation}
 \cos (\psi) \ket{10011000}  + \sin (\psi) \ket{11001000}  
\end{equation}

\subsubsection*{\(CR_{\mathbf{y}}(2 \phi)_{(2,3)}\)}
\begin{eqnarray}
  && \cos (\psi) \ket{10011000} + \cos(\phi) \sin (\psi) \ket{11001000}\nonumber \\
  && +
   \sin(\phi) \sin (\psi) \ket{11101000}
\end{eqnarray}

\subsubsection*{\( CX_{(3,2)}\)}
\begin{eqnarray}
   &&\cos (\psi) \ket{10011000} + \cos(\phi) \sin (\psi) \ket{11001000} \nonumber \\
   &&+
   \sin(\phi) \sin (\psi) \ket{10101000}
\end{eqnarray}

\subsubsection*{\( CX_{(4,3)}\)}
\begin{eqnarray}
  && \cos (\psi) \ket{10111000} + \cos(\phi) \sin (\psi) \ket{11001000}\nonumber \\
  && +
   \sin(\phi) \sin (\psi) \ket{10101000}
\end{eqnarray}

\subsubsection*{\(CR_{\mathbf{y}}(2 \theta)_{(3,4)}\)}
\begin{equation}
\begin{aligned}
&  -\sin(\theta) \cos (\psi) \ket{10101000} +
  \cos(\theta) \cos (\psi) \ket{10111000} \\
  &
  + \cos(\phi) \sin (\psi) \ket{11001000}  
   + \cos(\theta) \sin(\phi) \sin (\psi) \ket{10101000} \\
   &+
   \sin(\theta) \sin(\phi) \sin (\psi) \ket{10111000}
   \end{aligned}
\end{equation}

\subsubsection*{\( CX_{(4,3)}\)}
\begin{equation}
\begin{aligned}
&  (\cos(\theta) \sin(\phi) \sin (\psi) -\sin(\theta) \cos (\psi)) \ket{10101000}\\
& +
  (\cos(\theta) \cos (\psi) + \sin(\theta) \sin(\phi) \sin (\psi) \ket{10011000}  \\
  & + \cos(\phi) \sin (\psi) \ket{11001000} 
   \end{aligned}
\end{equation}

\section{Distortion within \ce{C_1} geometry (from $\Delta x_2 = 0.1~\textrm{\AA}$ and $\Delta y_3 = 0.05~\textrm{\AA}$): impact of the initial guess states}\label{app:initial_guess}

\begin{figure}
\begin{subfigure}[t]{\columnwidth}
\centering
    \includegraphics[width=\columnwidth]{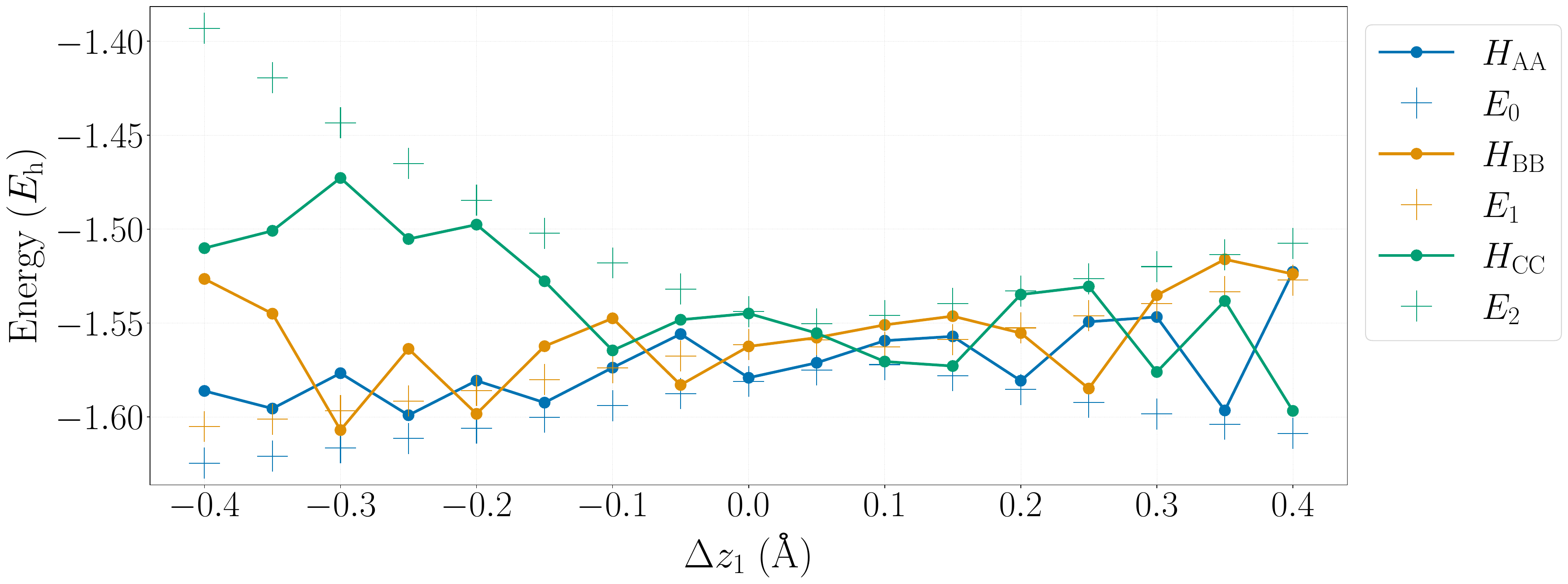}
    
   \caption{After determining the subspace of minimal ensemble-energy}
    \label{fig:solving_for_eigenstates_block_diagonalization_initial_states_impact}
  \end{subfigure}
  \hfill
  \begin{subfigure}[t]{\columnwidth}
  \centering
   \includegraphics[width=\columnwidth]{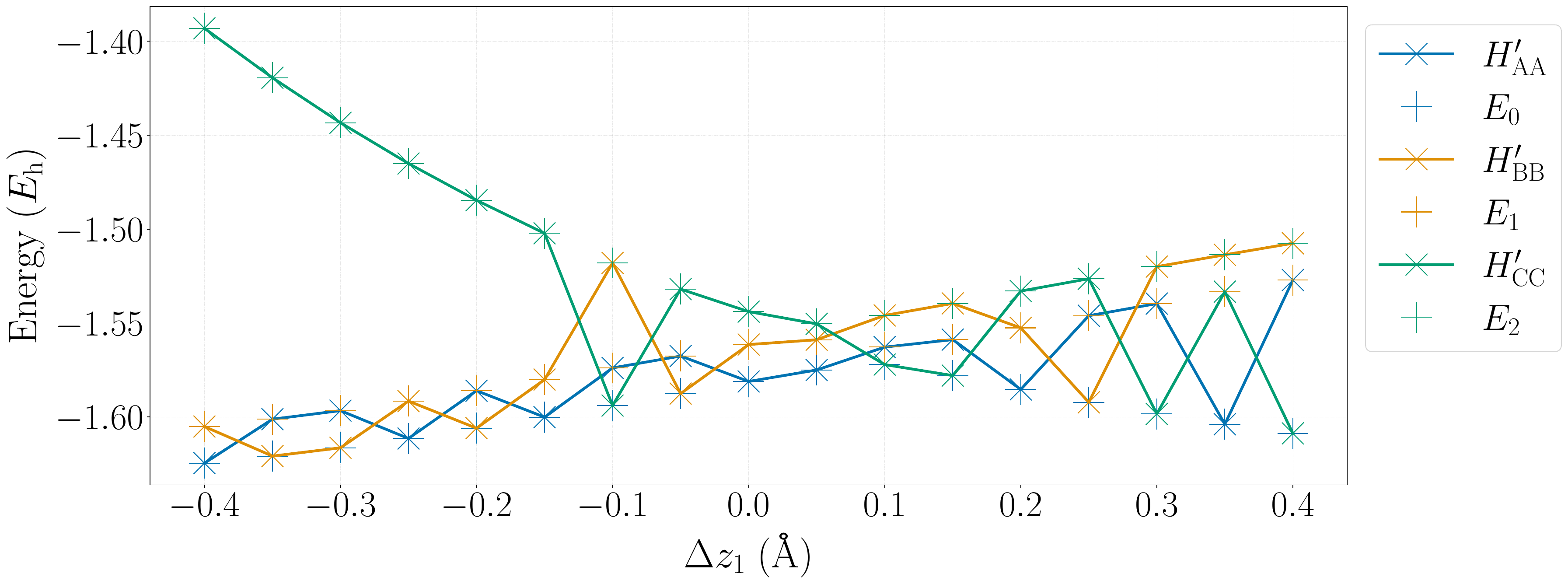}
    
    \caption{After eigenstate resolution}
     \label{fig:solving_for_eigenstates_diagonalization_initial_states_impact}
  \end{subfigure}
  \caption{ Energies of the ensemble-VQE states: (a) \( H_{\mathrm{AA}}, \, H_{\mathrm{BB}}, \, H_{\mathrm{CC}} \) after determining the subspace of minimal ensemble-energy; (b) \( H^{\prime}_{\mathrm{AA}}, \, H^{\prime}_{\mathrm{BB}}, \, H^{\prime}_{\mathrm{CC}} \) after eigenstate resolution (see Fig.~\ref{fig:circuit_phi_prime}, now using the initial state \(\ket{10000001}\) and Eq.~(\ref{eq:similarity_transformation})). FCI energies of the first three eigenstates: \( E_0 , \,E_1, \, E_2 \). Canonical MOs are used.}
  \label{fig:initial_states_impact}
\end{figure}

For illustration purposes, we deliberately tried an optimization from a ``wrong guess'' (doubly excited configurations):
\(\ket{\Phi^{0}_{\mathrm {A}}} = \ket{11000001}\), \(\ket{\Phi^{0}_{\mathrm {B}}} = \ket{10100001}\), and \(\ket{\Phi^{0}_{\mathrm {C}}} = \ket{10010001}\).
In these, the $\beta$-electron now occupies the third excited orbital and the totally-bonding orbital is singly-occupied by an $\alpha$-electron (the second $\alpha$-electron navigates among the three excited orbitals). It occurs that our procedure is robust enough (and the ansatz expressible enough) that it did not end up at a local minimum,
see Fig.~\ref{fig:initial_states_impact}.
Indeed, it yields correct values of the three lowest targeted eigenenergies at convergence, but now sorted erratically along consecutive molecular geometries.
While this went well here, a more systematic analysis should be made for assessing convergence properties under general circumstances.

\section{Distortion within \ce{C_1} geometry (from $\Delta x_2 = 0.1~\textrm{\AA}$ and $\Delta y_3 = 0.05~\textrm{\AA}$): impact of the reference geometry}\label{app:impact_geometry}

\subsection{Achieving a quasi-diabatic representation: after determining the subspace of minimal ensemble-energy}

\begin{figure}
\begin{subfigure}[t]{\columnwidth}
\centering
    \includegraphics[width=\columnwidth]{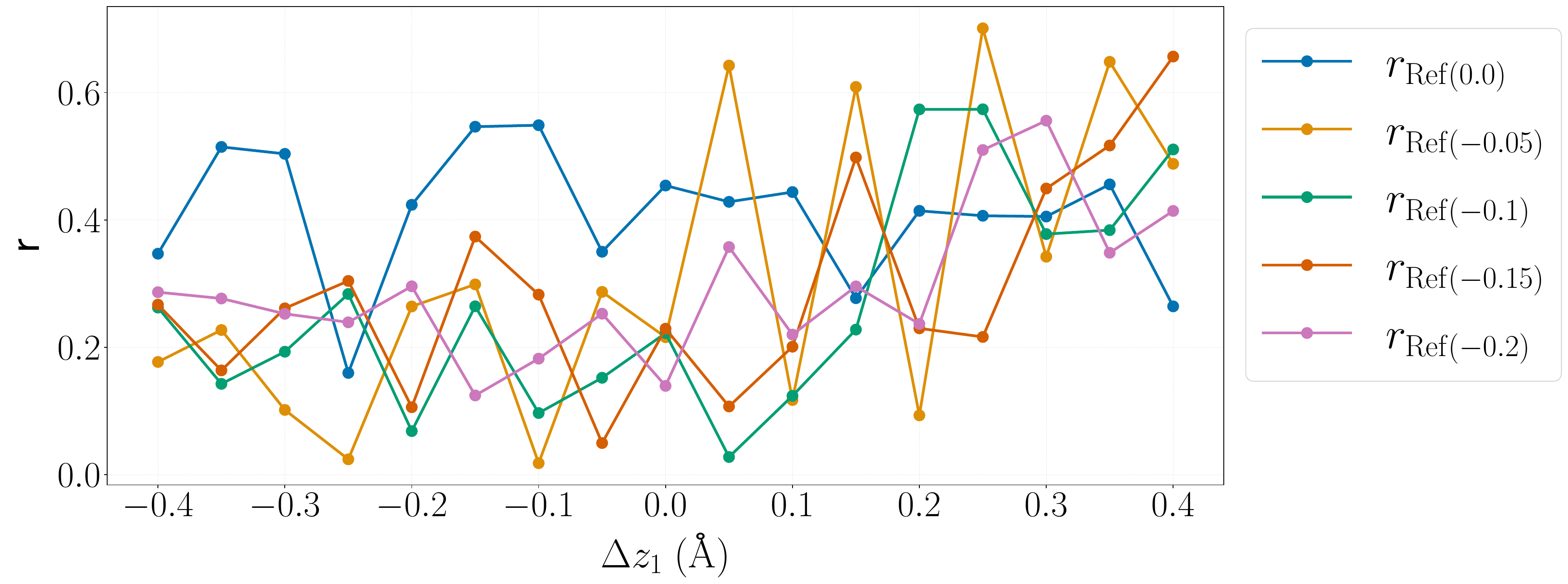}
  \end{subfigure}
  \hfill
  \begin{subfigure}[t]{\columnwidth}
  \centering
    \includegraphics[width=\columnwidth]{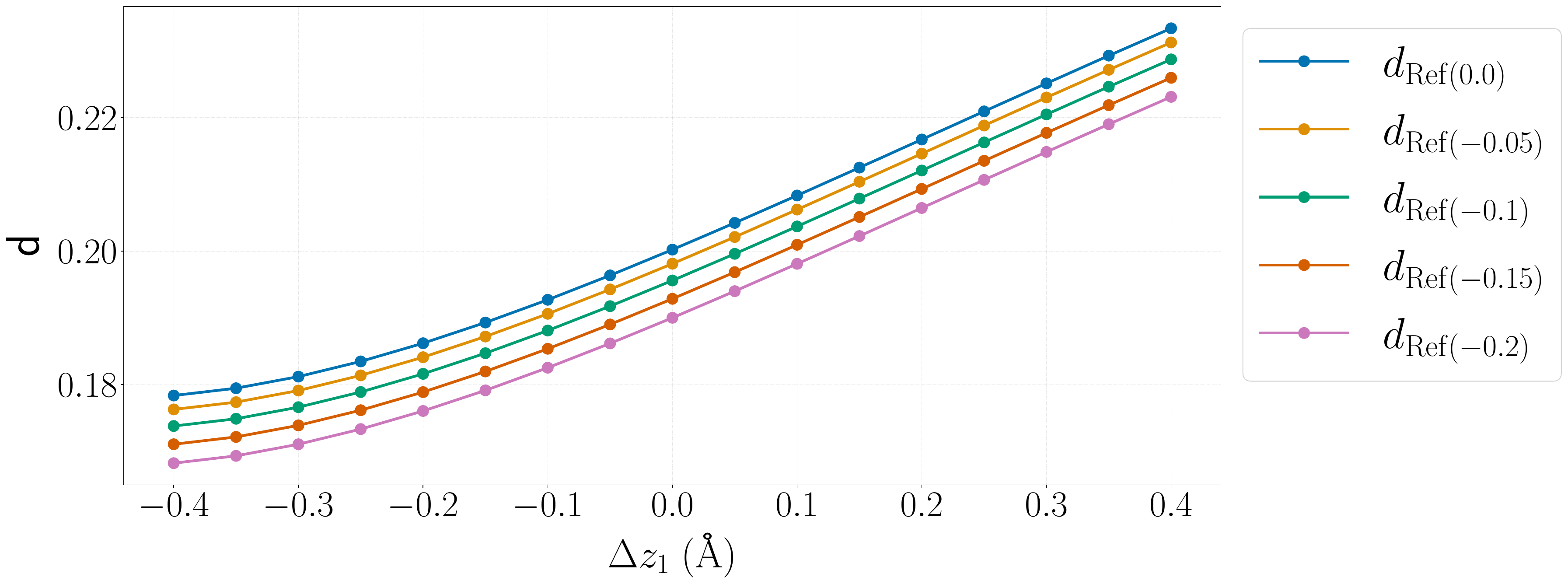}
  \end{subfigure}

  \caption{Diabaticity descriptors \(r\) and \(d\) using reference orbitals from various \ce{C_s} geometries: \(\Delta x_2^0 = 0.1~\textrm{\AA}\) for all, together with \(\Delta z_1^0 = 0.0~\textrm{\AA}\) (Ref(0.0)), \(\Delta z_1^0 = -0.05~\textrm{\AA}\) (Ref(0.05)), \(\Delta z_1^0 = -0.1~\textrm{\AA}\) (Ref(-0.1)), \(\Delta z_1^0 = -0.15~\textrm{\AA}\) (Ref(-0.15)), and \(\Delta z_1^0 = -0.2~\textrm{\AA}\) (Ref(-0.2))}
   \label{fig:descriptor_variations}
\end{figure}

Fig.~\ref{fig:descriptor_variations} shows that the descriptor \(r\) does not vary smoothly for any reference geometry \(\bm{R}^{0}\) used to construct the diabatic MO basis. In contrast, the descriptor \(d\) varies smoothly and depends only slightly on the choice of \(\bm{R}^{0}\). Moreover, similar behavior is observed for all reference geometries considered, and the differences between the corresponding curves remain nearly constant along the deformation coordinate. A more detailed analysis of these aspects will be provided in future work.

\subsection{Achieving a quasi-diabatic representation: after descriptor \(r\) minimization}

\begin{figure}
\begin{subfigure}[t]{\columnwidth}
\centering
    \includegraphics[width=\columnwidth]{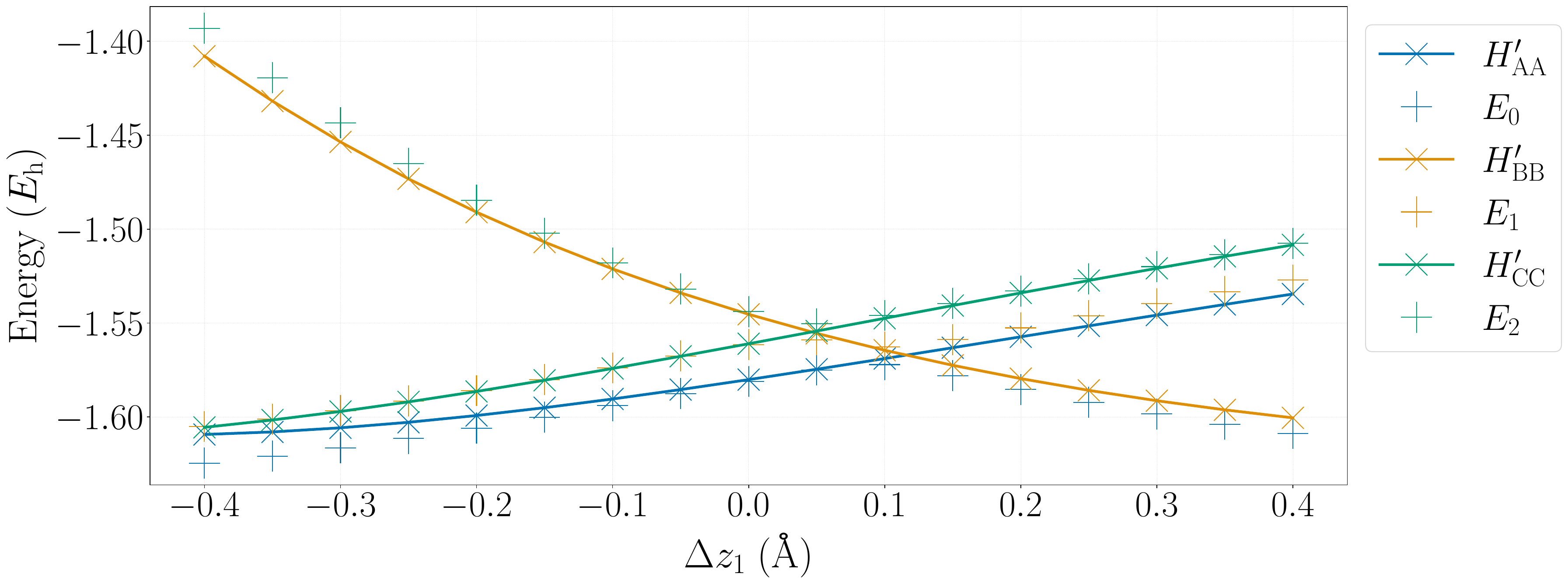}
   \caption{Reference orbitals from \ce{Cs} geometry: \(\Delta x_2^0 = 0.1~\textrm{\AA}\) and \(\Delta z_1^0 = 0.0~\textrm{\AA}\) }
  \end{subfigure}
  \hfill
  \begin{subfigure}[t]{\columnwidth}
  \centering
   \includegraphics[width=\columnwidth]{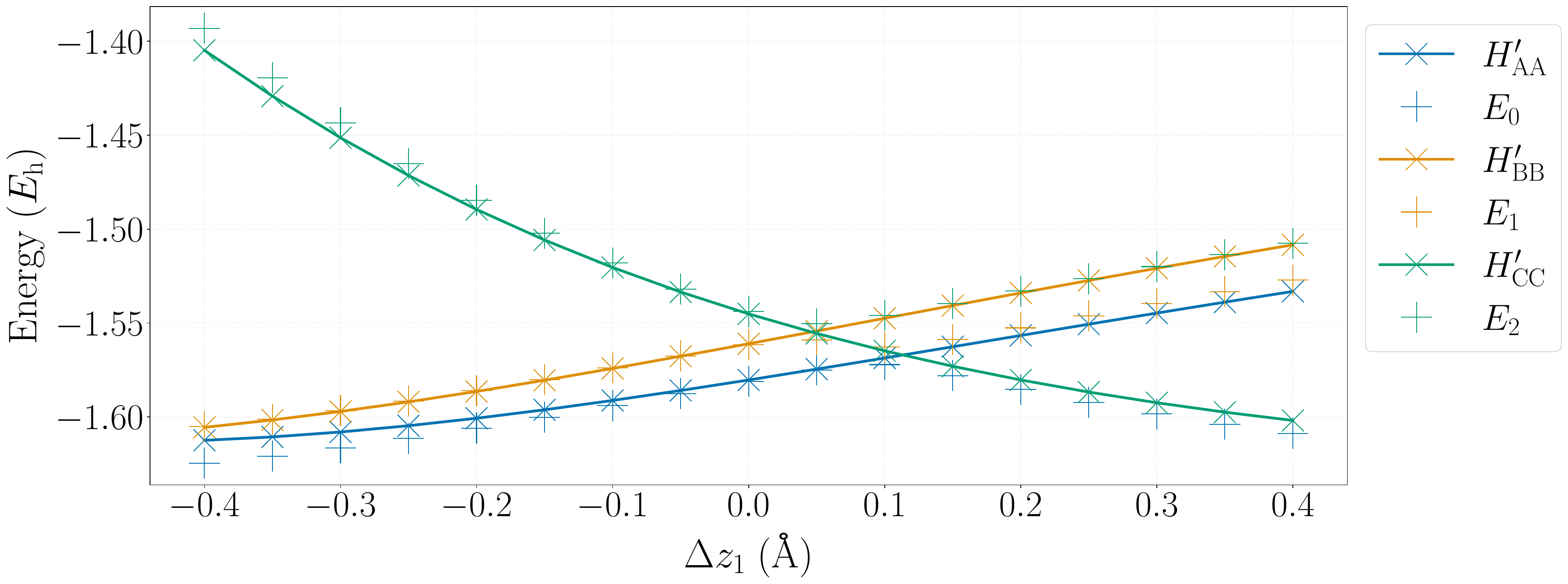}
    \caption{Reference orbitals from \ce{Cs} geometry: \(\Delta x_2^0 = 0.1~\textrm{\AA}\) and \(\Delta z_1^0 = -0.05~\textrm{\AA}\)}
 \end{subfigure}
  \caption{Energies of the ensemble-VQE states, \( H^{\prime}_{\mathrm{AA}}, \, H^{\prime}_{\mathrm{BB}}, \, H^{\prime}_{\mathrm{CC}} \), and FCI energies of the first three eigenstates, \( E_0 , \,E_1, \, E_2 \). Diabatic orbitals are used.}
  \label{fig:diabatization_r_minimisation_impact_reference_1}
\end{figure}

\begin{figure}
\begin{subfigure}[t]{\columnwidth}
\centering
    \includegraphics[width=\columnwidth]{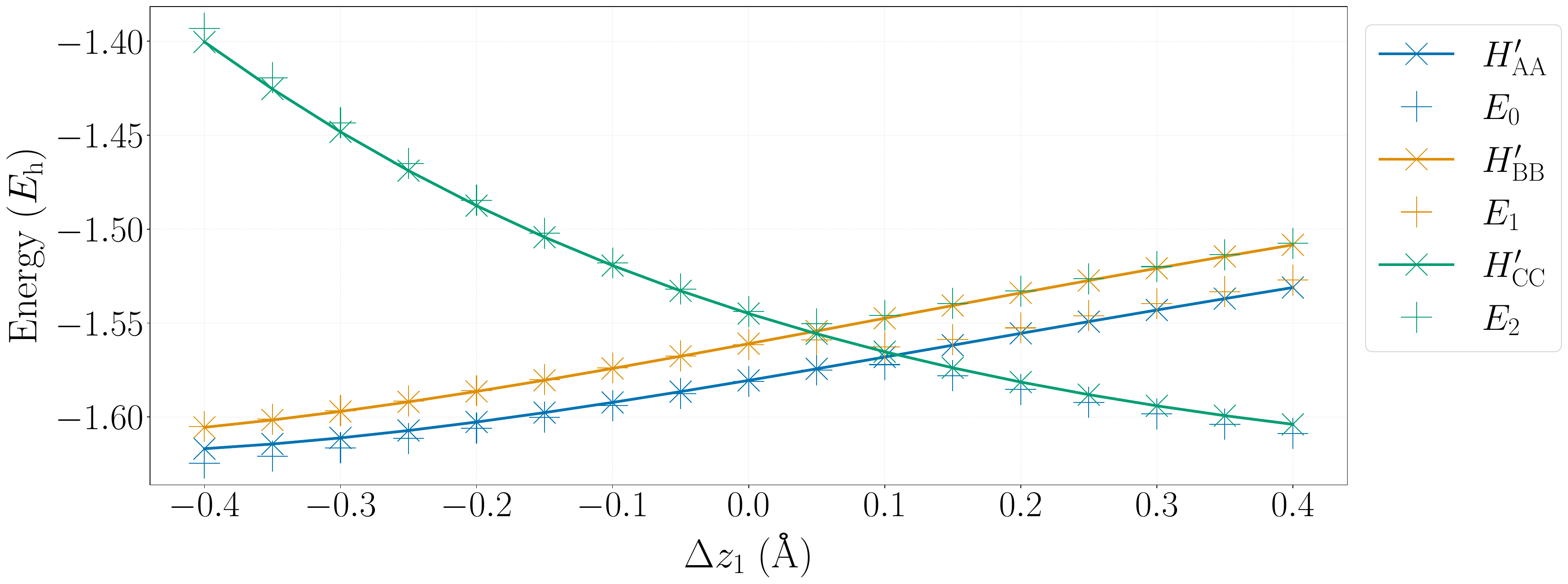}
   \caption{Reference orbitals from \ce{Cs} geometry: \(\Delta x_2^0 = 0.1~\textrm{\AA}\) and \(\Delta z_1^0 = -0.15~\textrm{\AA}\) }
  \end{subfigure}
  \hfill
  \begin{subfigure}[t]{\columnwidth}
  \centering
   \includegraphics[width=\columnwidth]{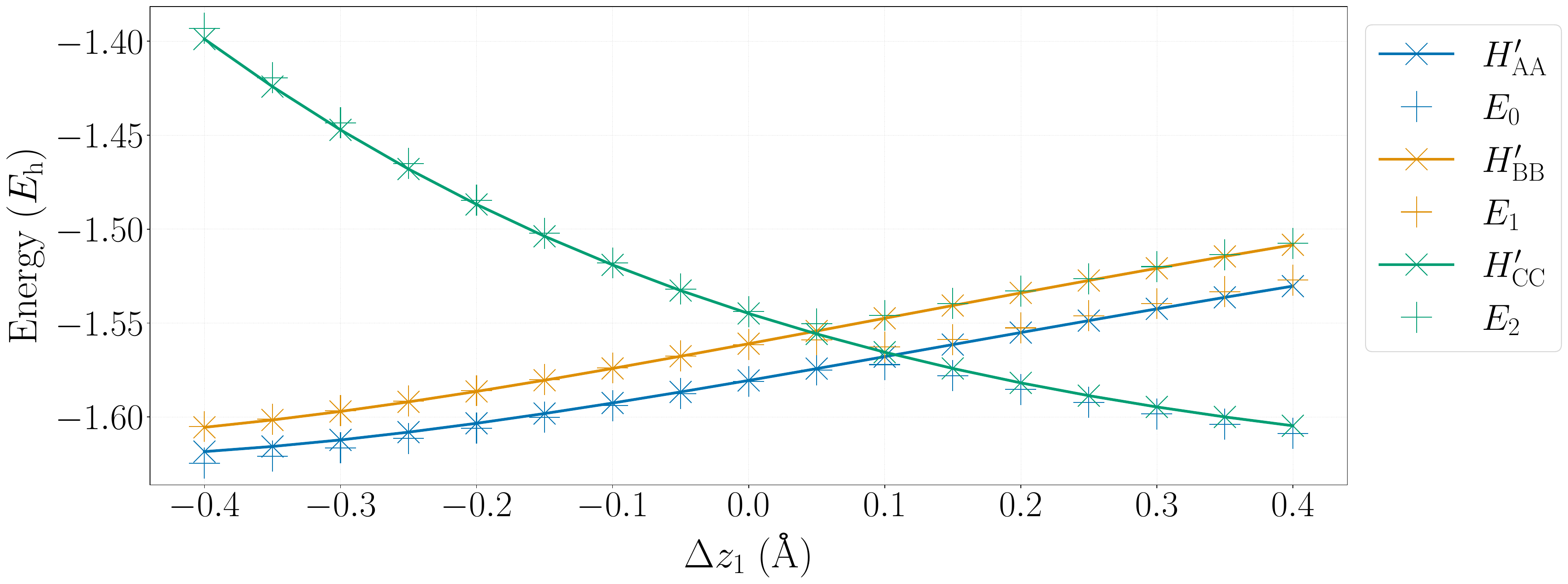}
    \caption{Reference orbitals from \ce{Cs} geometry: \(\Delta x_2^0 = 0.1~\textrm{\AA}\) and \(\Delta z_1^0 = -0.2~\textrm{\AA}\)}
 \end{subfigure}
  \caption{Energies of the ensemble-VQE states, \( H^{\prime}_{\mathrm{AA}}, \, H^{\prime}_{\mathrm{BB}}, \, H^{\prime}_{\mathrm{CC}} \), and FCI energies of the first three eigenstates, \( E_0 , \,E_1, \, E_2 \). Diabatic orbitals are used.}
  \label{fig:diabatization_r_minimisation_impact_reference_2}
\end{figure}

Fig.~\ref{fig:diabatization_r_minimisation_impact_reference_1} and~\ref{fig:diabatization_r_minimisation_impact_reference_2} show that small variations of the reference geometry \(\bm{R}^{0}\) used to construct the diabatic MO basis induce slight variations in the optimized diabatic states, consistent with the behavior of the descriptor $d$ observed in Fig.~\ref{fig:descriptor_variations}.
The case where \(\Delta x_2^0 = 0.1~\textrm{\AA}\) and \(\Delta z_1^0 = -0.1~\textrm{\AA}\) is shown in the main text.

\section{Distortion within \ce{C_s} geometry (from $\Delta x_2 = 0.1~\textrm{\AA}$ and $\Delta y_3 = 0.0~\textrm{\AA}$)}
\label{si_subsubsec:distortion_cs}

\subsection{Solving for eigenstates}

\begin{figure}
\begin{subfigure}[t]{\columnwidth}
\centering
    \includegraphics[width=\columnwidth]{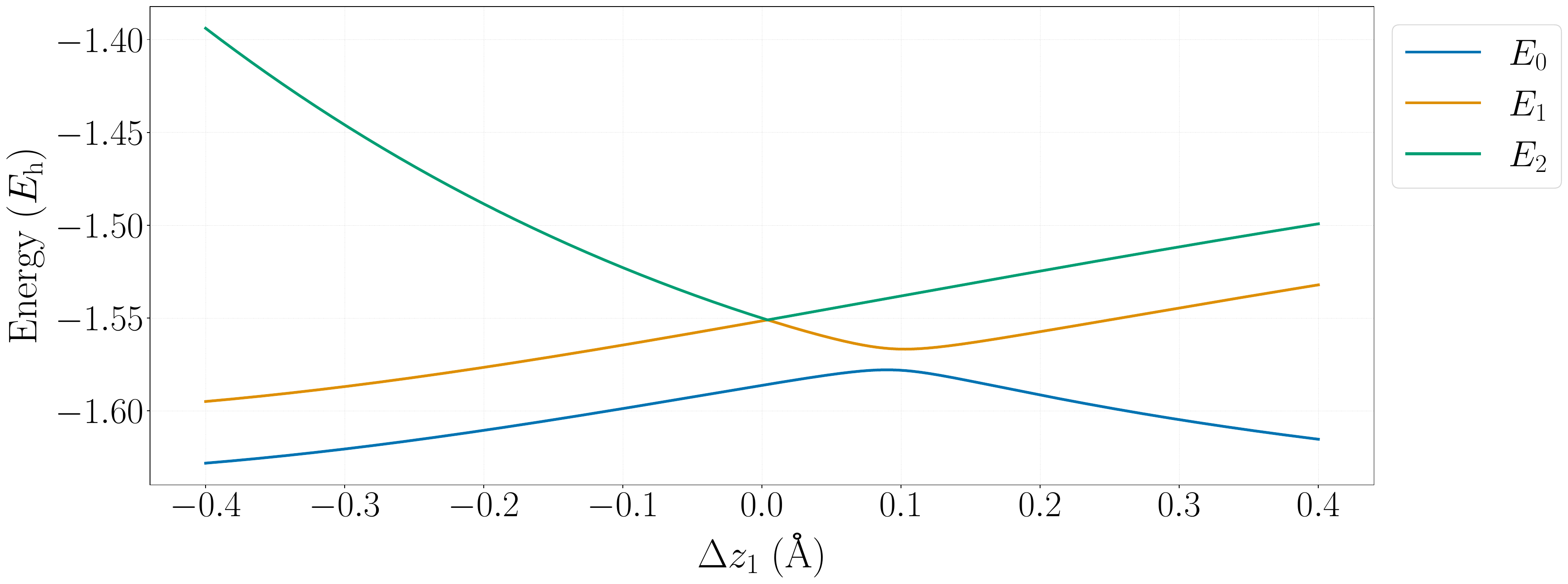}
    
   \caption{(a)}
  \end{subfigure}
  \hfill
  \begin{subfigure}[t]{\columnwidth}
  \centering
   \includegraphics[width=\columnwidth]{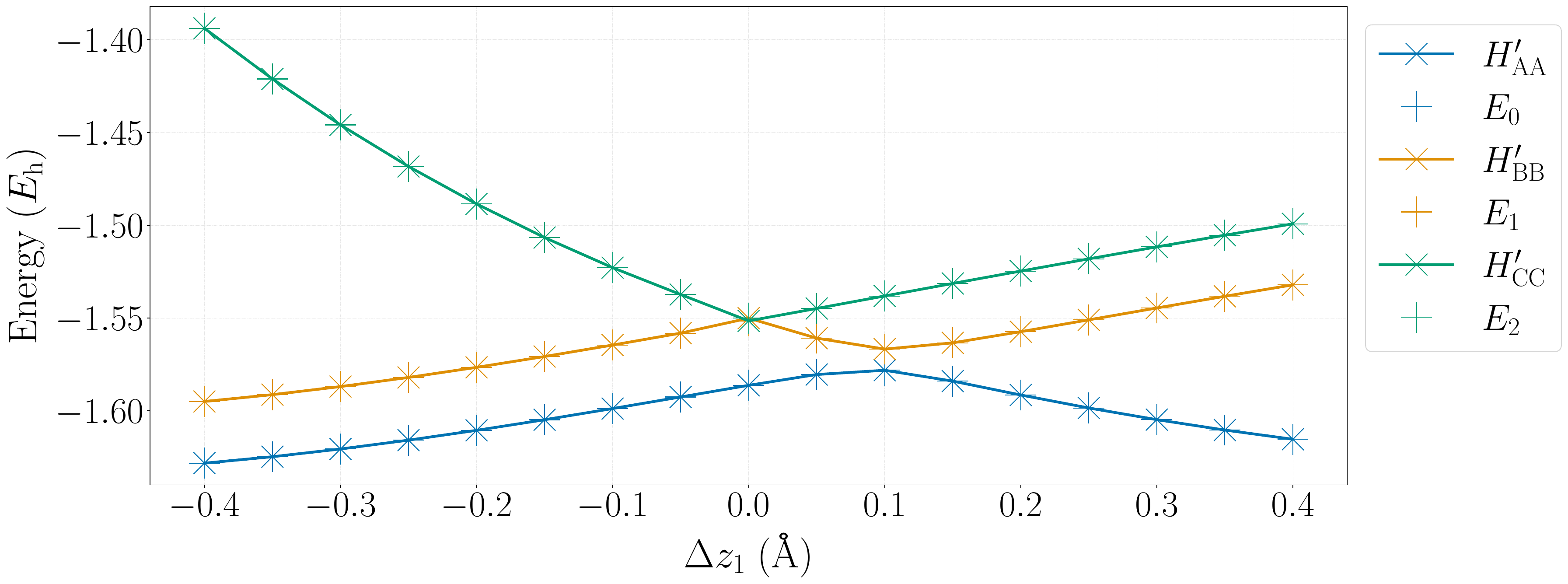}
    \caption{(b)}
  \end{subfigure}
   \caption{ (a) Adiabatic potential energies of the first three eigenstates, \(  E_0 ,E_1, E_2 \), along $\Delta z_1$ obtained from a 'restricted open-shell Hartree--Fock' (ROHF) FCI/STO-3G calculation with Psi4~\cite{turney_psi4_2012}. (b) Energies of the ensemble-VQE states: \( H^{\prime}_{\mathrm{AA}}, \, H^{\prime}_{\mathrm{BB}}, \, H^{\prime}_{\mathrm{CC}} \) after eigenstate resolution (see Fig.~\ref{fig:circuit_phi_prime} and Eq.~(\ref{eq:similarity_transformation})). FCI energies of the first three eigenstates: \( E_0 , \,E_1, \, E_2 \). Canonical MOs are used.}
   \label{fig:solving_for_eigenstates_diagonalization_cs}
\end{figure}

Fig.~\ref{fig:solving_for_eigenstates_diagonalization_cs} shows that the ensemble retro-variational energies, \( H^{\prime}_{\mathrm{AA}}, \, H^{\prime}_{\mathrm{BB}}, \, H^{\prime}_{\mathrm{CC}} \), fully match the eigenenergies as expected.

\subsection{Achieving a quasi-diabatic representation: after determining the subspace of minimal ensemble-energy}

\begin{figure}
\begin{subfigure}{\columnwidth}
\centering
   \includegraphics[width=\columnwidth]{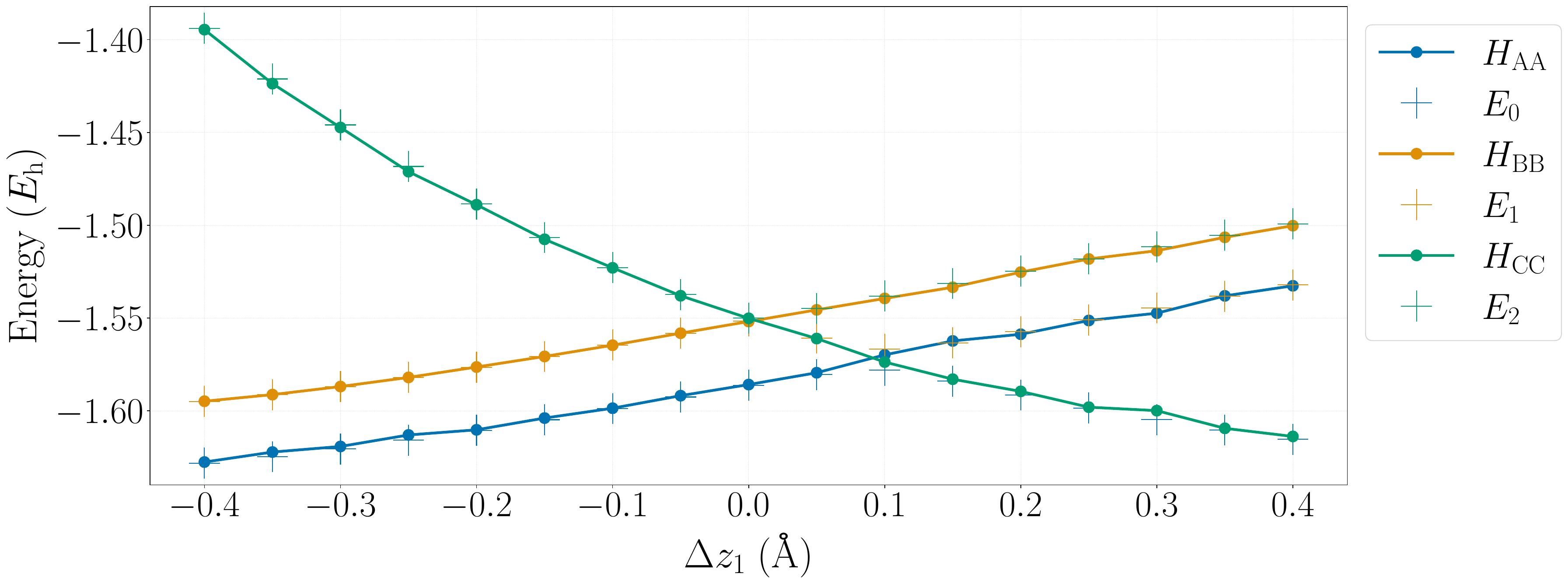}
    \caption{}
    \label{fig:diabatization_block_diagonalization_minus_zero_one_cs}
  \end{subfigure}
  \hfill
\begin{subfigure}{\columnwidth}
\centering
   \includegraphics[width=\columnwidth]{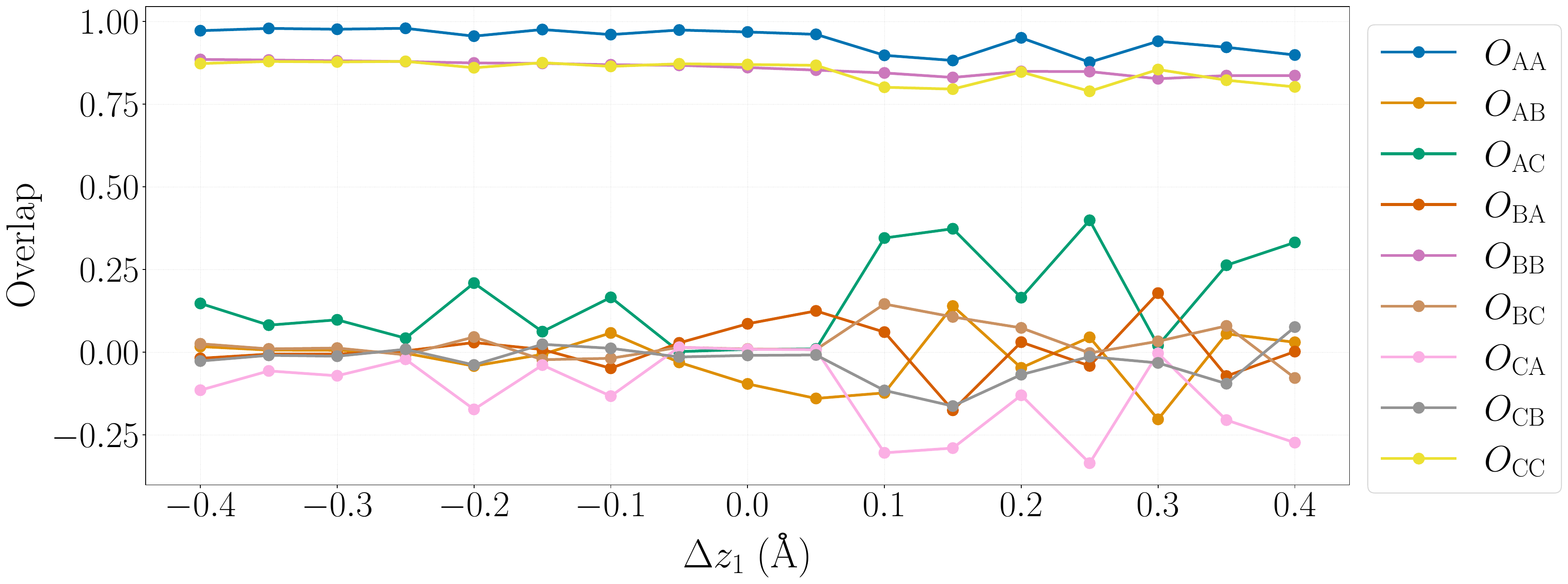}
    \caption{}
     \label{fig:o_matrix_minus_zero_one_cs}
  \end{subfigure}
  \caption{(a) Energies of the ensemble-VQE states, \( H_{\mathrm{AA}}, \, H_{\mathrm{BB}}, \, H_{\mathrm{CC}} \), and FCI energies of the first three eigenstates, \( E_0 , \,E_1, \, E_2 \). (b) CI-coefficient (overlap) submatrix elements of the ensemble-VQE states $\lbrace \ket{\Phi_I({\bm{t}^*})}\rbrace_{I\in\lbrace \text{A},\text{B},\text{C}\rbrace}$. Diabatic orbitals are used.}
  \label{fig:diabatization_block_diagonalization_cs}
\end{figure}

As in the case of the distortion within \ce{C_1} geometry, we chose, in the following, a $\ce{C_s}$ reference geometry, $\bm{R}^0$, for defining the diabatic orbitals~\cite{illesova_transformation-free_2025}, such that $\Delta x_2^0 = 0.1~\textrm{\AA}$ and $\Delta z_1^0 = -0.1~\textrm{\AA}$. 
As seen in Fig.~\ref{fig:diabatization_block_diagonalization_minus_zero_one}, the energies of the states in Fig.~\ref{fig:diabatization_block_diagonalization_minus_zero_one_cs} appear to follow a diabatic behavior, although they are not entirely smooth functions of the deformation coordinate. Fig.~\ref{fig:o_matrix_minus_zero_one_cs} indicates that the elements of the overlap matrix exhibit an irregular behavior similar to what is observed in Fig.~\ref{fig:o_matrix_minus_zero_one}.

\subsection{Achieving a quasi-diabatic representation: after descriptor \(r\) minimization}

\begin{figure}
\begin{subfigure}[t]{\columnwidth}
\centering
    \includegraphics[width=\columnwidth]{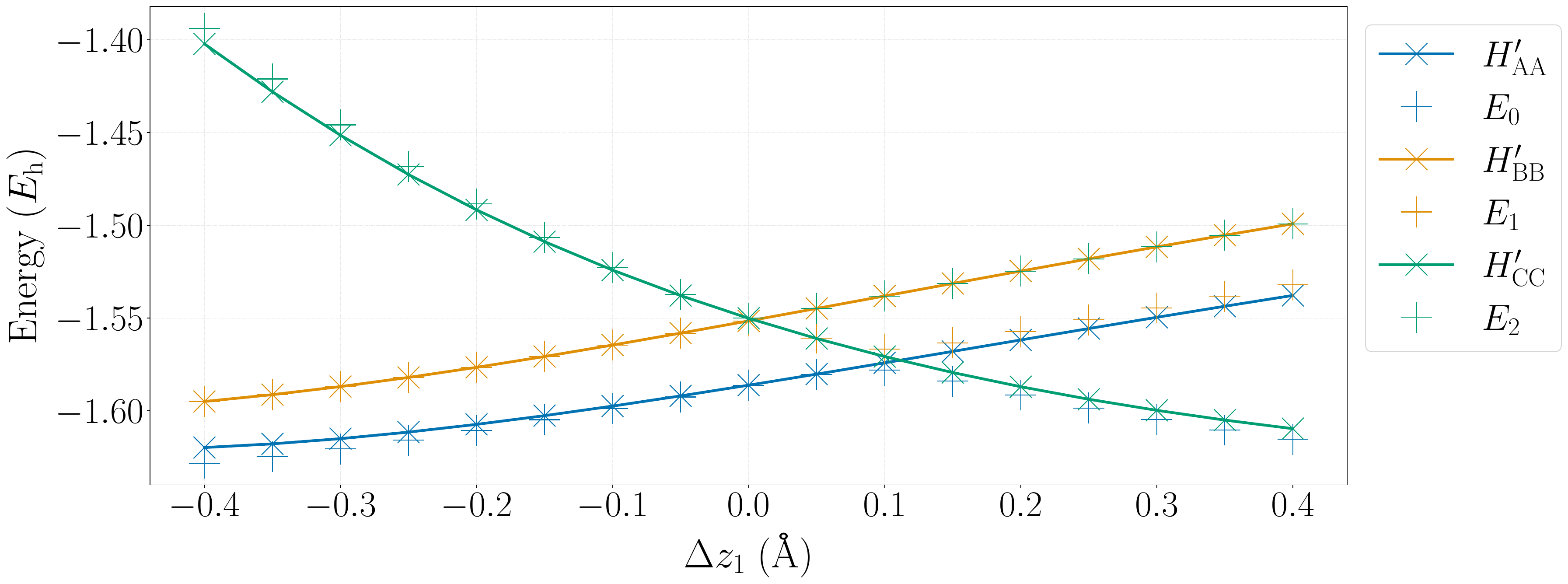}
   \caption{}
   \label{fig:energies_optimally_diabatic_cs}
  \end{subfigure}
  \hfill
  \begin{subfigure}[t]{\columnwidth}
  \centering
   \includegraphics[width=\columnwidth]{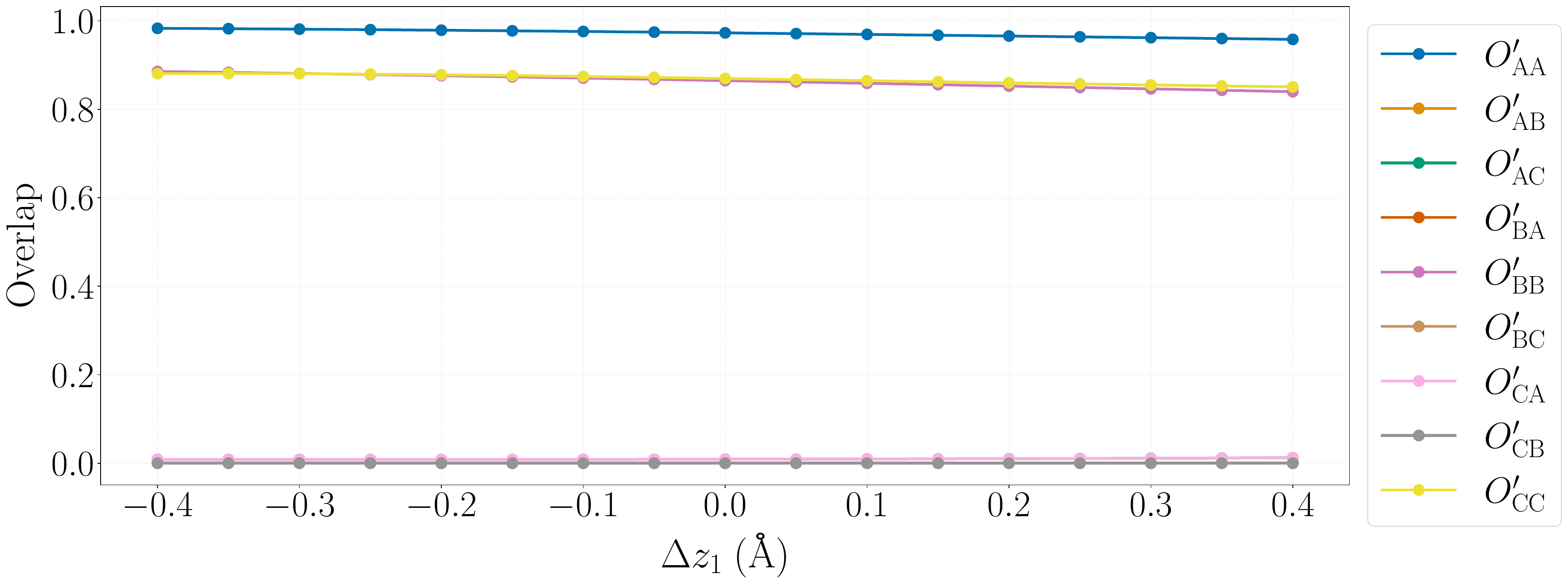}
    \caption{}
  \label{fig:o_star_matrix_minus_zero_one_cs}
  \end{subfigure}
  \caption{(a) Energies of the ensemble-VQE states, \( H^{\prime}_{\mathrm{AA}}, \, H^{\prime}_{\mathrm{BB}}, \, H^{\prime}_{\mathrm{CC}} \), and FCI energies of the first three eigenstates, \( E_0 , \,E_1, \, E_2 \). (b) CI-coefficient (overlap) submatrix elements of the ensemble-VQE states $\lbrace \ket{\Phi'_I({\bm{t}^*,\theta_\star, \phi_\star, \psi_\star)})}\rbrace_{I\in\lbrace \text{A},\text{B},\text{C}\rbrace}$. Diabatic orbitals are used.}
  \label{fig:diabatization_r_minimisation_cs}
\end{figure}

From Fig.~\ref{fig:energies_optimally_diabatic_cs}, the diabatic energies obtained after minimization of the descriptor $r$ appear smooth over the entire geometry profile. Moreover, the elements of the overlap matrix in Fig.~\ref{fig:o_star_matrix_minus_zero_one_cs} follow a behavior similar to that observed in Fig.~\ref{fig:o_star_matrix_minus_zero_one}.

\begin{figure}
 \begin{subfigure}[t]{\columnwidth}
\centering
    \includegraphics[width=\columnwidth]{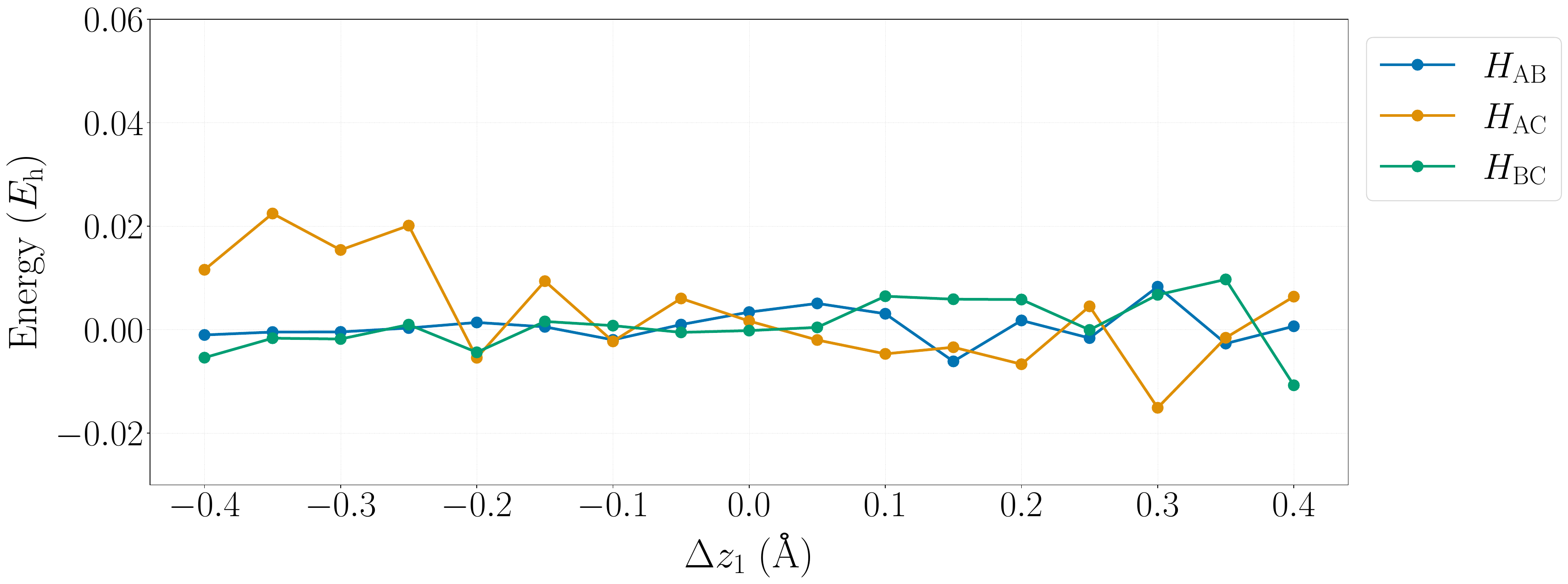}
    \caption{}
  \end{subfigure}
  \hfill
\begin{subfigure}[t]{\columnwidth}
\centering
    \includegraphics[width=\columnwidth]{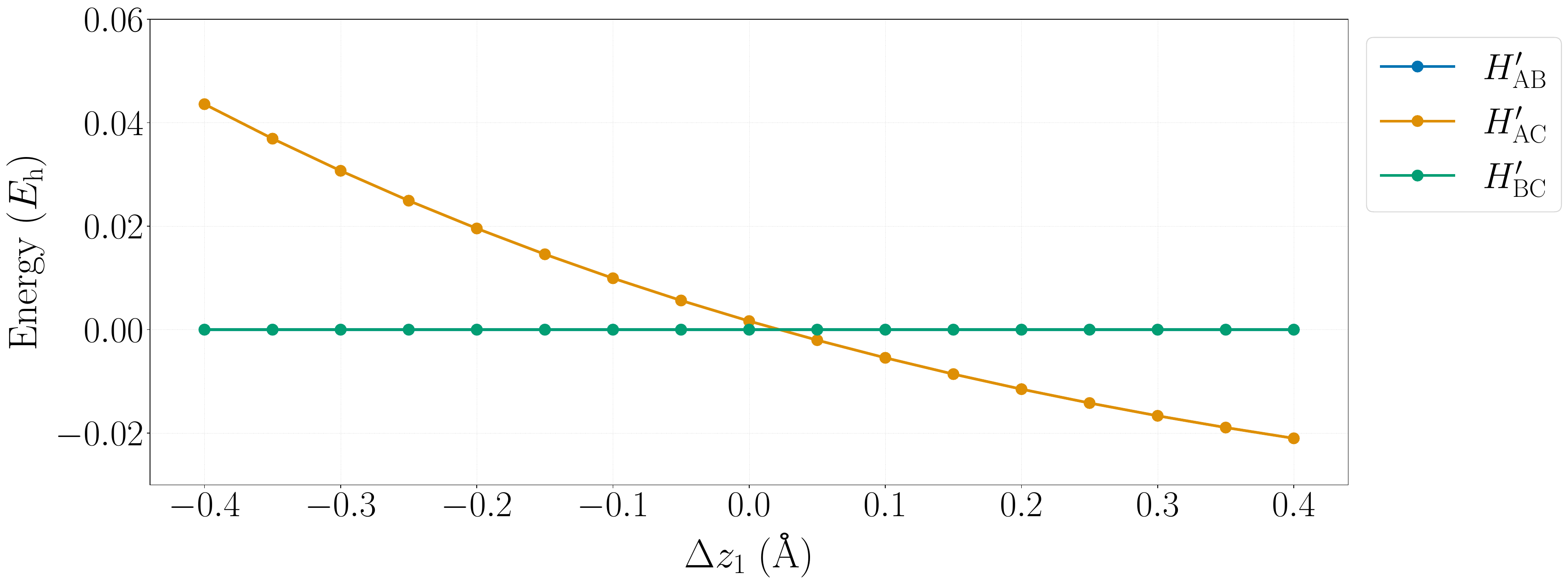}
    \caption{}
  \end{subfigure}
   \caption{(a) Off-diagonal Hamiltonian matrix elements in the basis of the ensemble-VQE states \(\lbrace \ket{\Phi_I(\boldsymbol{t}^*)} \rbrace _{I \in \lbrace \mathrm {A},\mathrm {B},\mathrm {C}\rbrace }\) (see Figs.~\ref{fig:diabatization_block_diagonalization_minus_zero_one_cs} and~\ref{fig:o_matrix_minus_zero_one_cs}). (b) Off-diagonal Hamiltonian matrix elements in the basis of the ensemble-VQE states \( \lbrace \ket{\Phi^{\prime}_I(\boldsymbol{t}^*, \theta_\star, \phi_\star, \psi_\star)} \rbrace _{I \in \lbrace \mathrm {A},\mathrm {B},\mathrm {C}\rbrace }  \) (see Figs.~\ref{fig:energies_optimally_diabatic_cs} and~\ref{fig:o_star_matrix_minus_zero_one_cs}). Diabatic orbitals are used.}
 \label{fig:off_diagonal_hamiltonian_matrix_cs_optimally_diabatic}
\end{figure}

Fig.~\ref{fig:off_diagonal_hamiltonian_matrix_cs_optimally_diabatic} indicates that the off-diagonal Hamiltonian matrix elements before the change of basis are
not completely smooth, similarly to those shown in Fig.~\ref{fig:off_diagonal_hamiltonian_matrix_c1_optimally_diabatic}. After the change of basis, two out of the three off-diagonal Hamiltonian matrix elements become zero as expected from symmetry considerations.

\section{Distortion from the \ce{T_d} point group (from $\Delta x_2 = 0.0~\textrm{\AA}$ and $\Delta y_3 = 0.0~\textrm{\AA}$) to the \ce{C_{3v}} point group}

\label{si_subsubsec:distortion_c3v}

\subsection{Solving for eigenstates}

\begin{figure}
\begin{subfigure}[t]{\columnwidth}
\centering
    \includegraphics[width=\columnwidth]{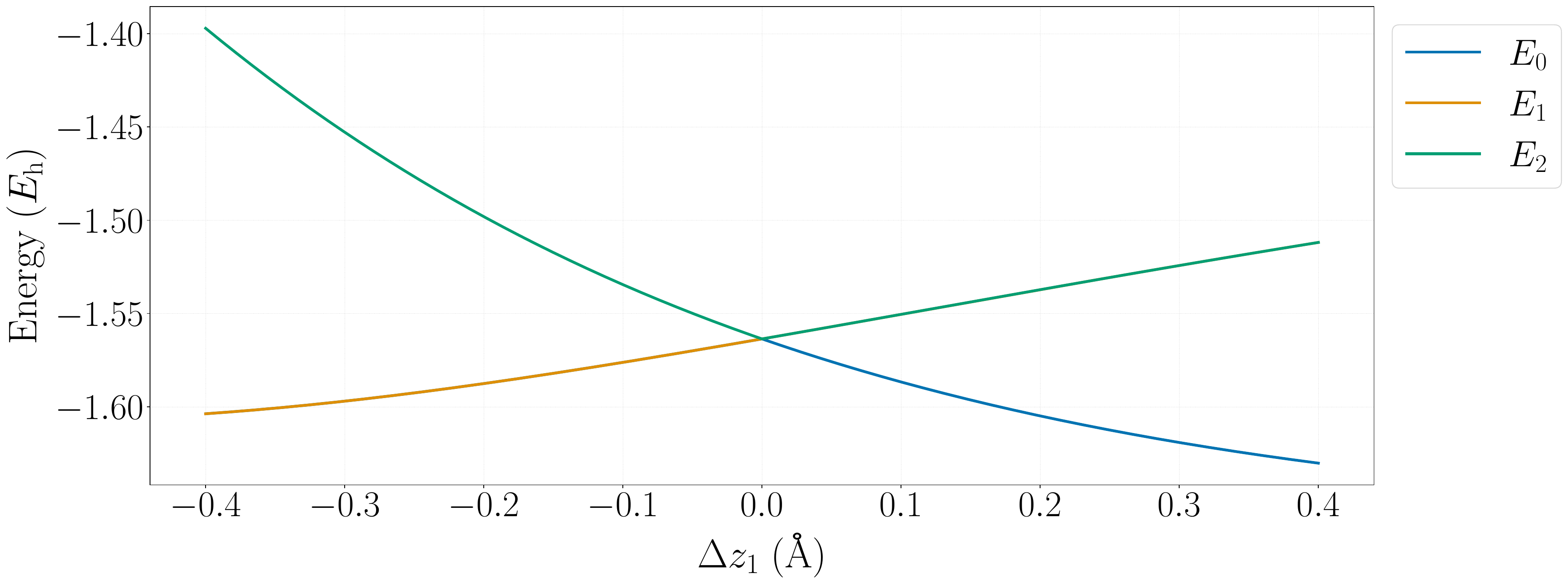}
    
   \caption{}
  \end{subfigure}
  \hfill
  \begin{subfigure}[t]{\columnwidth}
  \centering
   \includegraphics[width=\columnwidth]{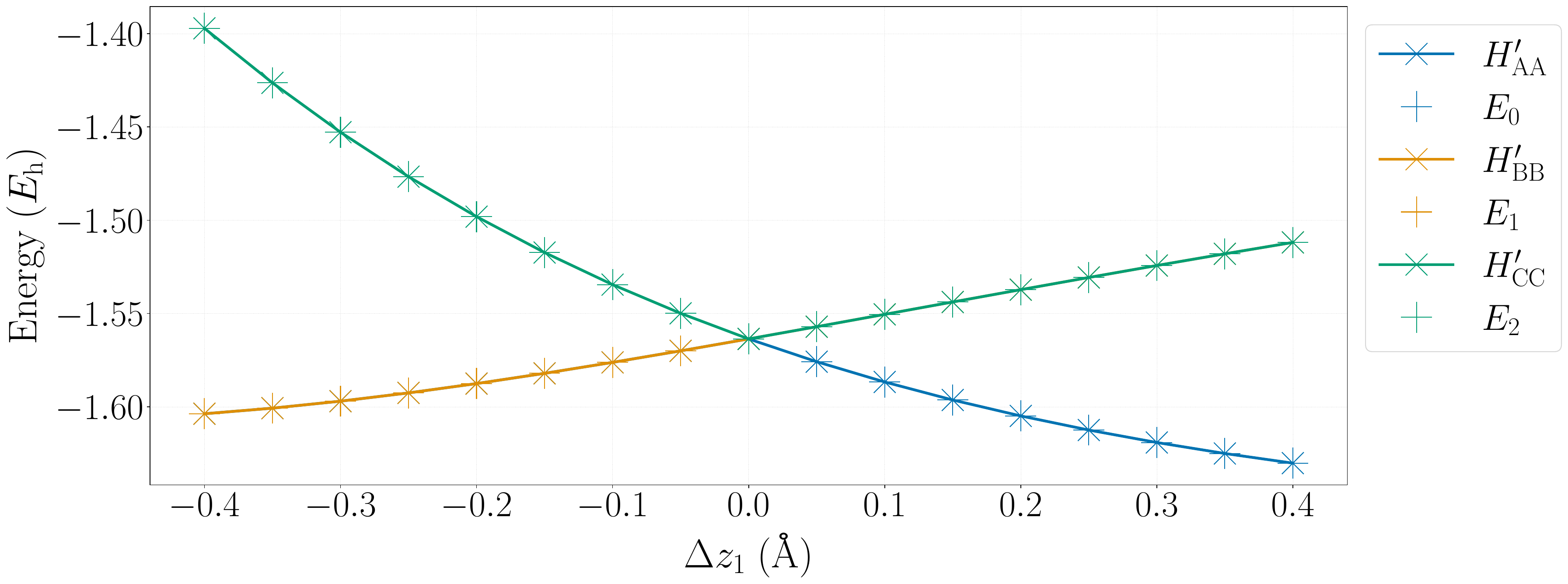}
    \caption{}
  \end{subfigure}
   \caption{ (a) Adiabatic potential energies of the first three eigenstates, \(  E_0 ,E_1, E_2 \), along $\Delta z_1$ obtained from a 'restricted open-shell Hartree--Fock' (ROHF) FCI/STO-3G calculation with Psi4~\cite{turney_psi4_2012}. (b) Energies of the ensemble-VQE states: \( H^{\prime}_{\mathrm{AA}}, \, H^{\prime}_{\mathrm{BB}}, \, H^{\prime}_{\mathrm{CC}} \) after eigenstate resolution (see Fig.~\ref{fig:circuit_phi_prime} and Eq.~(\ref{eq:similarity_transformation})). FCI energies of the first three eigenstates: \( E_0 , \,E_1, \, E_2 \). Canonical MOs are used.}
     \label{fig:solving_for_eigenstates_diagonalization_c3v}
\end{figure}

Fig.~\ref{fig:solving_for_eigenstates_diagonalization_c3v} shows that the ensemble retro-variational energies, \( H^{\prime}_{\mathrm{AA}}, \, H^{\prime}_{\mathrm{BB}}, \, H^{\prime}_{\mathrm{CC}} \), fully match the eigenenergies as expected.

\subsection{Achieving a quasi-diabatic representation: after determining the subspace of minimal ensemble-energy}

\begin{figure}
\begin{subfigure}{\columnwidth}
\centering
   \includegraphics[width=\columnwidth]{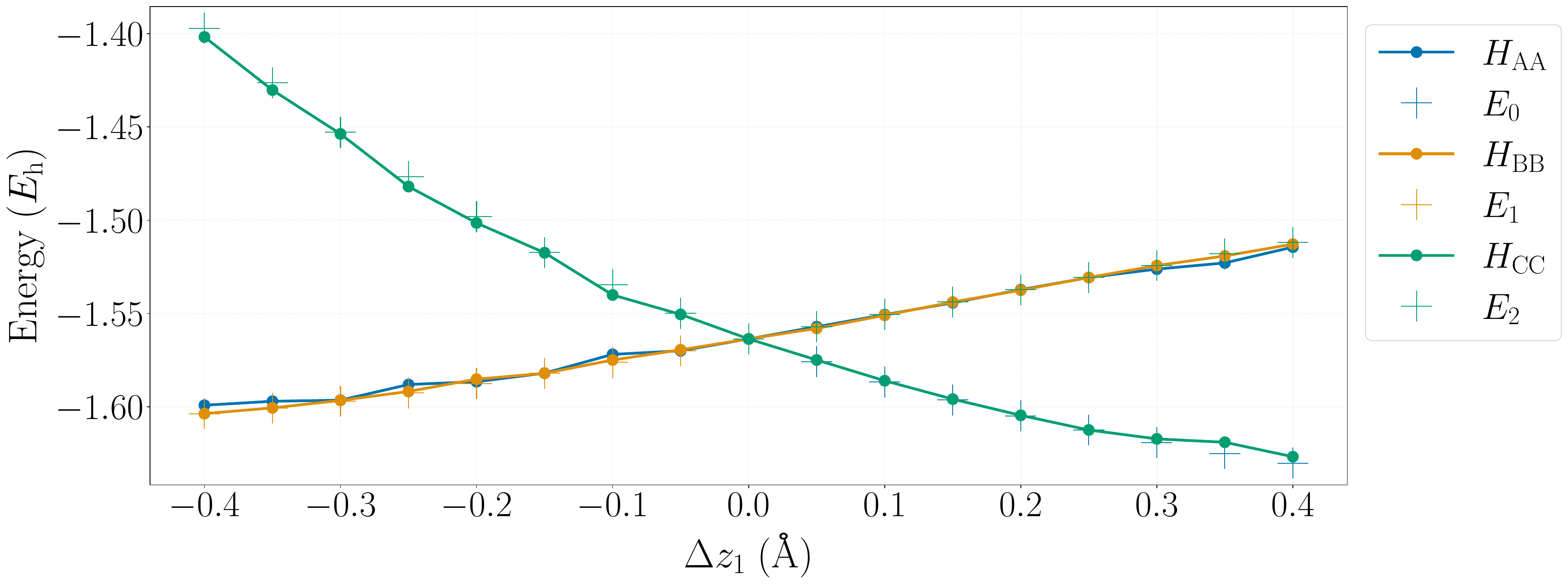}
    \caption{}
    \label{fig:diabatization_block_diagonalization_minus_zero_one_c3v}
  \end{subfigure}
  \hfill
\begin{subfigure}{\columnwidth}
\centering
   \includegraphics[width=\columnwidth]{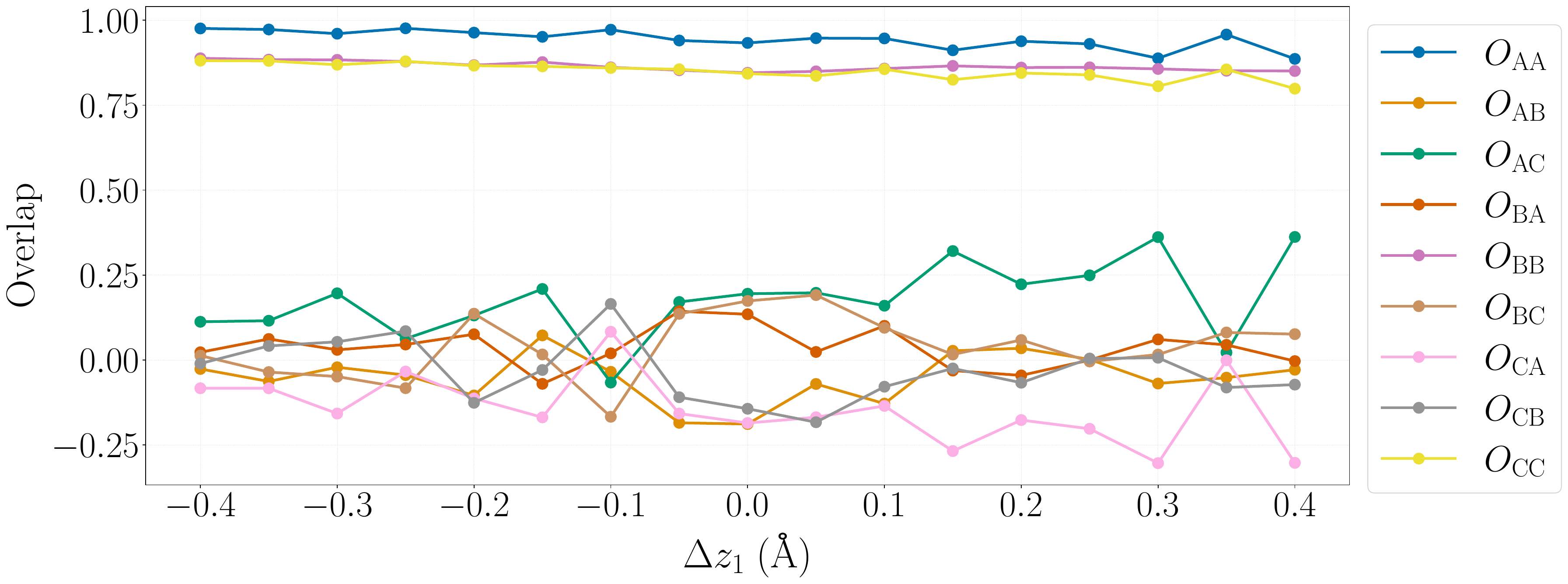}
    \caption{}
     \label{fig:o_matrix_minus_zero_one_c3v}
  \end{subfigure}
  \caption{(a) Energies of the ensemble-VQE states, \( H_{\mathrm{AA}}, \, H_{\mathrm{BB}}, \, H_{\mathrm{CC}} \), and FCI energies of the first three eigenstates, \( E_0 , \,E_1, \, E_2 \). (b) CI-coefficient (overlap) submatrix elements of the ensemble-VQE states $\lbrace \ket{\Phi_I({\bm{t}^*})}\rbrace_{I\in\lbrace \text{A},\text{B},\text{C}\rbrace}$. Diabatic orbitals are used.}
  \label{fig:diabatization_block_diagonalization_c3v}
\end{figure}

Similar observations as in Fig.~\ref{fig:diabatization_block_diagonalization} and Fig.~\ref{fig:diabatization_block_diagonalization_cs} apply to Fig.~\ref{fig:diabatization_block_diagonalization_c3v}, both regarding the state energies and the behavior of the overlap matrix elements. Moreover, it can be seen that for \(\Delta z_1 =0.0~\textrm{\AA}\), corresponding to the \ce{T_d} point, the energies of the three  states are equal, as expected from symmetry considerations. In this case, the convergence of the ensemble energy imply that the energies of the ensemble-VQE states coincide with those of the eigenstates.

\subsection{Achieving a quasi-diabatic representation: after descriptor \(r\) minimization}

\begin{figure}
\begin{subfigure}[t]{\columnwidth}
\centering
    \includegraphics[width=\columnwidth]{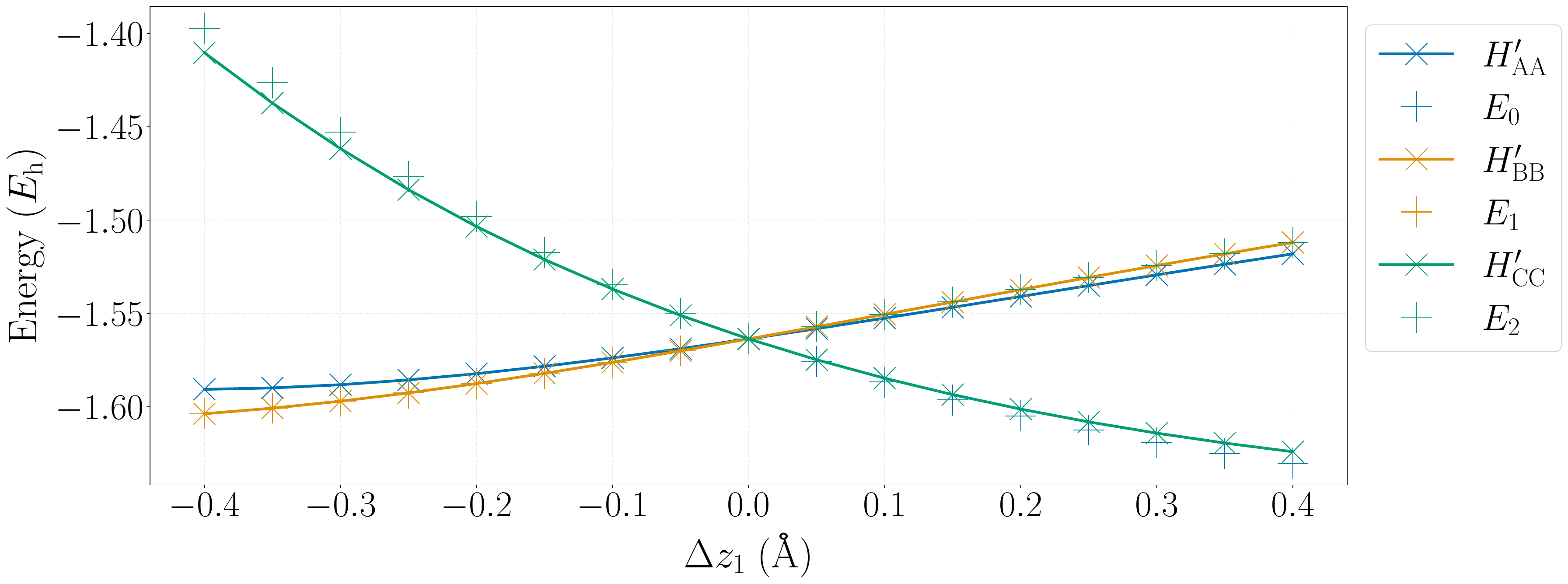}
   \caption{}
   \label{fig:energies_optimally_diabatic_c3v}
  \end{subfigure}
  \hfill
  \begin{subfigure}[t]{\columnwidth}
  \centering
   \includegraphics[width=\columnwidth]{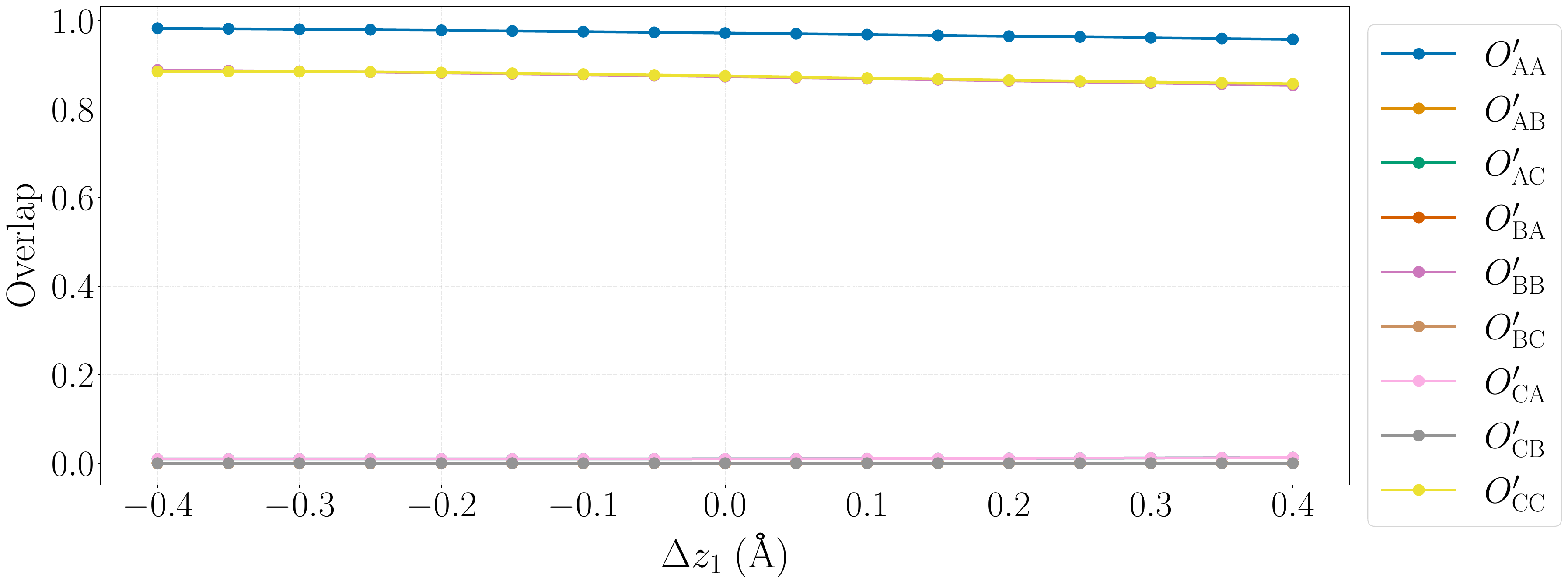}
    \caption{}
  \label{fig:o_star_matrix_minus_zero_one_c3v}
  \end{subfigure}
  \caption{(a) Energies of the ensemble-VQE states, \( H^{\prime}_{\mathrm{AA}}, \, H^{\prime}_{\mathrm{BB}}, \, H^{\prime}_{\mathrm{CC}} \), and FCI energies of the first three eigenstates, \( E_0 , \,E_1, \, E_2 \). (b) CI-coefficient (overlap) submatrix elements of the ensemble-VQE states $\lbrace \ket{\Phi'_I({\bm{t}^*,\theta_\star, \phi_\star, \psi_\star)})}\rbrace_{I\in\lbrace \text{A},\text{B},\text{C}\rbrace}$. Diabatic orbitals are used.}
  \label{fig:diabatization_r_minimisation_c3v}
\end{figure}

Similar remarks to those made for Fig.~\ref{fig:diabatization_r_minimisation} and Fig.~\ref{fig:diabatization_r_minimisation_cs} also apply to Fig.~\ref{fig:diabatization_r_minimisation_c3v}, both regarding the state energies and the overlap matrix elements.

\begin{figure}
 \begin{subfigure}[t]{\columnwidth}
\centering
    \includegraphics[width=\columnwidth]{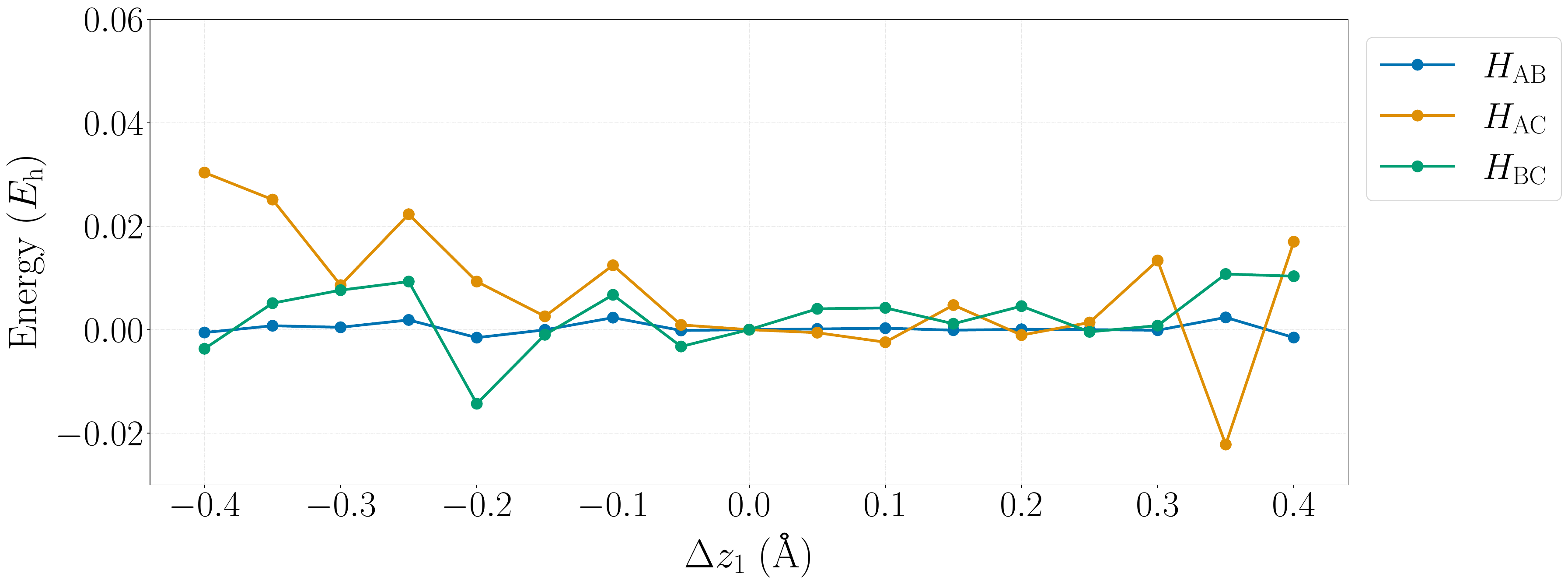}
    \caption{}
  \end{subfigure}
  \hfill
\begin{subfigure}[t]{\columnwidth}
\centering
    \includegraphics[width=\columnwidth]{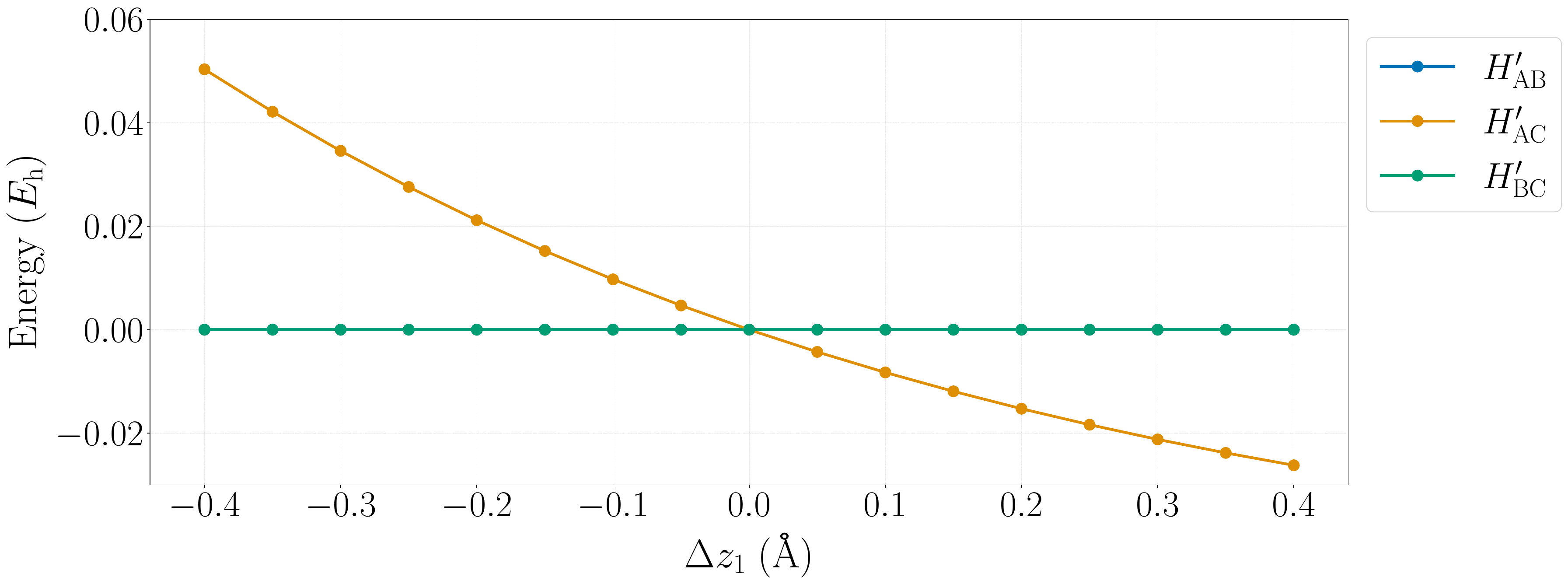}
    \caption{}
  \end{subfigure}
   \caption{(a) Off-diagonal Hamiltonian matrix elements in the basis of the ensemble-VQE states \(\lbrace \ket{\Phi_I(\boldsymbol{t}^*)} \rbrace _{I \in \lbrace \mathrm {A},\mathrm {B},\mathrm {C}\rbrace }\) (see Figs.~\ref{fig:diabatization_block_diagonalization_minus_zero_one_c3v} and~\ref{fig:o_matrix_minus_zero_one_c3v}). (b) Off-diagonal Hamiltonian matrix elements in the basis of the ensemble-VQE states \( \lbrace \ket{\Phi^{\prime}_I(\boldsymbol{t}^*, \theta_\star, \phi_\star, \psi_\star)} \rbrace _{I \in \lbrace \mathrm {A},\mathrm {B},\mathrm {C}\rbrace }  \) (see Figs.~\ref{fig:energies_optimally_diabatic_c3v} and~\ref{fig:o_star_matrix_minus_zero_one_c3v}). Diabatic orbitals are used.}
 \label{fig:off_diagonal_hamiltonian_matrix_c3v_optimally_diabatic}
\end{figure}

As observed Fig.~\ref{fig:off_diagonal_hamiltonian_matrix_c1_optimally_diabatic} and Fig.~\ref{fig:off_diagonal_hamiltonian_matrix_cs_optimally_diabatic},
the off-diagonal Hamiltonian matrix elements before the change of basis are
not entirely smooth in Fig.~\ref{fig:off_diagonal_hamiltonian_matrix_c3v_optimally_diabatic}. The three off-diagonal Hamiltonian matrix elements vanish at \(\Delta z_1 = 0.0~\textrm{\AA}\) which is consistent with the degeneracy of the energies of the three states observed Fig.~\ref{fig:diabatization_block_diagonalization_minus_zero_one_c3v} and expected from symmetry considerations. Moreover, after the change of basis, two out of the three Hamiltonian matrix elements become zero, in agreement with symmetry considerations.


\begin{thebibliography}{100}

\bibitem{johnson_local_2015}
Philip J.~M. Johnson, Alexei Halpin, Takefumi Morizumi, Valentyn~I.
  Prokhorenko, Oliver~P. Ernst, and R.~J.~Dwayne Miller.
\newblock Local vibrational coherences drive the primary photochemistry of
  vision.
\newblock {\em Nature Chemistry}, 7(12):980--986, December 2015.

\bibitem{improta_quantum_2016}
Roberto Improta, Fabrizio Santoro, and Lluís Blancafort.
\newblock Quantum {Mechanical} {Studies} on the {Photophysics} and the
  {Photochemistry} of {Nucleic} {Acids} and {Nucleobases}.
\newblock {\em Chemical Reviews}, 116(6):3540--3593, March 2016.

\bibitem{berkovic_spiropyrans_2000}
Garry Berkovic, Valeri Krongauz, and Victor Weiss.
\newblock Spiropyrans and {Spirooxazines} for {Memories} and {Switches}.
\newblock {\em Chemical Reviews}, 100(5):1741--1754, May 2000.

\bibitem{domcke_conical_2004}
Wolfgang Domcke, David~R Yarkony, and Horst Köppel.
\newblock {\em Conical {Intersections}: {Electronic} {Structure}, {Dynamics}
  and {Spectroscopy}}, volume~15 of {\em Advanced {Series} in {Physical}
  {Chemistry}}.
\newblock WORLD SCIENTIFIC, July 2004.

\bibitem{baer_beyond_2006}
Michael Baer.
\newblock {\em Beyond {Born}–{Oppenheimer}: {Conical} {Intersections} and
  {Electronic} {Nonadiabatic} {Coupling} {Terms}}.
\newblock Wiley, 1 edition, March 2006.

\bibitem{gatti_applications_2017}
Fabien Gatti, Benjamin Lasorne, Hans-Dieter Meyer, and André Nauts.
\newblock {\em Applications of {Quantum} {Dynamics} in {Chemistry}}, volume~98
  of {\em Lecture {Notes} in {Chemistry}}.
\newblock Springer International Publishing, Cham, 2017.

\bibitem{meyer_multi-configurational_1990}
H.-D. Meyer, U.~Manthe, and L.S. Cederbaum.
\newblock The multi-configurational time-dependent {Hartree} approach.
\newblock {\em Chemical Physics Letters}, 165(1):73--78, January 1990.

\bibitem{beck_multiconfiguration_2000}
M~Beck.
\newblock The multiconfiguration time-dependent {Hartree} ({MCTDH}) method: a
  highly efficient algorithm for propagating wavepackets.
\newblock {\em Physics Reports}, 324(1):1--105, January 2000.

\bibitem{worth_quantics_2020}
G.A. Worth.
\newblock Quantics: {A} general purpose package for {Quantum} molecular
  dynamics simulations.
\newblock {\em Computer Physics Communications}, 248:107040, March 2020.

\bibitem{tully_molecular_1990}
John~C. Tully.
\newblock Molecular dynamics with electronic transitions.
\newblock {\em The Journal of Chemical Physics}, 93(2):1061--1071, July 1990.

\bibitem{subotnik_understanding_2016}
Joseph~E. Subotnik, Amber Jain, Brian Landry, Andrew Petit, Wenjun Ouyang, and
  Nicole Bellonzi.
\newblock Understanding the {Surface} {Hopping} {View} of {Electronic}
  {Transitions} and {Decoherence}.
\newblock {\em Annual Review of Physical Chemistry}, 67(1):387--417, May 2016.

\bibitem{barbatti_newton-x_2022}
Mario Barbatti, Mattia Bondanza, Rachel Crespo-Otero, Baptiste Demoulin,
  Pavlo~O. Dral, Giovanni Granucci, Fábris Kossoski, Hans Lischka, Benedetta
  Mennucci, Saikat Mukherjee, Marek Pederzoli, Maurizio Persico, Max
  Pinheiro~Jr, Jiří Pittner, Felix Plasser, Eduarda Sangiogo~Gil, and
  Ljiljana Stojanovic.
\newblock Newton-{X} {Platform}: {New} {Software} {Developments} for {Surface}
  {Hopping} and {Nuclear} {Ensembles}.
\newblock {\em Journal of Chemical Theory and Computation}, 18(11):6851--6865,
  November 2022.

\bibitem{tully_mixed_1998}
John~C. Tully.
\newblock Mixed quantum–classical dynamics.
\newblock {\em Faraday Discussions}, 110:407--419, 1998.

\bibitem{mai_nonadiabatic_2018}
Sebastian Mai, Philipp Marquetand, and Leticia González.
\newblock Nonadiabatic dynamics: {The} {SHARC} approach.
\newblock {\em WIREs Computational Molecular Science}, 8(6):e1370, November
  2018.

\bibitem{crespo-otero_recent_2018}
Rachel Crespo-Otero and Mario Barbatti.
\newblock Recent {Advances} and {Perspectives} on {Nonadiabatic} {Mixed}
  {Quantum}–{Classical} {Dynamics}.
\newblock {\em Chemical Reviews}, 118(15):7026--7068, August 2018.

\bibitem{richings_practical_2015}
Gareth~W. Richings and Graham~A. Worth.
\newblock A {Practical} {Diabatisation} {Scheme} for {Use} with the
  {Direct}-{Dynamics} {Variational} {Multi}-{Configuration} {Gaussian}
  {Method}.
\newblock {\em The Journal of Physical Chemistry A}, 119(50):12457--12470,
  December 2015.

\bibitem{vie21:084302}
Alexandra Viel, David M.~G. Williams, and Wolfgang Eisfeld.
\newblock {Accurate quantum dynamics simulation of the photodetachment spectrum
  of the nitrate anion based on an artificial neural network diabatic potential
  model}.
\newblock {\em J. Chem. Phys.}, 154(8):084302, 02 2021.

\bibitem{zhang_diabatization_2021}
Yang Zhang, Wei Wang, Benjamin Lasorne, Peifeng Su, and Wei Wu.
\newblock Diabatization around {Conical} {Intersections} with a {New}
  {Phase}-{Corrected} {Valence}-{Bond}-{Based} {Compression} {Approach}.
\newblock {\em The Journal of Physical Chemistry Letters}, 12(7):1885--1892,
  February 2021.

\bibitem{shu_diabatic_2022}
Yinan Shu, Zoltan Varga, Siriluk Kanchanakungwankul, Linyao Zhang, and
  Donald~G. Truhlar.
\newblock Diabatic {States} of {Molecules}.
\newblock {\em The Journal of Physical Chemistry A}, 126(7):992--1018, February
  2022.

\bibitem{yalouz_state-averaged_2021}
Saad Yalouz, Bruno Senjean, Jakob Günther, Francesco Buda, Thomas~E O’Brien,
  and Lucas Visscher.
\newblock A state-averaged orbital-optimized hybrid quantum–classical
  algorithm for a democratic description of ground and excited states.
\newblock {\em Quantum Science and Technology}, 6(2):024004, April 2021.

\bibitem{yalouz_analytical_2022}
Saad Yalouz, Emiel Koridon, Bruno Senjean, Benjamin Lasorne, Francesco Buda,
  and Lucas Visscher.
\newblock Analytical {Nonadiabatic} {Couplings} and {Gradients} within the
  {State}-{Averaged} {Orbital}-{Optimized} {Variational} {Quantum}
  {Eigensolver}.
\newblock {\em Journal of Chemical Theory and Computation}, 18(2):776--794,
  February 2022.

\bibitem{beseda2024state}
Martin Beseda, Silvie Illésová, Saad Yalouz, and Bruno Senjean.
\newblock State-averaged orbital-optimized vqe: A quantum algorithm for the
  democratic description of ground and excited electronic states.
\newblock {\em J. Open Source Softw.}, 9(101):6036, 2024.

\bibitem{illesova_transformation-free_2025}
Silvie Illésová, Martin Beseda, Saad Yalouz, Benjamin Lasorne, and Bruno
  Senjean.
\newblock Transformation-{Free} {Generation} of a {Quasi}-{Diabatic}
  {Representation} from the {State}-{Average} {Orbital}-{Optimized}
  {Variational} {Quantum} {Eigensolver}.
\newblock {\em Journal of Chemical Theory and Computation}, 21(11):5457--5480,
  June 2025.

\bibitem{higgott2019variational}
Oscar Higgott, Daochen Wang, and Stephen Brierley.
\newblock Variational quantum computation of excited states.
\newblock {\em Quantum}, 3:156, 2019.

\bibitem{jones2019variational}
Tyson Jones, Suguru Endo, Sam McArdle, Xiao Yuan, and Simon~C Benjamin.
\newblock Variational quantum algorithms for discovering hamiltonian spectra.
\newblock {\em Phys. Rev. A}, 99(6):062304, 2019.

\bibitem{jouzdani2019method}
Pejman Jouzdani, Stefan Bringuier, and Mark Kostuk.
\newblock A method of determining excited-states for quantum computation.
\newblock {\em arXiv:1908.05238}, 2019.

\bibitem{ibe2020calculating}
Yohei Ibe, Yuya~O Nakagawa, Takahiro Yamamoto, Kosuke Mitarai, Qi~Gao, and
  Takao Kobayashi.
\newblock Calculating transition amplitudes by variational quantum
  eigensolvers.
\newblock {\em arXiv:2002.11724}, 2020.

\bibitem{chan2021molecular}
Hans Hon~Sang Chan, Nathan Fitzpatrick, Javier Segarra-Mart{\'\i}, Michael~J
  Bearpark, and David~P Tew.
\newblock Molecular excited state calculations with adaptive wavefunctions on a
  quantum eigensolver emulation: reducing circuit depth and separating spin
  states.
\newblock {\em Physical Chemistry Chemical Physics}, 23(46):26438--26450, 2021.

\bibitem{quiroga2025quantum}
David Quiroga, Jason Han, and Anastasios Kyrillidis.
\newblock Quantum eigengame for excited state calculation.
\newblock {\em arXiv:2503.13644}, 2025.

\bibitem{motta2024subspace}
Mario Motta, William Kirby, Ieva Liepuoniute, Kevin~J Sung, Jeffrey Cohn,
  Antonio Mezzacapo, Katherine Klymko, Nam Nguyen, Nobuyuki Yoshioka, and
  Julia~E Rice.
\newblock Subspace methods for electronic structure simulations on quantum
  computers.
\newblock {\em Electronic Structure}, 6(1):013001, 2024.

\bibitem{ruiz2022accessing}
Edgar~Andres Ruiz~Guzman and Denis Lacroix.
\newblock Accessing ground-state and excited-state energies in a many-body
  system after symmetry restoration using quantum computers.
\newblock {\em Physical Review C}, 105(2):024324, 2022.

\bibitem{cortes2022quantum}
Cristian~L Cortes and Stephen~K Gray.
\newblock Quantum krylov subspace algorithms for ground-and excited-state
  energy estimation.
\newblock {\em Physical Review A}, 105(2):022417, 2022.

\bibitem{de2026new}
Xeno De~Vriendt, Jacob Bringewatt, Nik~O Gjonbalaj, Stefan Ostermann, Davide
  Vodola, Johannes Borregaard, Michael K{\"u}hn, and Susanne~F Yelin.
\newblock A new angle on quantum subspace diagonalization for quantum
  chemistry.
\newblock {\em arXiv:2602.11985}, 2026.

\bibitem{gentinetta2026quantum}
Gian Gentinetta, Friederike Metz, William Kirby, and Giuseppe Carleo.
\newblock Quantum finite temperature lanczos method.
\newblock {\em arXiv:2603.25394}, 2026.

\bibitem{tkachenko2024quantum}
Nikolay~V Tkachenko, Lukasz Cincio, Alexander~I Boldyrev, Sergei Tretiak,
  Pavel~A Dub, and Yu~Zhang.
\newblock Quantum davidson algorithm for excited states.
\newblock {\em Quantum Science and Technology}, 9(3):035012, 2024.

\bibitem{berthusen2024multi}
Noah Berthusen, Faisal Alam, and Yu~Zhang.
\newblock Multi-reference quantum davidson algorithm for quantum dynamics.
\newblock {\em arXiv:2406.08675}, 2024.

\bibitem{motta2020determining}
Mario Motta, Chong Sun, Adrian~TK Tan, Matthew~J O’Rourke, Erika Ye, Austin~J
  Minnich, Fernando~GSL Brand{\~a}o, and Garnet Kin-Lic Chan.
\newblock Determining eigenstates and thermal states on a quantum computer
  using quantum imaginary time evolution.
\newblock {\em Nat. Phys.}, 16(2):205--210, 2020.

\bibitem{ollitrault2020quantum}
Pauline~J Ollitrault, Abhinav Kandala, Chun-Fu Chen, Panagiotis~Kl Barkoutsos,
  Antonio Mezzacapo, Marco Pistoia, Sarah Sheldon, Stefan Woerner, Jay~M
  Gambetta, and Ivano Tavernelli.
\newblock Quantum equation of motion for computing molecular excitation
  energies on a noisy quantum processor.
\newblock {\em Phys. Rev. Res.}, 2(4):043140, 2020.

\bibitem{kim2023two}
Yongbin Kim and Anna~I Krylov.
\newblock Two algorithms for excited-state quantum solvers: Theory and
  application to eom-uccsd.
\newblock {\em The Journal of Physical Chemistry A}, 127(31):6552--6566, 2023.

\bibitem{asthana2023quantum}
Ayush Asthana, Ashutosh Kumar, Vibin Abraham, Harper Grimsley, Yu~Zhang, Lukasz
  Cincio, Sergei Tretiak, Pavel~A Dub, Sophia~E Economou, Edwin Barnes, et~al.
\newblock Quantum self-consistent equation-of-motion method for computing
  molecular excitation energies, ionization potentials, and electron affinities
  on a quantum computer.
\newblock {\em Chemical Science}, 14(9):2405--2418, 2023.

\bibitem{bhatia2026quantum}
Amandeep~Singh Bhatia and Sabre Kais.
\newblock Quantum computing for electronic circular dichroism spectrum
  prediction of chiral molecules.
\newblock {\em arXiv:2602.03710}, 2026.

\bibitem{mcclean2017hybrid}
Jarrod~R. McClean, Mollie~E. Kimchi-Schwartz, Jonathan Carter, and Wibe~A.
  de~Jong.
\newblock Hybrid quantum-classical hierarchy for mitigation of decoherence and
  determination of excited states.
\newblock {\em Phys. Rev. A}, 95(4):042308, 2017.

\bibitem{colless2018computation}
James~I Colless, Vinay~V Ramasesh, Dar Dahlen, Machiel~S Blok,
  ME~Kimchi-Schwartz, JR~McClean, J~Carter, WA~De~Jong, and I~Siddiqi.
\newblock Computation of molecular spectra on a quantum processor with an
  error-resilient algorithm.
\newblock {\em Phys. Rev. X}, 8(1):011021, 2018.

\bibitem{castellanos2024quantum}
Maria~A Castellanos, Mario Motta, and Julia~E Rice.
\newblock Quantum computation of $\pi$→ $\pi$* and n→ $\pi$* excited states
  of aromatic heterocycles.
\newblock {\em Molecular Physics}, 122(7-8):e2282736, 2024.

\bibitem{motta2023quantum}
Mario Motta, Gavin~O Jones, Julia~E Rice, Tanvi~P Gujarati, Rei Sakuma, Ieva
  Liepuoniute, Jeannette~M Garcia, and Yu-ya Ohnishi.
\newblock Quantum chemistry simulation of ground-and excited-state properties
  of the sulfonium cation on a superconducting quantum processor.
\newblock {\em Chemical Science}, 14(11):2915--2927, 2023.

\bibitem{akande2025variational}
Hamzat~A Akande, Alexandre Perrin, Bruno Senjean, and Matthieu Sauban{\`e}re.
\newblock Variational quantum subspace construction via symmetry-preserving
  cost functions.
\newblock {\em Phys. Rev. A}, 112(3):032623, 2025.

\bibitem{klymko2022real}
Katherine Klymko, Carlos Mejuto-Zaera, Stephen~J Cotton, Filip Wudarski,
  Miroslav Urbanek, Diptarka Hait, Martin Head-Gordon, K~Birgitta Whaley,
  Jonathan Moussa, Nathan Wiebe, Wibe~A de~Jong, and Norm~M Tubman.
\newblock Real-time evolution for ultracompact hamiltonian eigenstates on
  quantum hardware.
\newblock {\em PRX Quantum}, 3(2):020323, 2022.

\bibitem{wang2023electronic}
Yuchen Wang and David~A Mazziotti.
\newblock Electronic excited states from a variance-based contracted quantum
  eigensolver.
\newblock {\em Physical Review A}, 108(2):022814, 2023.

\bibitem{wang2024quantum}
Yuchen Wang and David~A Mazziotti.
\newblock Quantum simulation of conical intersections.
\newblock {\em Phys. Chem. Chem. Phys.}, 26(15):11491--11497, 2024.

\bibitem{cadi2024folded}
Lila Cadi~Tazi and Alex~JW Thom.
\newblock Folded spectrum vqe: A quantum computing method for the calculation
  of molecular excited states.
\newblock {\em Journal of Chemical Theory and Computation}, 20(6):2491--2504,
  2024.

\bibitem{endo2020calculation}
Suguru Endo, Iori Kurata, and Yuya~O Nakagawa.
\newblock Calculation of the green's function on near-term quantum computers.
\newblock {\em Phys. Rev. Res.}, 2(3):033281, 2020.

\bibitem{yoshikura2023quantum}
Takahiro Yoshikura, Seiichiro~L Ten-No, and Takashi Tsuchimochi.
\newblock Quantum inverse algorithm via adaptive variational quantum linear
  solver: applications to general eigenstates.
\newblock {\em J. Phys. Chem. A}, 127(31):6577--6592, 2023.

\bibitem{nykanen2024toward}
Anton Nykanen, Leander Thiessen, Elsi-Mari Borrelli, Vijay Krishna, Stefan
  Knecht, and Fabijan Pavosevic.
\newblock Toward accurate calculation of excitation energies on quantum
  computers with $\delta$adapt-vqe: a case study of bodipy derivatives.
\newblock {\em J. Phys. Chem. Lett.}, 15(28):7111--7117, 2024.

\bibitem{tilly2020computation}
Jules Tilly, Glenn Jones, Hongxiang Chen, Leonard Wossnig, and Edward Grant.
\newblock Computation of molecular excited states on ibm quantum computers
  using a discriminative variational quantum eigensolver.
\newblock {\em Physical Review A}, 102(6):062425, 2020.

\bibitem{xie2022orthogonal}
Qing-Xing Xie, Sheng Liu, and Yan Zhao.
\newblock Orthogonal state reduction variational eigensolver for the
  excited-state calculations on quantum computers.
\newblock {\em Journal of Chemical Theory and Computation}, 18(6):3737--3746,
  2022.

\bibitem{shen2023real}
Yizhi Shen, Katherine Klymko, James Sud, David~B Williams-Young, Wibe~A
  de~Jong, and Norm~M Tubman.
\newblock Real-time krylov theory for quantum computing algorithms.
\newblock {\em Quantum}, 7:1066, 2023.

\bibitem{parrish2019quantumfilter}
Robert~M Parrish and Peter~L McMahon.
\newblock Quantum filter diagonalization: Quantum eigendecomposition without
  full quantum phase estimation.
\newblock {\em arXiv:1909.08925}, 2019.

\bibitem{cohn2021quantum}
Jeffrey Cohn, Mario Motta, and Robert~M Parrish.
\newblock Quantum filter diagonalization with compressed double-factorized
  hamiltonians.
\newblock {\em PRX Quantum}, 2(4):040352, 2021.

\bibitem{stroschein2025ground}
Timothy Stroschein, Davide Castaldo, and Markus Reiher.
\newblock Ground and excited-state energies with analytic errors and short time
  evolution on a quantum computer.
\newblock {\em arXiv:2507.15148}, 2025.

\bibitem{tamiya2021calculating}
Shiro Tamiya, Sho Koh, and Yuya~O Nakagawa.
\newblock Calculating nonadiabatic couplings and berry's phase by variational
  quantum eigensolvers.
\newblock {\em Phys. Rev. A}, 3(2):023244, 2021.

\bibitem{omiya2022analytical}
Keita Omiya, Yuya~O Nakagawa, Sho Koh, Wataru Mizukami, Qi~Gao, and Takao
  Kobayashi.
\newblock Analytical energy gradient for state-averaged orbital-optimized
  variational quantum eigensolvers and its application to a photochemical
  reaction.
\newblock {\em Journal of Chemical Theory and Computation}, 18(2):741--748,
  2022.

\bibitem{obrien2022efficient}
Thomas~E O'Brien, Michael Streif, Nicholas~C Rubin, Raffaele Santagati, Yuan
  Su, William~J Huggins, Joshua~J Goings, Nikolaj Moll, Elica Kyoseva, Matthias
  Degroote, et~al.
\newblock Efficient quantum computation of molecular forces and other energy
  gradients.
\newblock {\em Phys. Rev. Res.}, 4(4):043210, 2022.

\bibitem{koridon2024hybrid}
Emiel Koridon, Joana Fraxanet, Alexandre Dauphin, Lucas Visscher, Thomas~E
  O'Brien, and Stefano Polla.
\newblock A hybrid quantum algorithm to detect conical intersections.
\newblock {\em Quantum}, 8:1259, 2024.

\bibitem{hirai2022non}
Hirotoshi Hirai and Sho Koh.
\newblock Non-adiabatic quantum wavepacket dynamics simulation based on
  electronic structure calculations using the variational quantum eigensolver.
\newblock {\em Chem. Phys.}, 556:111460, 2022.

\bibitem{gil2025sharc}
Eduarda~Sangiogo Gil, Markus Oppel, Jakob~S Kottmann, and Leticia Gonz{\'a}lez.
\newblock Sharc meets tequila: mixed quantum-classical dynamics on a quantum
  computer using a hybrid quantum-classical algorithm.
\newblock {\em Chem. Sci.}, 16(2):596--609, 2025.

\bibitem{gandon2024nonadiabatic}
Anthony Gandon, Alberto Baiardi, Pauline Ollitrault, and Ivano Tavernelli.
\newblock Nonadiabatic molecular dynamics with fermionic subspace-expansion
  algorithms on quantum computers.
\newblock {\em J. Chem. Theory Comput.}, 20(14):5951--5963, 2024.

\bibitem{li2026efficient}
Tianyi Li, Yumeng Zeng, Qiming Ding, Zixuan Huo, Xiaosi Xu, Jiajun Ren,
  Diandong Tang, Xiaoxia Cai, and Xiao Yuan.
\newblock Efficient quantum simulation of non-adiabatic molecular dynamics with
  precise electronic structure.
\newblock {\em Digital Discovery}, 2026.

\bibitem{sangiogo2025nonadiabatic}
Eduarda Sangiogo-Gil and Leticia Gonz{\'a}lez.
\newblock Nonadiabatic molecular dynamics on quantum computers: challenges and
  opportunities.
\newblock {\em Pure and Applied Chemistry}, 97(11):1647--1665, 2025.

\bibitem{bultrini2023mixed}
Daniel Bultrini and Oriol Vendrell.
\newblock Mixed quantum-classical dynamics for near term quantum computers.
\newblock {\em Communications Physics}, 6(1):328, 2023.

\bibitem{helgaker_molecular_2000}
Trygve Helgaker, Poul Jørgensen, and Jeppe Olsen.
\newblock {\em Molecular {Electronic}‐{Structure} {Theory}}.
\newblock Wiley, 1 edition, August 2000.

\bibitem{preskill_quantum_2018}
John Preskill.
\newblock Quantum {Computing} in the {NISQ} era and beyond.
\newblock {\em Quantum}, 2:79, August 2018.

\bibitem{aspuru-guzik_simulated_2005}
Alán Aspuru-Guzik, Anthony~D. Dutoi, Peter~J. Love, and Martin Head-Gordon.
\newblock Simulated {Quantum} {Computation} of {Molecular} {Energies}.
\newblock {\em Science}, 309(5741):1704--1707, September 2005.

\bibitem{peruzzo_variational_2014}
Alberto Peruzzo, Jarrod McClean, Peter Shadbolt, Man-Hong Yung, Xiao-Qi Zhou,
  Peter~J. Love, Alán Aspuru-Guzik, and Jeremy~L. O’Brien.
\newblock A variational eigenvalue solver on a photonic quantum processor.
\newblock {\em Nature Communications}, 5(1):4213, July 2014.

\bibitem{kandala_hardware-efficient_2017}
Abhinav Kandala, Antonio Mezzacapo, Kristan Temme, Maika Takita, Markus Brink,
  Jerry~M. Chow, and Jay~M. Gambetta.
\newblock Hardware-efficient variational quantum eigensolver for small
  molecules and quantum magnets.
\newblock {\em Nature}, 549(7671):242--246, September 2017.

\bibitem{bittel2021training}
Lennart Bittel and Martin Kliesch.
\newblock Training variational quantum algorithms is np-hard.
\newblock {\em Phys. Rev. Lett.}, 127(12):120502, 2021.

\bibitem{larocca2025barren}
Martin Larocca, Supanut Thanasilp, Samson Wang, Kunal Sharma, Jacob Biamonte,
  Patrick~J Coles, Lukasz Cincio, Jarrod~R McClean, Zo{\"e} Holmes, and Marco
  Cerezo.
\newblock Barren plateaus in variational quantum computing.
\newblock {\em Nat. Rev. Phys.}, 7(4):174--189, 2025.

\bibitem{cerezo2025does}
Marco Cerezo, Martin Larocca, Diego Garc{\'\i}a-Mart{\'\i}n, Nelson~L Diaz,
  Paolo Braccia, Enrico Fontana, Manuel~S Rudolph, Pablo Bermejo, Aroosa Ijaz,
  Supanut Thanasilp, et~al.
\newblock Does provable absence of barren plateaus imply classical
  simulability?
\newblock {\em Nat. Commun.}, 16(1):7907, 2025.

\bibitem{nakanishi_subspace-search_2019}
Ken~M. Nakanishi, Kosuke Mitarai, and Keisuke Fujii.
\newblock Subspace-search variational quantum eigensolver for excited states.
\newblock {\em Physical Review Research}, 1(3):033062, October 2019.

\bibitem{jordan1928p}
P.~Jordan and E.~Wigner.
\newblock {\"U}ber das paulische {\"a}quivalenzverbot.
\newblock {\em Z. Phys.}, 47:631, 1928.

\bibitem{lee_generalized_2019}
Joonho Lee, William~J. Huggins, Martin Head-Gordon, and K.~Birgitta Whaley.
\newblock Generalized {Unitary} {Coupled} {Cluster} {Wave} functions for
  {Quantum} {Computation}.
\newblock {\em Journal of Chemical Theory and Computation}, 15(1):311--324,
  January 2019.

\bibitem{gross_rayleigh-ritz_1988}
E.~K.~U. Gross, L.~N. Oliveira, and W.~Kohn.
\newblock Rayleigh-{Ritz} variational principle for ensembles of fractionally
  occupied states.
\newblock {\em Physical Review A}, 37(8):2805--2808, April 1988.

\bibitem{A_K_Theophilou_1979}
A~K Theophilou.
\newblock The energy density functional formalism for excited states.
\newblock {\em Journal of Physics C: Solid State Physics}, 12(24):5419, dec
  1979.

\bibitem{tsuchimochi_spin-projection_2020}
Takashi Tsuchimochi, Yuto Mori, and Seiichiro~L. Ten-no.
\newblock Spin-projection for quantum computation: {A} low-depth approach to
  strong correlation.
\newblock {\em Physical Review Research}, 2(4):043142, October 2020.

\bibitem{ryabinkin_constrained_2019}
Ilya~G. Ryabinkin, Scott~N. Genin, and Artur~F. Izmaylov.
\newblock Constrained {Variational} {Quantum} {Eigensolver}: {Quantum}
  {Computer} {Search} {Engine} in the {Fock} {Space}.
\newblock {\em Journal of Chemical Theory and Computation}, 15(1):249--255,
  January 2019.

\bibitem{Nocedal2006Numerical}
Jorge Nocedal and Stephen~J. Wright.
\newblock {\em {Numerical Optimization}}.
\newblock Springer Series in Operations Research and Financial Engineering.
  Springer, New York, 2 edition, 2006.

\bibitem{kraft1988slsqp}
Dieter Kraft.
\newblock A software package for sequential quadratic programming.
\newblock Technical Report DFVLR-FB 88-28, Institut für Dynamik der
  Flugsysteme, DFVLR, Oberpfaffenhofen, July 1988.
\newblock Technical Report.

\bibitem{2020SciPy-NMeth}
Pauli Virtanen, Ralf Gommers, Travis~E. Oliphant, Matt Haberland, Tyler Reddy,
  David Cournapeau, Evgeni Burovski, Pearu Peterson, Warren Weckesser, Jonathan
  Bright, St{\'e}fan~J. {van der Walt}, Matthew Brett, Joshua Wilson, K.~Jarrod
  Millman, Nikolay Mayorov, Andrew R.~J. Nelson, Eric Jones, Robert Kern, Eric
  Larson, C~J Carey, {\.I}lhan Polat, Yu~Feng, Eric~W. Moore, Jake
  {VanderPlas}, Denis Laxalde, Josef Perktold, Robert Cimrman, Ian Henriksen,
  E.~A. Quintero, Charles~R. Harris, Anne~M. Archibald, Ant{\^o}nio~H. Ribeiro,
  Fabian Pedregosa, Paul {van Mulbregt}, and {SciPy 1.0 Contributors}.
\newblock {{SciPy} 1.0: Fundamental Algorithms for Scientific Computing in
  Python}.
\newblock {\em Nature Methods}, 17:261--272, 2020.

\bibitem{golub_matrix_2013}
Gene~H. Golub and Charles~F. Van~Loan.
\newblock {\em Matrix computations}.
\newblock Johns {Hopkins} studies in the mathematical sciences. Johns Hopkins
  Univ. Press, Baltimore, MD, 4. ed edition, 2013.

\bibitem{lasorne_use_2014}
Benjamin Lasorne.
\newblock On the {Use} of {Lie} {Group} {Homomorphisms} for {Treating}
  {Similarity} {Transformations} in {Nonadiabatic} {Photochemistry}.
\newblock {\em Advances in Mathematical Physics}, 2014:1--14, 2014.

\bibitem{bernstein_matrix_2009}
Dennis~S. Bernstein.
\newblock {\em Matrix mathematics: theory, facts, and formulas}.
\newblock Princeton university press, 2nd ed edition, 2009.

\bibitem{piovan_coordinate-free_2012}
Giulia Piovan and Francesco Bullo.
\newblock On {Coordinate}-{Free} {Rotation} {Decomposition}: {Euler} {Angles}
  {About} {Arbitrary} {Axes}.
\newblock {\em IEEE Transactions on Robotics}, 28(3):728--733, June 2012.

\bibitem{hong2023refining}
Cheng-Lin Hong, Luis Colmenarez, Lexin Ding, Carlos~L Benavides-Riveros, and
  Christian Schilling.
\newblock Refining the weighted subspace-search variational quantum
  eigensolver: compression of ansatze into a single pure state and optimization
  of weights.
\newblock {\em arXiv:2306.11844}, 2023.

\bibitem{ding2024ground}
Lexin Ding, Cheng-Lin Hong, and Christian Schilling.
\newblock Ground and excited states from ensemble variational principles.
\newblock {\em Quantum}, 8:1525, 2024.

\bibitem{rajamani2025equi}
Akilan Rajamani, Martin Beseda, Benjamin Lasorne, and Bruno Senjean.
\newblock How an equi-ensemble description systematically outperforms the
  weighted-ensemble variational quantum eigensolver.
\newblock {\em arXiv:2509.17982}, 2025.

\bibitem{werner_adiabatic_1988}
Hans-Joachim Werner, Bernd Follmeg, and Millard~H. Alexander.
\newblock Adiabatic and diabatic potential energy surfaces for collisions of
  {CN}( \textit{{X}} {2$\Sigma$}+, \textit{{A}} {2$\Pi$}) with {He}.
\newblock {\em The Journal of Chemical Physics}, 89(5):3139--3151, September
  1988.

\bibitem{simah_photodissociation_1999}
David Simah, Bernd Hartke, and Hans-Joachim Werner.
\newblock Photodissociation dynamics of {H2S} on new coupled \textit{ab initio}
  potential energy surfaces.
\newblock {\em The Journal of Chemical Physics}, 111(10):4523--4534, September
  1999.

\bibitem{pacher1988approximately}
T~Pacher, LS~Cederbaum, and H~K{\"o}ppel.
\newblock Approximately diabatic states from block diagonalization of the
  electronic hamiltonian.
\newblock {\em J. Chem. Phys.}, 89(12):7367--7381, 1988.

\bibitem{cederbaum1989block}
LS~Cederbaum, J~Schirmer, and H-D Meyer.
\newblock Block diagonalisation of hermitian matrices.
\newblock {\em Journal of physics A: Mathematical and General},
  22(13):2427--2439, 1989.

\bibitem{richings2020a}
Gareth~W. Richings and Scott Habershon.
\newblock {A new diabatization scheme for direct quantum dynamics: Procrustes
  diabatization}.
\newblock {\em J. Chem. Phys.}, 152(15):154108, 04 2020.

\bibitem{kirchner_first_1984}
Nicholas~J. Kirchner, James~R. Gilbert, and Michael~T. Bowers.
\newblock The first experimental observation of stable {H4}+ ions.
\newblock {\em Chemical Physics Letters}, 106(1-2):7--12, April 1984.

\bibitem{jiang_geometrical_1998}
G.~Jiang, H.Y. Wang, and Z.H. Zhu.
\newblock Geometrical configurations of {H4}+ and the {Jahn}–{Teller} effect.
\newblock {\em Chemical Physics Letters}, 284(3-4):267--272, February 1998.

\bibitem{turney_psi4_2012}
Justin~M. Turney, Andrew~C. Simmonett, Robert~M. Parrish, Edward~G. Hohenstein,
  Francesco~A. Evangelista, Justin~T. Fermann, Benjamin~J. Mintz, Lori~A.
  Burns, Jeremiah~J. Wilke, Micah~L. Abrams, Nicholas~J. Russ, Matthew~L.
  Leininger, Curtis~L. Janssen, Edward~T. Seidl, Wesley~D. Allen, Henry~F.
  Schaefer, Rollin~A. King, Edward~F. Valeev, C.~David Sherrill, and T.~Daniel
  Crawford.
\newblock Psi4: an open‐source \textit{ab initio} electronic structure
  program.
\newblock {\em WIREs Computational Molecular Science}, 2(4):556--565, July
  2012.

\bibitem{SAOOVQE}
M.~Beseda, S.~Illésová, S.~Yalouz, and B.~Senjean.
\newblock State-averaged orbital-optimized vqe: A quantum algorithm for the
  democratic description of ground and excited electronic states (1.2.0), 2025.

\bibitem{nielsen_quantum_2012}
Michael~A. Nielsen and Isaac~L. Chuang.
\newblock {\em Quantum {Computation} and {Quantum} {Information}: 10th
  {Anniversary} {Edition}}.
\newblock Cambridge University Press, 1 edition, June 2012.

\bibitem{laporte_numerical_2003}
Emmanuel Laporte and Patrick Le~Tallec.
\newblock {\em Numerical {Methods} in {Sensitivity} {Analysis} and {Shape}
  {Optimization}}.
\newblock Modeling and {Simulation} in {Science}, {Engineering} and
  {Technology}. Birkhäuser Boston, Boston, MA, 2003.

\end{thebibliography}
\end{document}